\pdfoutput=1
\documentclass{jfm}
\usepackage[utf8x]{inputenc}
\usepackage[dvips]{graphicx}
\usepackage{float}
\usepackage{amsmath}
\usepackage{amssymb}
\usepackage{natbib}
\usepackage{bm}
\usepackage{color}
\usepackage{soul}
\usepackage{multirow}
\usepackage[breaklinks=true]{hyperref}
\usepackage{breakcites}
\hypersetup{
    colorlinks=true,
    linkcolor=black,
    citecolor=black,
    filecolor=black,
    urlcolor=black,
}

\ifCUPmtlplainloaded \else
  \checkfont{eurm10}
  \iffontfound
    \IfFileExists{upmath.sty}
      {\typeout{^^JFound AMS Euler Roman fonts on the system,
                   using the 'upmath' package.^^J}%
       \usepackage{upmath}}
      {\typeout{^^JFound AMS Euler Roman fonts on the system, but you
                   dont seem to have the}%
       \typeout{'upmath' package installed. JFM.cls can take advantage
                 of these fonts,^^Jif you use 'upmath' package.^^J}%
      }
  \else
  \fi
\fi

\ifCUPmtlplainloaded \else
  \checkfont{msam10}
  \iffontfound
    \IfFileExists{amssymb.sty}
      {\typeout{^^JFound AMS Symbol fonts on the system, using the
                'amssymb' package.^^J}%
       \usepackage{amssymb}%
       \let\le=\leqslant  \let\leq=\leqslant
         \let\geq=\geqslant
      }{}
  \fi
\fi

\ifCUPmtlplainloaded \else
  \IfFileExists{amsbsy.sty}
    {\typeout{^^JFound the 'amsbsy' package on the system, using it.^^J}%
     \usepackage{amsbsy}}
    {\providecommand\boldsymbol[1]{\mbox{\boldmath $##1$}}}
\fi

\newcommand{\figs}{./figs}

\title[Reynolds-number effects on inertial particle dynamics, Part II.]
      {The effect of Reynolds number on inertial particle dynamics in isotropic turbulence.
       Part II: Simulations with gravitational effects.}

\author[P. J. Ireland et al.]%
{Peter J. Ireland,\ns Andrew D. Bragg%
  \thanks{Present address:  Now with the Applied Mathematics \& Plasma Physics Group, Los Alamos National Laboratory, Los Alamos, NM  87545, USA.},\ns and Lance R. Collins%
  \thanks{Email address for correspondence: lc246@cornell.edu}}

\affiliation{Sibley School of Mechanical and Aerospace Engineering, Cornell University, Ithaca, NY  14853, USA\\[\affilskip]
International Collaboration for Turbulence Research}

\pubyear{2014}
\volume{}
\pagerange{}
\date{?; revised ?; accepted ?. - To be entered by editorial office}

\begin{document}

\maketitle

\begin{abstract}
In Part I of this study \citep{ireland15a}, we analyzed the motion of inertial particles in isotropic turbulence
in the absence of gravity using direct numerical simulation (DNS).
Here, in Part II, we introduce gravity and study its effect on single-particle and 
particle-pair dynamics over a wide range of flow Reynolds numbers, Froude numbers, and particle
Stokes numbers. The overall goal of this study is to explore the mechanisms affecting
particle collisions, and to thereby improve our understanding of droplet interactions in atmospheric clouds.
We find that the dynamics of heavy particles falling under gravity can be artificially
influenced by the finite domain size and the periodic boundary conditions,
and we therefore perform our simulations on larger domains to reduce these effects.
We first study single-particle statistics which influence
the relative positions and velocities of inertial particles.
We see that gravity causes particles to sample the flow more uniformly
and reduces the time particles can spend interacting with the underlying turbulence.
We also find that gravity tends to increase inertial particle
accelerations, and we introduce a model to explain that effect.

We then analyze the particle relative velocities and radial distribution functions (RDFs),
which are generally seen to be independent of Reynolds number for low and moderate 
Kolmogorov-scale Stokes numbers $St$.
We see that gravity causes particle relative velocities
to decrease by reducing the degree of preferential sampling and the importance of path-history interactions,
and that the relative velocities have higher scaling exponents with gravity.
We observe that gravity has a non-trivial effect on clustering, acting to decrease clustering at low $St$ 
and to increase clustering at high $St$.  By considering the effect of gravity on the clustering 
mechanisms described in the theory of \cite{zaichik09}, we provide an explanation for this 
non-trivial effect of gravity.  We also 
show that when the effects of gravity are accounted for in
the theory of \cite{zaichik09}, the results compare favorably with DNS.
The relative velocities and RDFs exhibit considerable anisotropy at small separations,
and this anisotropy is quantified using spherical harmonic functions.
We use the relative velocities and the RDFs to compute the particle collision kernels, 
and find that the collision kernel remains as it was for the case without gravity, namely
nearly independent of Reynolds number 
for low and moderate $St$. We conclude by discussing practical implications 
of the results for the cloud physics and turbulence communities and by suggesting possible 
avenues for future research.
\end{abstract}

\section{Introduction}
\label{sec:intro}
This is the second part of a two-part paper in which we consider the Reynolds-number dependence of
inertial particle statistics using direct numerical simulations (DNS)
of homogeneous, isotropic turbulence (HIT).
In Part I of this study \citep{ireland15a}, 
we used high-Reynolds-number DNS
to explore the motion of inertial particles in the absence of gravity.
We saw that particles with weak inertia 
preferentially sampled certain regions of the turbulence 
\citep[a phenomenon known as `preferential sampling,' see][]{maxey87,squires91a,eaton94}.
By exploring the specific regions of the flow contributing to this preferential sampling
and using the theory in \cite{chun05},
we were able to understand and model the resulting trends.
Particles with higher inertia had a modulated response
to the underlying turbulence (a phenomenon known as `inertial filtering'), 
decreasing the particle kinetic energies and accelerations, and we found our DNS data for 
these quantities to be in excellent agreement the models of \cite{abrahamson75} and \cite{zaichik08}.
Such particles also exhibited increased relative velocities and `caustics' \citep{wilkinson05,wilkinson06}, 
which occur as a result of the particles' memory of their path-history interactions with the turbulence.

A primary goal of our analysis in Part I was to determine the effect of Reynolds number on particle collision 
rates. It is well-known that droplet growth and precipitation
in warm, cumulus clouds occurs faster than 
current microphysical models can predict, and the discrepancies
are generally linked to turbulence effects \citep[see][]{shaw03,devenish12,grabowski13}.
We explored droplet motions in turbulence at the highest Reynolds numbers to date,
and used the results to extrapolate to Reynolds numbers representative of those in atmospheric clouds.
A secondary motivation was to understand the extent to which protoplanetary nebula formation, 
which depends on the collision and coalescence of small dust grains, is affected
by turbulence. (A more complete explanation of the physical processes involved
in cloud and protoplanetary nebula formation is provided in Part I.)

To determine the collision rates, we computed particle radial distributions and relative velocities 
and used the theory of \cite{sundaram4} to calculate the kinematic collision kernel from these
quantities. We observed that the collision rates of weakly inertial particles 
(such as those that would be present in the early stage of cloud formation)
are almost entirely insensitive to the flow Reynolds number. This suggests that particle collisions
are determined by the small-scale turbulence, and that DNS at low Reynolds numbers
should capture the essential physics responsible for particle collisions in highly turbulent
clouds.

One major simplification in Part I was the neglect of gravitational forces on the particles.
However, as noted in Part I, gravity is not negligible for many 
particle-laden environmental flows. For example, in warm cumulus clouds,
the gravitational settling speeds of droplets may be an order of magnitude larger
than the Kolmogorov velocity, suggesting that such droplets fall quickly 
through the turbulence and may therefore have a substantially 
modified response to the underlying flow \citep{ayala08a}.
Therefore, in Part II of our study, we systematically explore the effect of gravity on inertial particle statistics.

The inclusion of gravity, while superficially trivial, adds richness and complexity to both 
the physical interpretation of the results as well as the numerics. 
Physically, the dimensionality of the parameter space is augmented by one to quantify 
the strength of gravity relative to the turbulence parameters. 
Here, we introduce a Froude number, $Fr$, which is defined as the ratio of the r.m.s. of the fluid 
acceleration to the gravitational constant \citep[e.g., see][]{bec14}. Even for terrestrial clouds 
(for which gravity may be regarded as a constant), the turbulent acceleration may vary by orders 
of magnitude \citep[see][]{prupp97}, leading to significant variations in $Fr$. To quantify the role 
of gravity, therefore, we must analyze particle statistics over a wide range of $St$, $R_\lambda$, 
and $Fr$. In addition to the expanded parameter space, the gravitational vector causes a reduction 
in symmetry for the particle field, from isotropic to axisymmetric, with the axis aligned with the 
gravitational vector \citep{ayala08a,woittiez09,bec14,gustavsson14}. 
We will investigate the degree of anisotropy as a function of the system parameters. 
Finally, there is a numerical concern that relates to the settling speed of the particles. 
With periodic boundary conditions, particles exiting the bottom of the box are reintroduced at the top. 
If the time required to traverse the box becomes smaller than the correlation time of the turbulence, 
the reflected particles' preexisting correlation with the flow will introduce an unphysical effect 
\citep{woittiez09}. This effect can be mitigated by increasing the domain size. 
We will explore different domain sizes to quantify this effect (see Appendix~\ref{sec:periodicity} 
for a complete discussion).

Previous DNS of inertial particles subjected to gravity have primarily focused on how
turbulence alters the terminal velocity 
\citep{wang93,yang98,yang05,ireland12,bec14,good14}
and the collision frequency \citep{franklin07,ayala08a,woittiez09,onishi09,rosa13,bec14}.
Our work extends those studies by performing simulations on larger domains, over
a wider range of Reynolds numbers, and using more particle classes. We also consider the effect of gravity on 
additional particle statistics (such as fluid velocity gradients, Lagrangian timescales, and particle accelerations)
to provide new insight into the physical mechanisms responsible for particle collisions,
and we specifically address the influence of anisotropy on these and other statistics.
To understand the trends in many of these statistics, we introduce theoretical models
and compare them with the DNS data.

The organization of this paper is as follows. 
In \textsection \ref{sec:dns}, we discuss the numerical methods and parameters for our simulations. 
Single-particle statistics are presented in \textsection \ref{sec:single_particle}.
Within this section, we analyze particle velocity gradients (\textsection \ref{sec:topology}), 
mean settling speeds (\textsection \ref{sec:settling}), and accelerations (\textsection \ref{sec:accelerations}).
These statistics provide insight into
the mechanisms contributing to the clustering and near-contact motion of heavy, inertial particles.
To further understand these and other relevant collision mechanisms,
we directly compute particle-pair statistics in \textsection \ref{sec:two_particle}. 
We consider the particle relative velocities in \textsection \ref{sec:relative_velocity} and
radial distribution functions in \textsection \ref{sec:clustering}, and use
these data to compute the particle kinematic collision kernel in \textsection \ref{sec:collision_kernel}.
We conclude in \textsection \ref{sec:conclusions}
by summarizing our findings and suggesting some practical implications for the cloud physics 
and turbulence communities.
\section{Overview of simulations}
\label{sec:dns}

\subsection{Fluid phase}
\label{sec:fluid_phase}

As in Part I, we perform pseudospectral DNS of HIT on a cubic, tri-periodic domain of length $\mathcal{L}$ with $N^3$ grid points, 
solving the continuity and Navier-Stokes equations for an incompressible flow
\begin{equation}
 \nabla \cdot \bm{u} = 0 \mathrm{,}
 \label{eq:continuity}
\end{equation}
\begin{equation}
 \frac{\partial \bm{u}}{\partial t} + \boldsymbol{\omega} \times \bm{u} 
   + \nabla \left( \frac{p}{\rho_f} + \frac{u^2}{2} \right) = \nu \nabla^2 \bm{u} + \bm{f} \mathrm{,}
 \label{eq:momentum}
\end{equation}
where $\bm{u}$ is the fluid velocity, $\boldsymbol{\omega}$ is the vorticity,
$p$ is the pressure, $\rho_f$ is the fluid density, $\nu$ is the kinematic viscosity, 
and $\bm{f}$ is a large-scale forcing term that is added to make the 
flow field statistically stationary. In these simulations, deterministic forcing
is applied to wavenumbers with magnitude $\kappa=\sqrt{2}$.
Note that the gravity term in the Navier-Stokes equation is not shown because it
is precisely canceled by the mean pressure gradient, and so it has no dynamical 
consequence on the turbulent flow field.
More details of the numerical methods can be found in \cite{ireland13}.

In Part I, $\mathcal{L}=2\pi$ for all the simulations performed.
To reduce artificial periodicity effects in these simulations with gravity, 
the domain lengths are here extended to $\mathcal{L}=16\pi$ (for $R_\lambda = 90$), 
$\mathcal{L}=8\pi$ (for $R_\lambda = 147$), 
and $\mathcal{L}=4\pi$ (for $R_\lambda = 230$).
($R_\lambda \equiv 2 k \sqrt{5/\left(3 \nu \epsilon \right)}$ denotes the Taylor-scale Reynolds number,
where $k$ is the kinetic energy and $\epsilon$ is the turbulent energy dissipation rate.)
The grid spacing is kept constant as the domain size is increased, 
and so the small-scale resolution $\kappa_\mathrm{max} \eta$
is approximately constant between the different domain sizes
(where $\kappa_\mathrm{max}$ is the maximum resolved wavenumber 
and $\eta \equiv (\nu^3/\epsilon)^{1/4}$ is the Kolmogorov length scale).
In increasing the domain size, we also keep the viscosity and forcing parameters 
the same, and thus both small-scale and large-scale flow parameters
are held approximately constant.
At the two highest Reynolds numbers, we expect periodicity effects to be minimal,
and the domain sizes are the same ($\mathcal{L}=2\pi$) both with and without gravity.
Refer to Appendix~\ref{sec:periodicity}
for a detailed examination of the effect of the domain size on fluid and particle
statistics.

The simulation parameters are given in table~\ref{tab:parameters}.
The simulation results from this study will be frequently compared to those from Part I.
These comparisons are justified, since in all cases, the parameters are very close to those used in Part I,
and based on the results in Appendix~\ref{sec:periodicity},
we can safely assume 
that the differences in the particle statistics between these simulations and those in Part I
are due entirely to gravitational effects and not to any differences in the underlying flow fields.
In showing comparisons between the present data and those from Part I, 
we will refer to both fields by the Reynolds numbers from table~\ref{tab:parameters}, keeping
in mind that the corresponding Reynolds numbers from Part I may differ slightly (but by no more than 5\%).

\begin{table}
 \centering
 \caption{Simulation parameters for the DNS study. All dimensional parameters are in arbitrary units, 
 and all quantities are defined in the text in \textsection 
 \ref{sec:fluid_phase} and \textsection \ref{sec:particle_phase}.}
 \label{tab:parameters}
 \begin{tabular}{ l  l  l  l  l  l  l }
 Simulation& I & II & III & IV & V & IIIb \\
 $R_\lambda$ & 90 & 147 & 230 & 398 & 597 & 227\\
 $\mathcal{L}$ & $16 \pi$ & $8 \pi$ & $4 \pi$ & $2\pi$ & $2\pi$ & $2\pi$ \\
 $\nu$ & 0.005 & 0.002 & 0.0008289 & 0.0003 & 0.00013 & 0.0008289\\
 $\epsilon$ & 0.257 & 0.244 & 0.239 & 0.223 & 0.228 & 0.246 \\
 $\ell$ & 1.47 & 1.44 & 1.49 & 1.45 & 1.43 & 1.43 \\
 $\ell/\eta$ & 55.6 & 107 & 213 & 436 & 812 & 206 \\
 $u'$ & 0.912 & 0.914 & 0.914 & 0.915 & 0.915 & 0.915\\
 $u'/u_\eta$ & 4.82 & 6.15 & 7.70 & 10.1 & 12.4 & 7.65 \\
 $\kappa_\mathrm{max} \eta$ & 1.61 & 1.63 & 1.68 & 1.60 & 1.70 & 1.67\\
 $N$ & 1024 & 1024 & 1024 & 1024 & 2048 & 512 \\
 \end{tabular}
\end{table}

To perform a more complete parametric study of the effects of inertia and gravity on particle statistics,
we also conducted a simulation with similar flow parameters to simulation III, but with a smaller domain size
(due to computational limitations). The parameters
for this simulation (referred to as IIIb) are also given in table~\ref{tab:parameters}.

\subsection{Particle phase}
\label{sec:particle_phase}

We simulate the motion of spherical particles with finite inertia and gravitational forces. To model the dynamics of inertial particles, 
we make the following simplifying assumptions. The particles are assumed to be small ($d/\eta \ll 1$, where $d$ is the particle diameter) 
and dense ($\rho_p/\rho_f \gg 1$, where $\rho_p$ is the particle density), and subject to only linear drag forces. 
The last assumption is reasonable when the particle Reynolds number $Re_p \equiv \lvert \bm{u}^p(t)-\bm{v}^p(t) \rvert / \nu < 0.5$, 
where $\bm{u}^p(t)\equiv\bm{u}(\bm{x}^p(t),t)$ denotes the undisturbed fluid velocity at the particle position $\bm{x}^p$, 
and $\bm{v}^p$ denotes the velocity of the particle  \citep{elghobashi92}. 
(As in Part I, we use the superscript $p$ on $\bm{x}$, $\bm{u}$, and $\bm{v}$ to denote time-dependent, 
Lagrangian variables defined along particle trajectories. Phase-space positions and velocities will be denoted without the superscript $p$.)
In a recent study, \cite{good14} found that linear-drag models yield settling speeds 
that are inconsistent with experiments at larger settling velocities, 
demonstrating the breakdown of the linear-drag model for $St\gtrsim 3$. 
However, simply introducing a nonlinear drag coefficient does not accurately 
capture the nonlinear drag effects in a time-dependent flow like turbulence. 
A full wake-resolving, nonlinear model would be required, but is beyond the scope of this study. 
Despite the need for caution in the interpretation of the higher-Stokes simulations, 
linear drag yields the correct qualitative trends, thus making the physical arguments and discussions 
presented in this paper valid even if a more accurate nonlinear model were introduced. 
Moreover, the added value of using linear drag is that the majority of theoretical models make the same assumption, 
and so this facilitates comparisons between DNS and theory.

Under these assumptions, the governing equations for the inertial particles are \citep{maxey83}
\begin{equation}
 \frac{d^2 \bm{x}^p}{dt^2} = \frac{d \bm{v}^p}{dt} = 
 \frac{\bm{u}\left(\bm{x}^p(t),t\right)-\bm{v}^p(t)}{\tau_p} + \boldsymbol{\mathfrak{g}} \mathrm{,}
 \label{eq:maxey_riley}
\end{equation}
where $\boldsymbol{\mathfrak{g}} = (0,0,-\mathfrak{g})$ is the gravitational acceleration vector and $\tau_p\equiv\rho_p d^2/(18\rho_f\nu)$ 
is the response time of the particle. To compute $\bm{u}(\bm{x}^p(t),t)$, 
we employ an eight-point B-spline interpolation from the Eulerian grid \citep{vanhinsberg12}. 
As in Part I, we begin computing particle statistics once the particle distributions and velocities 
become statistically stationary and independent of their initial conditions. 
For a subset $N_\mathrm{tracked}$ of the total number of particles in each class $N_{p}$, we store particle positions, 
velocities, and velocity gradients every $0.1 \tau_\eta$ for a duration of about $100 \tau_\eta$.
These data are used to compute Lagrangian correlations, accelerations, and timescales.

The Stokes number $St\equiv\tau_p/\tau_\eta$ is a non-dimensional measure of the particle inertia, 
where $\tau_\eta\equiv (\nu/\epsilon)^{1/2}$ is the Kolmogorov timescale. 
Gravity can be parameterized in a number of ways \citep[e.g., see][]{good14}. 
Here, we define the Froude number as $Fr\equiv\epsilon^{3/4}/(\nu^{1/4}\mathfrak{g})$, 
where $\epsilon^{3/4}/\nu^{1/4}$ is the Kolmogorov estimate for the acceleration r.m.s. 
(Note that the Froude number is sometimes expressed as the inverse of our definition; 
we choose the present definition for consistency with the standard practice of defining the 
Froude number with $\mathfrak{g}$ in the denominator.) Alternatively, gravity can be parameterized 
by the settling parameter $Sv\equiv\tau_p \mathfrak{g} / u_\eta$, which is the ratio of the particle's settling velocity 
$\tau_p \mathfrak{g}$ to the Kolmogorov velocity of the turbulence $u_\eta \equiv \left( \nu \epsilon \right)^{1/4}$. 
Note that these two non-dimensional parameters are related through the Stokes number, $Fr = St / Sv$, 
and that the Froude number $Fr$ is independent of $\tau_p$.

Since a primary aim of this paper is to study the effect of gravity at conditions representative of those in cumulus clouds, 
we calculate $Fr$ and $Sv$ by assuming a dissipation rate 
$\epsilon = 10^{-2}\ \mathrm{m}^2/\mathrm{s}^3$ \citep{shaw03}, a kinematic viscosity 
$\nu = 1.5 \times 10^{-5}\ \mathrm{m}^2/\mathrm{s}$, and a gravitational acceleration 
$\mathfrak{g} = 9.8\ \mathrm{m}/\mathrm{s}^2$, yielding $Fr = 0.052$ or $Sv = 19.3 St$. 
Simulations I, II, III, and IV were run with $Fr = 0.052$ and particles with Stokes numbers in the range $0 \leq St \leq 3$; 
due to computational limitations, simulation V was only run with $0 \leq St \leq 0.3$.

However, experimental observations suggest that cloud dissipation rates can vary by orders of magnitude 
\citep[e.g., see][]{prupp97}, resulting in commensurate variations in $Fr$ and $Sv$. 
For example, a strongly turbulent cumulonimbus cloud with 
$\epsilon \sim 10^{-1}\ \mathrm{m}^2/\mathrm{s}^3$ yields $Fr \approx 0.3$ and $Sv \approx 3.4 St$, 
while a weakly turbulent stratiform cloud with $\epsilon \sim 10^{-3}\ \mathrm{m}^2/\mathrm{s}^3$ 
yields $Fr \approx 0.01$ and $Sv \approx 100 St$ \citep{pinsky07}. To study a larger $St$-$Sv$ parameter space, 
we analyzed $513$ different combinations of $St$ and $Sv$ in simulation IIIb, with $0 \leq St \leq 56.2$ and $0 \leq Sv \leq 100$. 
The results from this simulation will be used to study detailed trends in particle accelerations, clustering, relative velocities, 
and collision rates for different values of particle inertia and gravity. 
(The trends in the particle kinetic energies and settling velocities obtained from this data set are discussed in detail in \cite{good14}.)

\section{Single-particle statistics}
\label{sec:single_particle}
As in Part I, we begin with the statistics of individual inertial particles.
We present velocity gradient statistics in \textsection \ref{sec:topology},
gravitational settling velocity statistics in \textsection \ref{sec:settling},
and acceleration statistics in \textsection \ref{sec:accelerations}.

\subsection{Velocity gradient statistics}
\label{sec:topology}
We begin by considering velocity gradients sampled along inertial particle trajectories, 
here denoted by $\bm{A}(\bm{x}^p(t),t) \equiv \nabla \bm{u}(\bm{x}^p(t),t)$. 
These statistics affect the relative motion of particles with weak-to-moderate inertia
with separations in the dissipation range, 
and also quantify the degree to which the inertial particles preferentially sample the underlying turbulence \citep[e.g., see][]{chun05}. 
This information will prove useful in interpreting the relative
velocity and clustering statistics presented in \S\ref{sec:relative_velocity} and \S\ref{sec:clustering}. 
It is convenient to define the rate-of-strain and rate-of-rotation tensors as 
$\boldsymbol{\mathcal{S}}(\bm{x}^p(t),t) \equiv [\bm{A}(\bm{x}^p(t),t) + \bm{A}^\intercal(\bm{x}^p(t),t) ]/2$ 
and $\boldsymbol{\mathcal{R}}(\bm{x}^p(t),t) \equiv [\bm{A}(\bm{x}^p(t),t) - \bm{A}^\intercal(\bm{x}^p(t),t)]/2$, 
respectively.

We first consider the variance of the diagonal components of $\boldsymbol{\mathcal{S}}$ 
averaged along the particle trajectories. For fluid particles that uniformly sample the flow, 
their variances reduce to
$\langle [\mathcal{S}_{11}(\bm{x}^p(t),t)]^2 \rangle = 
 \langle [\mathcal{S}_{22}(\bm{x}^p(t),t)]^2 \rangle = 
 \langle [\mathcal{S}_{33}(\bm{x}^p(t),t)]^2 \rangle = 1/(15\tau_\eta^2)$.
In isotropic turbulence without gravity, these components are statistically equivalent; 
however, in the presence of gravity, the symmetry of the particle motion is reduced from 
isotropic to axisymmetric, and thus the statistics of $\boldsymbol{\mathcal{S}}$ 
along the direction of gravity, $x_3$, will differ from the transverse directions, $x_1$ and $x_2$. 
For the remainder of the paper, the $x_3$ direction will be denoted as the `vertical' direction 
and the $x_1$ and $x_2$ directions will be referred to as the `horizontal' directions. 
The data for the two horizontal directions will be averaged together whenever possible, 
and denoted with the subscript `1.'
 
\begin{figure}
 \centering
 \includegraphics[width=2.6in]{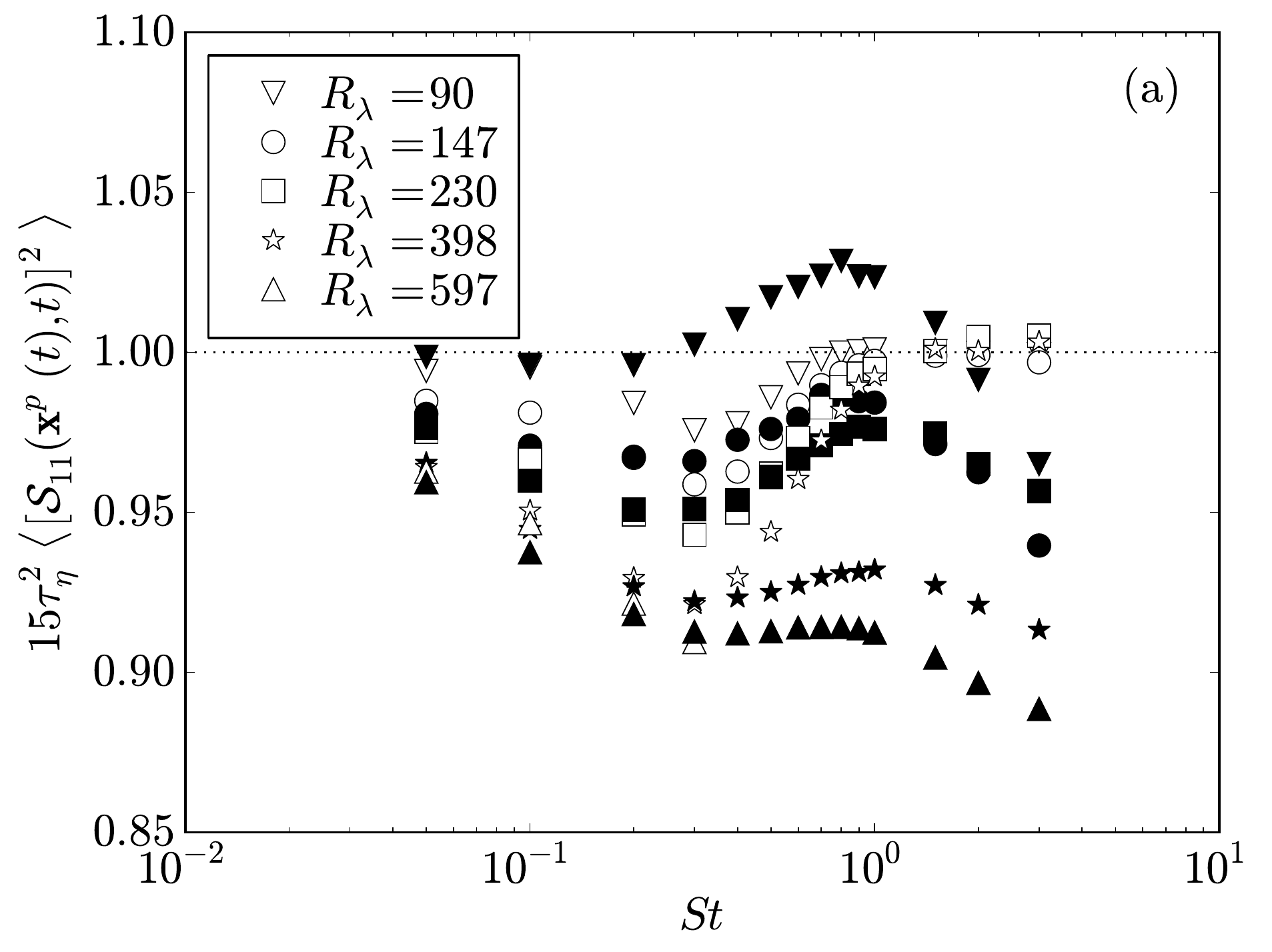}
 \includegraphics[width=2.6in]{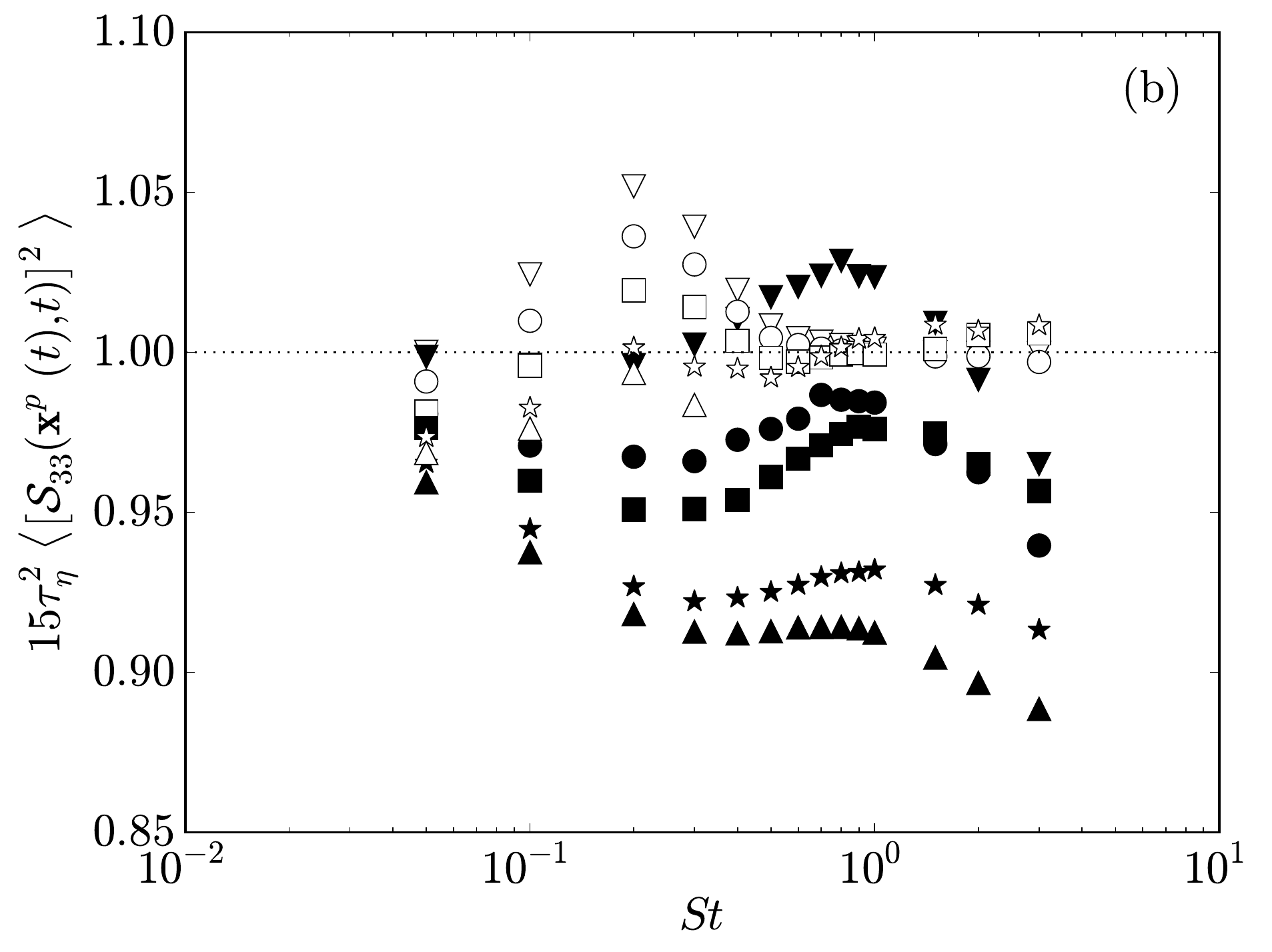}
 \caption{The normalized variance of the longitudinal velocity gradients sampled by inertial particles
 for different values of $R_\lambda$ and $St$.
 Open symbols denote data with gravity ($Fr = 0.052$), and filled symbols denote data
 without gravity ($Fr=\infty$). The horizontal dotted line indicates the expected value for uniformly distributed
 particles in isotropic turbulence.
 The gradients in the horizontal and vertical directions are shown in 
 (a) and (b), respectively.}
 \label{fig:AijAij}
\end{figure}

Figure~\ref{fig:AijAij} shows the variance of (a) horizontal 
($\langle\mathcal{S}_{11}^2\rangle^p\equiv\langle[\mathcal{S}_{11}(\bm{x}^p(t),t)]^2\rangle$) 
and (b) vertical ($\langle\mathcal{S}_{33}^2\rangle^p\equiv\langle[\mathcal{S}_{33}(\bm{x}^p(t),t)]^2\rangle$) 
components of the rate-of-strain tensor as a function of $St$ at different values of $R_\lambda$, 
both with gravity ($Fr=0.052$) and without gravity ($Fr=\infty$). 
Without gravity, the strain rates
decrease with increasing $St$ for $St \lesssim 0.3$. This is closely related
to the trend in the strain rates observed in Part I, which was attributed
to the fact that inertia causes particles to be ejected from vortex sheets.
With gravity, the strain rates in the horizontal direction 
also decrease with increasing $St$ for $St \lesssim 0.3$, as seen in figure~\ref{fig:AijAij}(a). 
In addition, they are quite
close to the corresponding values without gravity, suggesting
that gravity does not lead to significant changes in preferential
sampling in the horizontal direction at low $St$.
We see in figure~\ref{fig:AijAij}(b), however, that gravity causes 
low-$St$ particles to sample regions of larger vertical strain rates, 
and that these strain rates are considerably different from those when gravity is absent.
We also observe that gravity reduces the degree of preferential sampling
at high $St$, causing the strain rates to approach the values predicted for uniformly distributed 
particles in isotropic turbulence for $St \gtrsim 1$.

\begin{figure}
 \centering
 \includegraphics[width=2.6in]{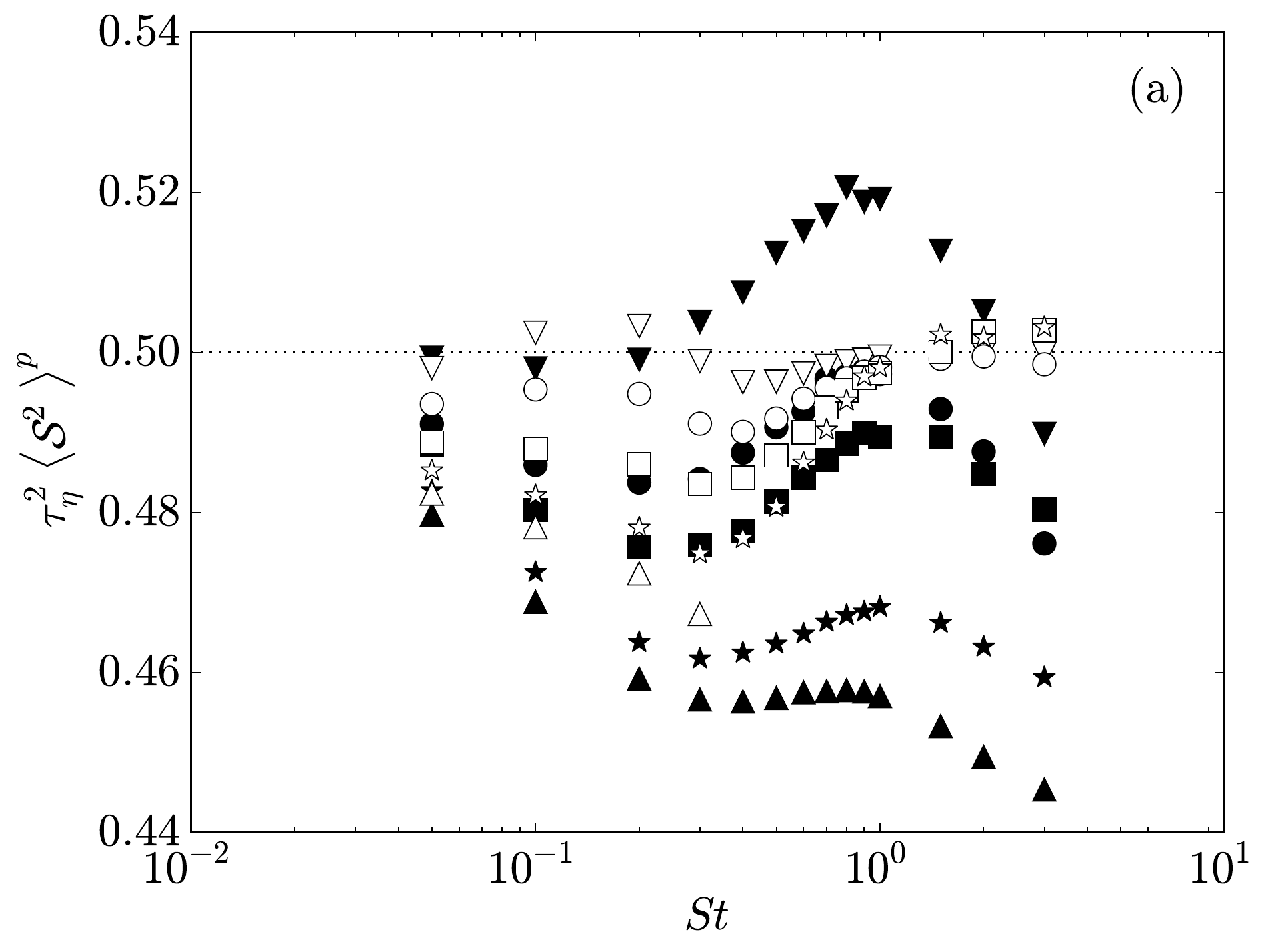}
 \includegraphics[width=2.6in]{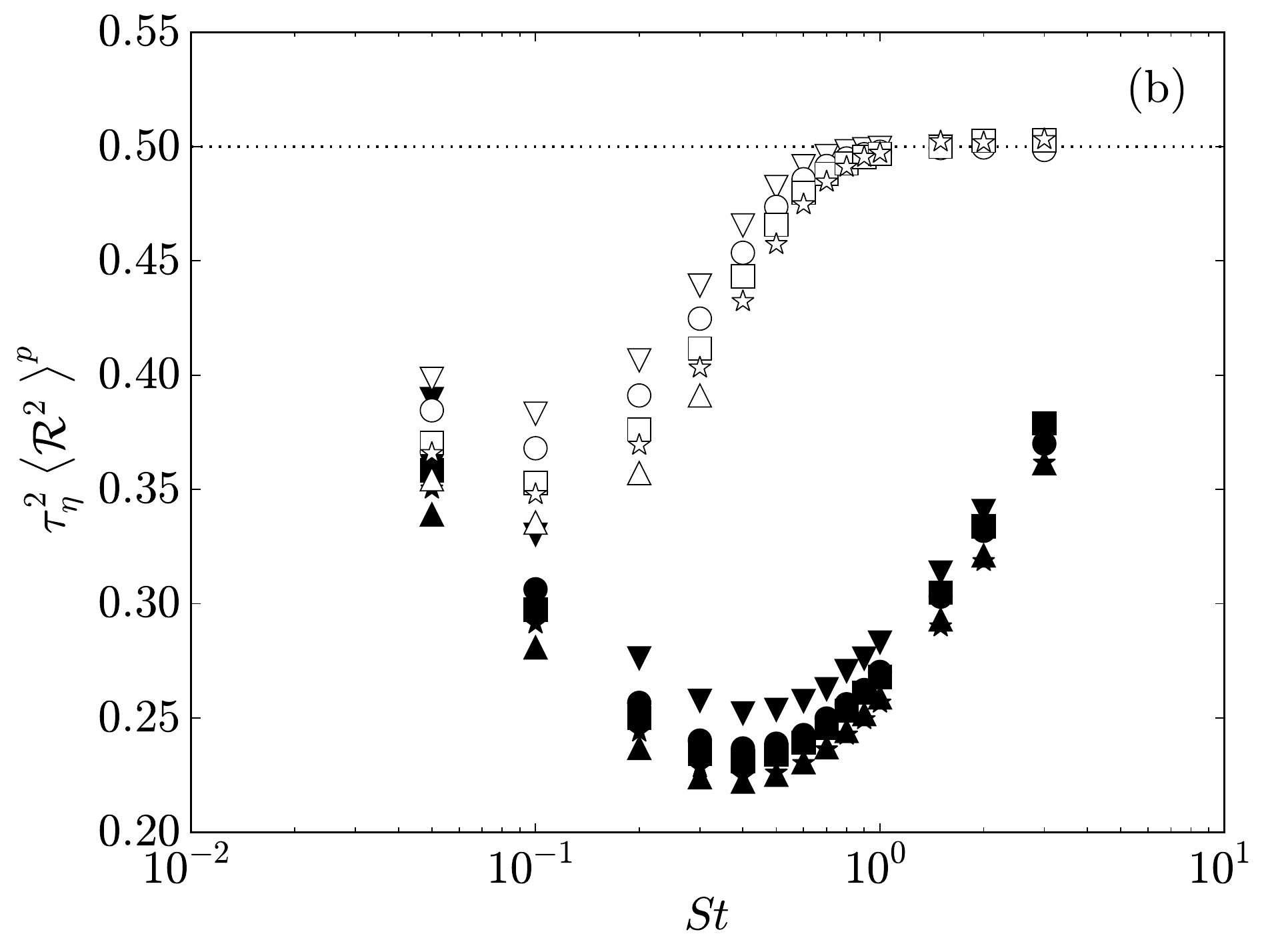}
 \includegraphics[width=2.6in]{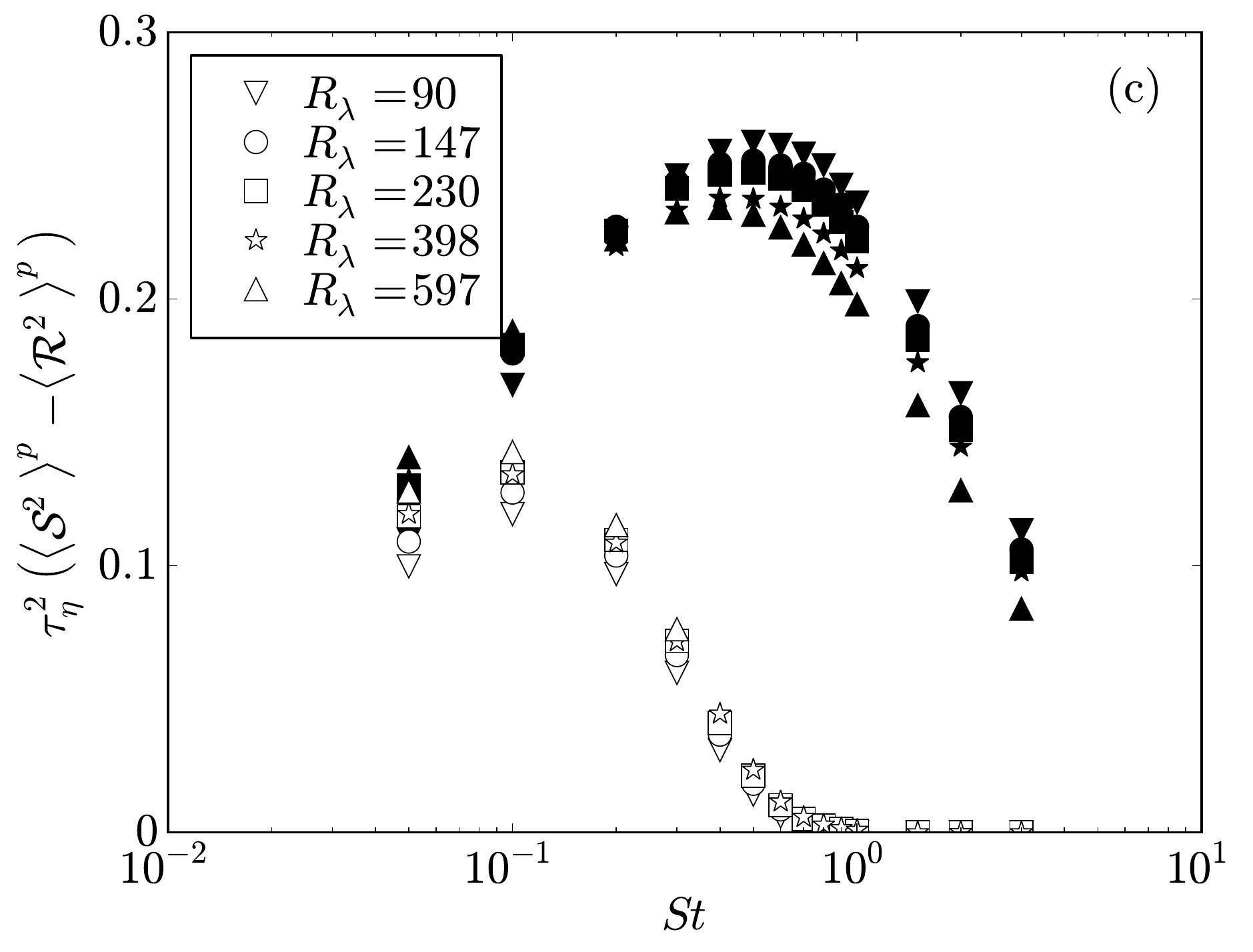}
 \caption{$\langle \mathcal{S}^2 \rangle^p$ (a), $\langle \mathcal{R}^2 \rangle^p$ (b),
 and $\langle \mathcal{S}^2 \rangle^p - \langle \mathcal{R}^2 \rangle^p$ (c)
 as function of $St$ for different values of $R_\lambda$.
 Open symbols denote data with gravity ($Fr = 0.052$), and filled symbols denote data
 without gravity ($Fr=\infty$).}
 \label{fig:S2_R2}
\end{figure}

In figure~\ref{fig:S2_R2}, we plot
$\tau_\eta^2 \langle \mathcal{S}^2 \rangle^p$, 
$\tau_\eta^2 \langle \mathcal{R}^2 \rangle^p$, and 
$\tau_\eta^2 (\langle \mathcal{S}^2 \rangle^p - \langle \mathcal{R}^2 \rangle^p)$, where
$\langle \mathcal{S}^2 \rangle^p \equiv \langle \boldsymbol{\mathcal{S}}(\bm{x}^p(t),t) :
\boldsymbol{\mathcal{S}}(\bm{x}^p(t),t) \rangle$ and 
$\langle \mathcal{R}^2 \rangle^p \equiv \langle \boldsymbol{\mathcal{R}}(\bm{x}^p(t),t) :
\boldsymbol{\mathcal{R}}(\bm{x}^p(t),t) \rangle$ are the second invariants of the rate-of-strain and rate-of-rotation tensors, respectively,
for particles with ($Fr = 0.052$) and without ($Fr=\infty$) gravity. As noted in Part I, when preferential sampling effects are absent,
$\tau_\eta^2 \langle \mathcal{S}^2 \rangle^p = \tau_\eta^2 \langle \mathcal{R}^2 \rangle^p = 0.5$.
We see from figure~\ref{fig:S2_R2} that gravity reduces the degree of preferential sampling,
causing $\tau_\eta^2 \langle \mathcal{S}^2 \rangle^p$,
$\tau_\eta^2 \langle \mathcal{R}^2 \rangle^p$, and 
$\tau_\eta^2 (\langle \mathcal{S}^2 \rangle^p - \langle \mathcal{R}^2 \rangle^p)$
to be closer to the corresponding values for uniformly distributed particles.
We also note that preferential sampling effects are eliminated altogether for $St \gtrsim 1$,
which is consistent with the results presented earlier. 
The trends in the mean strain and rotation rates with $R_\lambda$ 
are similar both with and without gravity, and are discussed in Part I.

To explain the reduction in preferential sampling with gravity, we note that 
gravity, by causing particles to settle relative to the underlying flow,
reduces the interaction times between the particles and the turbulent eddies.
As a result, particles have less time to be affected by the straining and rotating regions of the flow,
and therefore experience less preferential sampling \citep[see also][]{gustavsson14}.

We test this argument by computing Lagrangian rate-of-strain and rate-of-rotation timescales along 
inertial particle trajectories, both with and without gravity. 
These timescales will prove useful for extending the theoretical model of \cite{zaichik09} 
to account for gravitational effects. 
Their model, developed for an isotropic particle field, uses a single timescale for the rate-of-strain tensor 
and a single timescale for the rate-of-rotation tensor. 
To generalize the model for the case with gravity, directionally dependent times scales are required. 
We discuss the generalized model and its comparison with DNS in \textsection\ref{sec:clustering}.

We consider both the Lagrangian strain timescale $T^p_{\mathcal{S}_{ij} \mathcal{S}_{k m}}$, which we define as 
\begin{equation}
 T^p_{\mathcal{S}_{ij} \mathcal{S}_{k m}} \equiv \frac{
                             \displaystyle{\int\limits_0^\infty
                             \left \langle \mathcal{S}_{ij}(\bm{x}^p(0),0) 
                              \mathcal{S}_{k m}(\bm{x}^p(s),s)\right \rangle ds}}
                              {\left \langle \mathcal{S}_{ij}(\bm{x}^p(0),0) 
                              \mathcal{S}_{k m}(\bm{x}^p(0),0) \right \rangle}
                              \mathrm{,}
 \label{eq:timescales_strain}
\end{equation}
and the Lagrangian rotation timescale $T^p_{\mathcal{R}_{ij} \mathcal{R}_{k m}}$, defined as
\begin{equation}
 T^p_{\mathcal{R}_{ij} \mathcal{R}_{k m}} \equiv \frac{
                             \displaystyle{\int\limits_0^\infty
                             \left \langle \mathcal{R}_{ij}(\bm{x}^p(0),0) 
                              \mathcal{R}_{k m}(\bm{x}^p(s),s)\right \rangle ds}}
                              {\left \langle \mathcal{R}_{ij}(\bm{x}^p(0),0) 
                              \mathcal{R}_{k m}(\bm{x}^p(0),0) \right \rangle}
                              \mathrm{.}
 \label{eq:timescales_rotation}
\end{equation}
In both of these equations, no summation is implied by the repeated indices.

In the absence of gravity, the particle field is isotropic, which implies the rate-of-strain timescales $T^p_{\mathcal{S}_{11} \mathcal{S}_{11}}$, $T^p_{\mathcal{S}_{11} \mathcal{S}_{22}}$, 
$T^p_{\mathcal{S}_{11} \mathcal{S}_{33}}$, $T^p_{\mathcal{S}_{12} \mathcal{S}_{12}}$, 
$T^p_{\mathcal{S}_{13} \mathcal{S}_{13}}$, $T^p_{\mathcal{S}_{22} \mathcal{S}_{22}}$, 
$T^p_{\mathcal{S}_{22} \mathcal{S}_{33}}$, $T^p_{\mathcal{S}_{23} \mathcal{S}_{23}}$, 
and $T^p_{\mathcal{S}_{33} \mathcal{S}_{33}}$ are statistically equivalent,
as are the rotation timescales $T^p_{\mathcal{R}_{12} \mathcal{R}_{12}}$, 
$T^p_{\mathcal{R}_{13} \mathcal{R}_{13}}$, and 
$T^p_{\mathcal{R}_{23} \mathcal{R}_{23}}$.
However, with gravity, the nine rate-of-strain timescales are no longer equivalent, 
nor are the three rate-of-rotation timescales. 
Applying symmetry analysis, we can group the nine rate-of-strain timescales into six categories, 
and three rate-of-rotation timescales into two categories, as shown in 
table~\ref{tab:strain_rotation_axisymmetric}. 
Additionally, we introduce $T^p_{\mathcal{S} \mathcal{S}}$ and $T^p_{\mathcal{R} \mathcal{R}}$ 
as the average of the nine rate-of-strain timescales and three rate-of-rotation timescales, respectively.

\begin{table}
 \centering
 \caption{Categorization of the strain and rotation timescales
 with gravity. (Gravity is aligned with the $x_3$-direction.)
 The elements within a given row are statistically equivalent, while the elements in different
 rows may differ due to the anisotropy in the particle motion.}
 \label{tab:strain_rotation_axisymmetric}
 \begin{tabular}{lll}
 (1) & $T^p_{\mathcal{S}_{11} \mathcal{S}_{11}}$ & $T^p_{\mathcal{S}_{22} \mathcal{S}_{22}}$ \\
 (2) & $T^p_{\mathcal{S}_{11} \mathcal{S}_{22}}$ \\
 (3) & $T^p_{\mathcal{S}_{12} \mathcal{S}_{12}}$ \\
 (4) & $T^p_{\mathcal{S}_{13} \mathcal{S}_{13}}$ & $T^p_{\mathcal{S}_{23} \mathcal{S}_{23}}$ \\
 (5) & $T^p_{\mathcal{S}_{11} \mathcal{S}_{33}}$ & $T^p_{\mathcal{S}_{22} \mathcal{S}_{33}}$ \\
 (6) & $T^p_{\mathcal{S}_{33} \mathcal{S}_{33}}$ \\
 \\
 (1) & $T^p_{\mathcal{R}_{12} \mathcal{R}_{12}}$ \\
 (2) & $T^p_{\mathcal{R}_{13} \mathcal{R}_{13}}$ & $T^p_{\mathcal{R}_{23} \mathcal{R}_{23}}$ \\
 \end{tabular}
\end{table}

\begin{figure}
 \centering
 \includegraphics[width=2.6in]{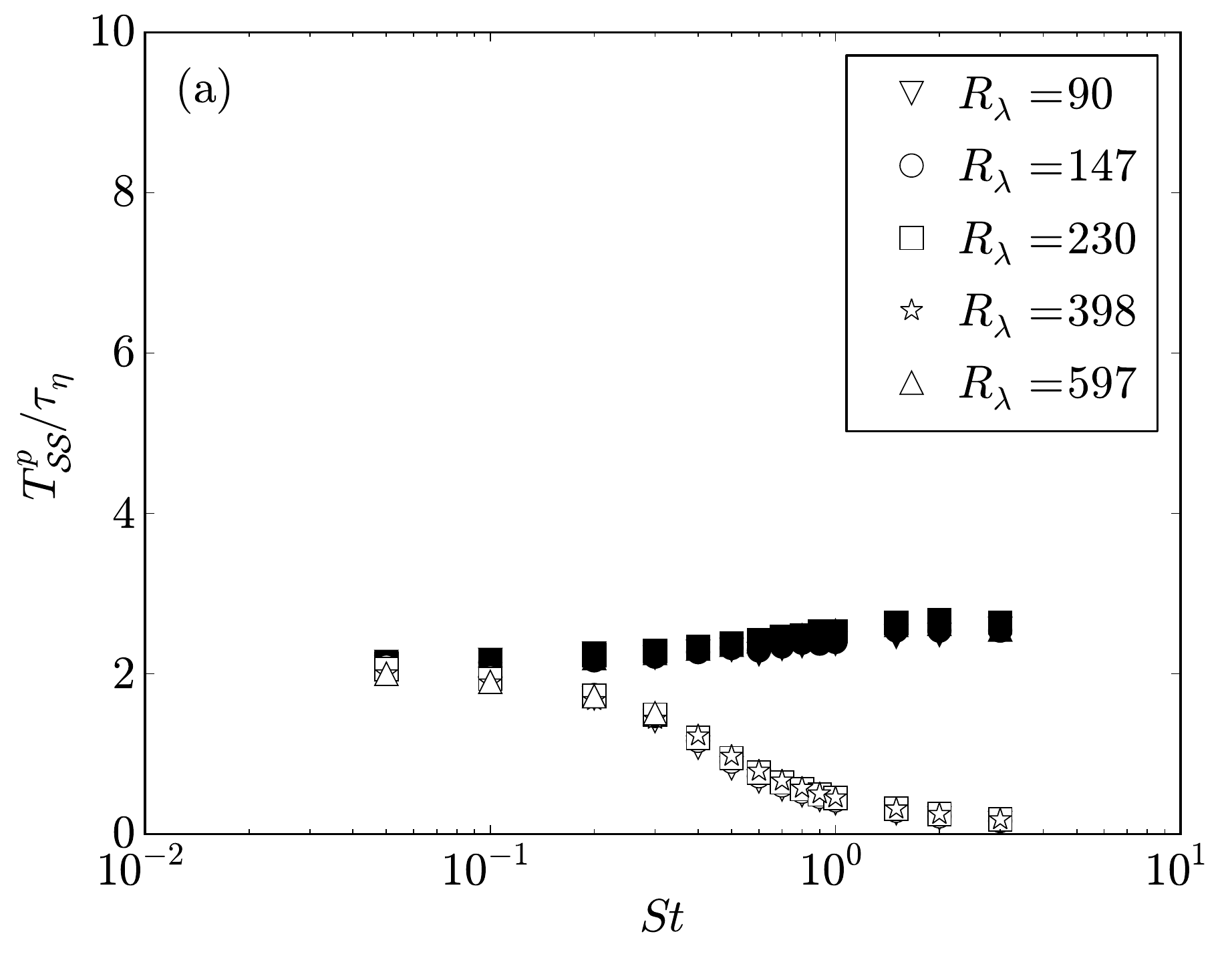}
 \includegraphics[width=2.6in]{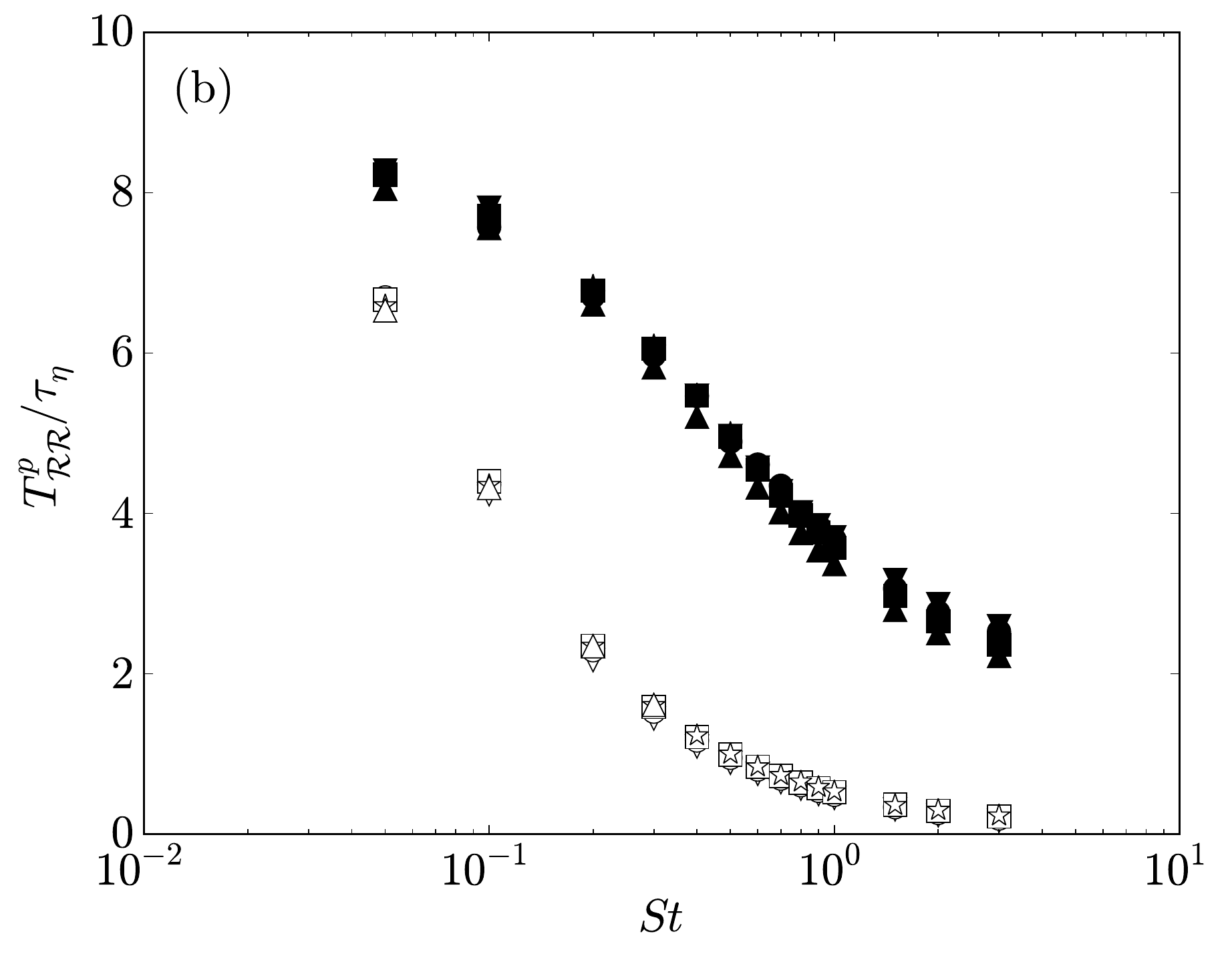}
 \caption{The averaged strain timescale $T^p_{\mathcal{S} \mathcal{S}}$ (a) and 
 the averaged rotation timescale $T^p_{\mathcal{R} \mathcal{R}}$ (b) plotted as a function
 of $St$ for different values of $R_\lambda$. Open symbols denote data with gravity ($Fr = 0.052$), 
 and filled symbols denote data without gravity ($Fr=\infty$).} 
 \label{fig:strain_rotation_times}
\end{figure}

Figure~\ref{fig:strain_rotation_times} shows the averaged time scales 
$T^p_{\mathcal{S} \mathcal{S}}$ and $T^p_{\mathcal{R} \mathcal{R}}$ 
as a function of the particle Stokes number at different values of $R_\lambda$, 
with and without gravity. In agreement with the physical explanations offered above, 
we see that the timescales without gravity always exceed those with gravity. 
We also note that both with and without gravity, the strain timescales are almost entirely 
insensitive to the Reynolds number.  In addition, while the rate-of-rotation timescales without gravity 
vary weakly with $R_\lambda$ (as noted in Part I), 
they are nearly independent of the Reynolds number with gravity.

\begin{figure}
 \centering
  \includegraphics[height=1.9in]{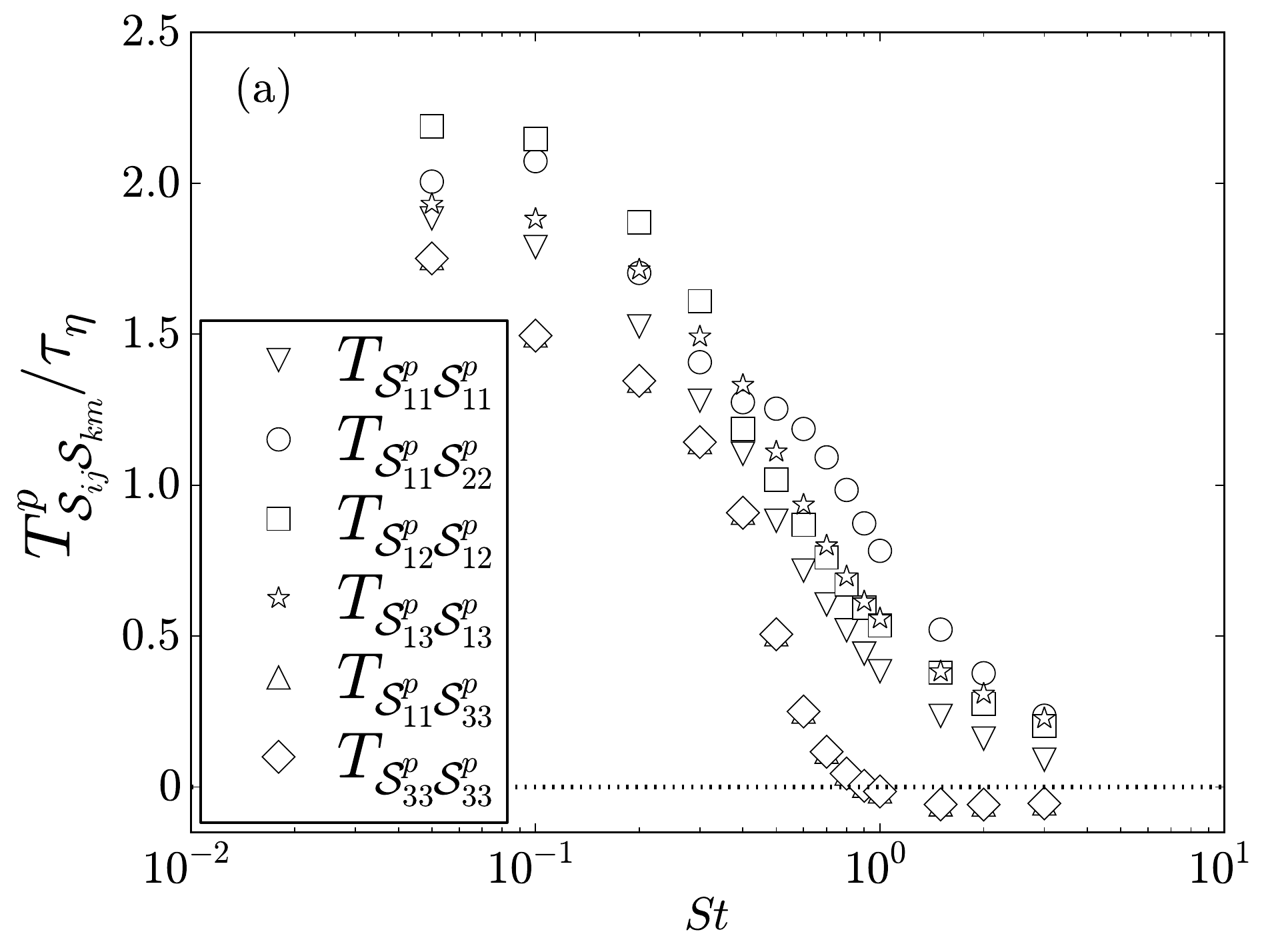}
  \includegraphics[height=1.9in]{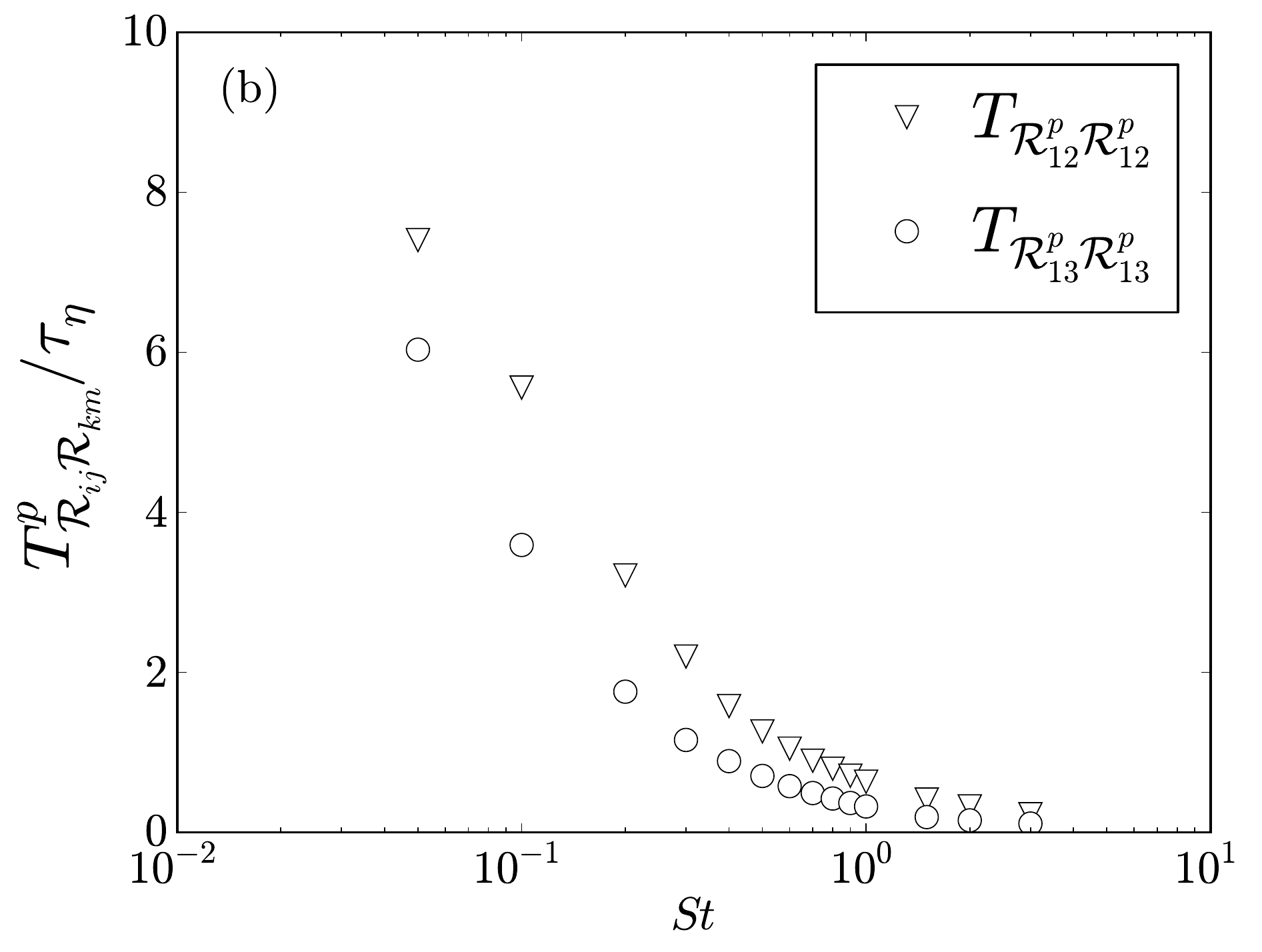}
 \caption{The Lagrangian strain (a) and rotation (b) timescales in different
 directions at $R_\lambda=398$, plotted as a function
 of $St$, for particles with gravity ($Fr = 0.052$).}
 \label{fig:strain_rotation_times_anisotropic}
\end{figure}

We now consider the anisotropy in the different timescales with gravity at $R_\lambda=398$ in 
figure~\ref{fig:strain_rotation_times_anisotropic}. 
In this plot, we have averaged together the statistically equivalent timescales 
grouped in table~\ref{tab:strain_rotation_axisymmetric}, 
and have denoted these timescales by the first (or only) item in these groups. 
As expected, the timescales show considerable anisotropy with gravity. 
If we consider the limit of strong gravity ($Sv \gg u'/u_\eta$), 
we can derive a simple model for the timescales (see Appendix~\ref{sec:high_g_models} 
for the derivation and estimates of droplet diameters required for the model to be valid in 
different atmospheric clouds). Using the relations in \textsection \ref{sec:high_g_timescales}, 
the Lagrangian rate-of-strain and rate-of-rotation timescales take the form
\begin{equation}
\label{eq:timescale_model_simplified_nondimensional_main}
 \hat{T}^p_{\mathcal{S}_{ij} \mathcal{S}_{km}}
 = \frac{\hat{\ell}_{\mathcal{S}_{ij} \mathcal{S}_{km},3}}{Sv} \mathrm{,}
\end{equation}
and
\begin{equation}
\label{eq:timescale_model_simplified_nondimensional_rotation_main}
\hat{T}^p_{\mathcal{R}_{ij} \mathcal{R}_{km}}
 = \frac{\hat{\ell}_{\mathcal{R}_{ij} \mathcal{R}_{km},3}}{Sv} \mathrm{,}
\end{equation}
respectively, where we have used the top-hat symbol to denote a variable normalized by 
Kolmogorov scales, and 
$\ell_{\mathcal{S}_{ij} \mathcal{S}_{km},3}$ ($\ell_{\mathcal{R}_{ij} \mathcal{R}_{km},3}$) 
is the integral lengthscale of $\mathcal{S}_{ij} \mathcal{S}_{km}$ 
($\mathcal{R}_{ij} \mathcal{R}_{km}$) evaluated along the $x_3$-direction.

Equations~(\ref{eq:timescale_model_simplified_nondimensional_main}) and (\ref{eq:timescale_model_simplified_nondimensional_rotation_main}) imply in the limit $Sv \gg u'/u_\eta$,
the Lagrangian integral timescales of the flow
experienced by the particles are directly proportional to the Eulerian integral
lengthscales of the strain and rotation fields.
We can therefore write each of the Lagrangian timescales in table~\ref{tab:strain_rotation_axisymmetric}
as functions of their corresponding Eulerian integral lengthscales.
This allows us to use tensor invariance theory and the properties of the strain and rotation rate tensors in 
HIT to predict the relative magnitude of the different timescales.
We therefore can show that
\begin{equation}
\label{eq:strain_timescales_directional}
 T^p_{\mathcal{S}_{11} \mathcal{S}_{33}} = T^p_{\mathcal{S}_{33} \mathcal{S}_{33}} < T^p_{\mathcal{S}_{11} \mathcal{S}_{11}} < T^p_{\mathcal{S}_{12} \mathcal{S}_{12}} = T^p_{\mathcal{S}_{13} \mathcal{S}_{13}} < T^p_{\mathcal{S}_{11} \mathcal{S}_{22}} \mathrm{,}
\end{equation}
and
\begin{equation}
\label{eq:rotation_timescales_directional}
 T^p_{\mathcal{R}_{13} \mathcal{R}_{13}} < T^p_{\mathcal{R}_{12} \mathcal{R}_{12}}\mathrm{.}
\end{equation}
The DNS data in figure~\ref{fig:strain_rotation_times_anisotropic} follow the trends
predicted by (\ref{eq:strain_timescales_directional}) and (\ref{eq:rotation_timescales_directional}) at high $St$.

\subsection{Mean particle settling velocities}
\label{sec:settling}
We now analyze the effect of turbulence on the mean settling speed of inertial particles.
The role of the mean settling velocity, aside from its importance to the process of cloud rain out, 
is also related to the evolution of the droplet size distribution through its effect on the collision 
kernel of different size particles, 
as discussed in \cite{davila01} and \cite{ghosh05}, and later modeled by \cite{grabowski13}.

Recently, \cite{good14} measured settling speeds of particles in turbulence and showed that 
turbulence causes particles with low-to-moderate Stokes numbers to settle more quickly 
than they would in a quiescent fluid. 
The so-called `preferential sweeping' or `fast tracking' of particles 
(particles sampling downward-moving fluid over upward-moving fluid) 
had been observed in DNS \citep{wang93,yang98,ireland12,bec14} 
and earlier experiments \citep{aliseda02,yang03,yang05}. 
For $St\gtrsim 1$, \cite{good14} observed a transition to so-called `loitering' 
(particles settling more slowly than they would in a quiescent fluid), 
again consistent with earlier experiments \citep{nielsen93,yang03,kawanisi08}, 
although \cite{good14} were the first to measure it for droplets in air. 
In contrast, DNS with linear drag shows 
preferential sweeping at all Stokes numbers, 
with no transition to loitering \citep{wang93,yang98,ireland12,bec14}. 
In order to observe loitering in DNS, the linear drag must be replaced by a nonlinear drag 
model \citep[e.g., ][]{clift78}. 
The nonlinear drag model yields settling speeds that are only in qualitative, 
but not quantitative, agreement with the experiments, 
most likely due to the simplistic nature of the nonlinear drag model that was used \citep{good14}.

\begin{figure}
 \centering
 \includegraphics[height=2.5in]{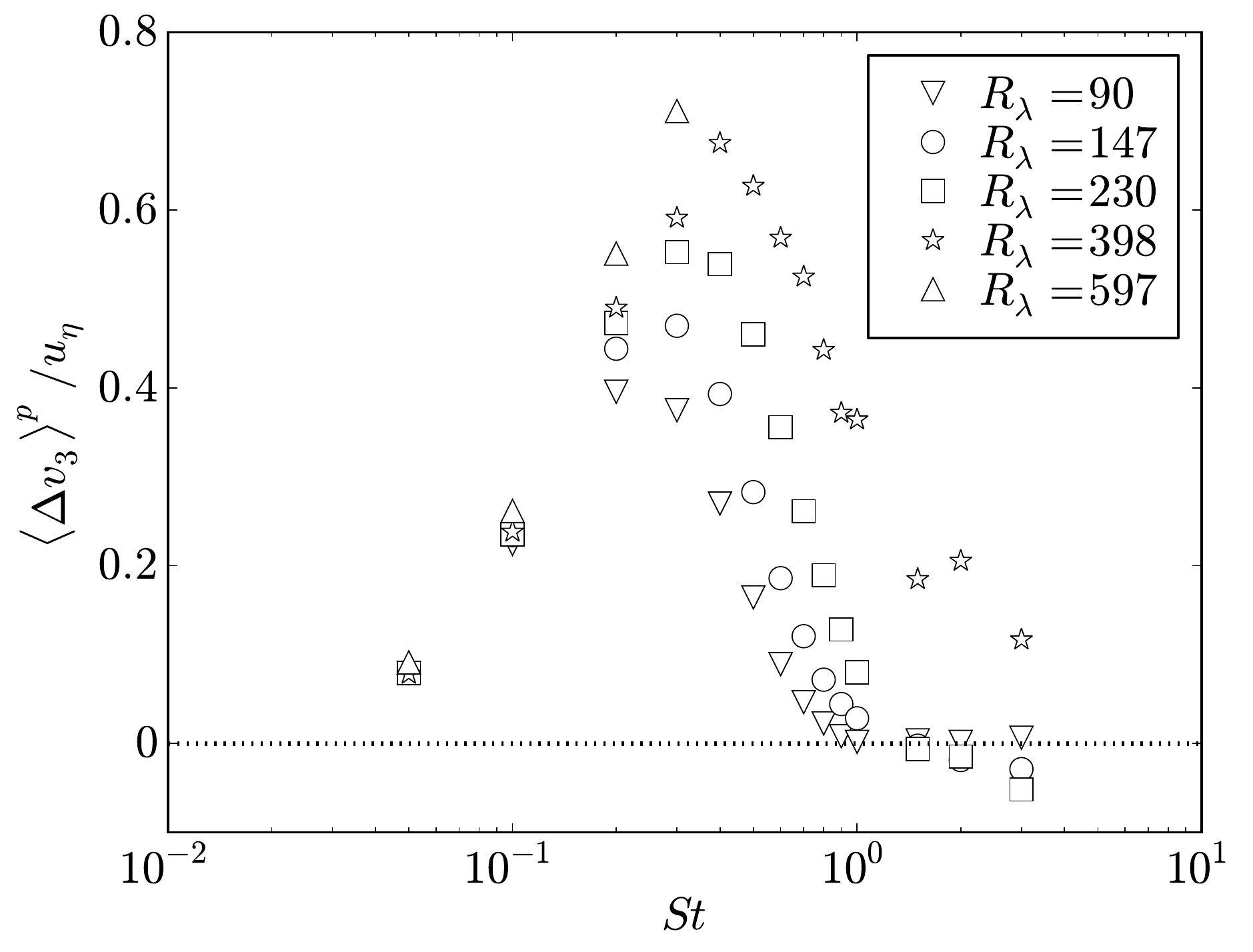}
 \caption{Turbulence-induced enhancements (positive) or reductions (negative) 
 in the mean settling velocities of inertial particles
 with gravity ($Fr = 0.052$), normalized by the Kolmogorov velocity $u_\eta$.
 The symbols denote different values of $R_\lambda$.}
 \label{fig:settling}
\end{figure}

Here, we extend the range of Reynolds numbers for $0 \leq St \leq 3$ and $Fr = 0.052$ 
using a linear drag model. From (\ref{eq:maxey_riley}) we can show that
\begin{equation}
 \langle \bm{u}(\bm{x}^p(t),t) \rangle = 
 \langle \bm{v}^p(t) \rangle - \tau_p \boldsymbol{\mathfrak{g}} \equiv - \langle \bm{\Delta v} \rangle^p \mathrm{,}
 \label{eq:settling}
\end{equation}
where $\tau_p \boldsymbol{\mathfrak{g}} = (0,0,-\tau_p \mathfrak{g})^\intercal$ is the gravitational 
settling velocity in a quiescent flow. Preferential sweeping corresponds to 
$\langle \Delta v_3 \rangle^p > 0$ and loitering to $\langle \Delta v_3 \rangle^p < 0$. 
Figure~\ref{fig:settling} shows $\langle \Delta v_3 \rangle^p$, normalized by the 
Kolmogorov velocity $u_\eta$, as a function of the particle Stokes number. 
Consistent with the DNS results presented in \cite{good14}, the DNS with linear drag shows no 
statistically significant loitering at any Reynolds number. 
Moreover, we see that the mean settling speeds are independent of 
$R_\lambda$ for $St \leq 0.1$, suggesting that in this limit, they are determined entirely 
by the small-scale turbulence, in agreement with \cite{bec14}. 
At higher $St$, the settling speeds are stronger functions of the Reynolds number; 
however, these results must be interpreted with some caution given the concerns 
raised about linear drag in this regime. Nevertheless, we see at least the potential for 
a significant Reynolds number effect on the mean settling velocities at higher Stokes number.

\subsection{Particle accelerations}
\label{sec:accelerations}

\begin{figure}
 \centering
 \includegraphics[width=2.6in]{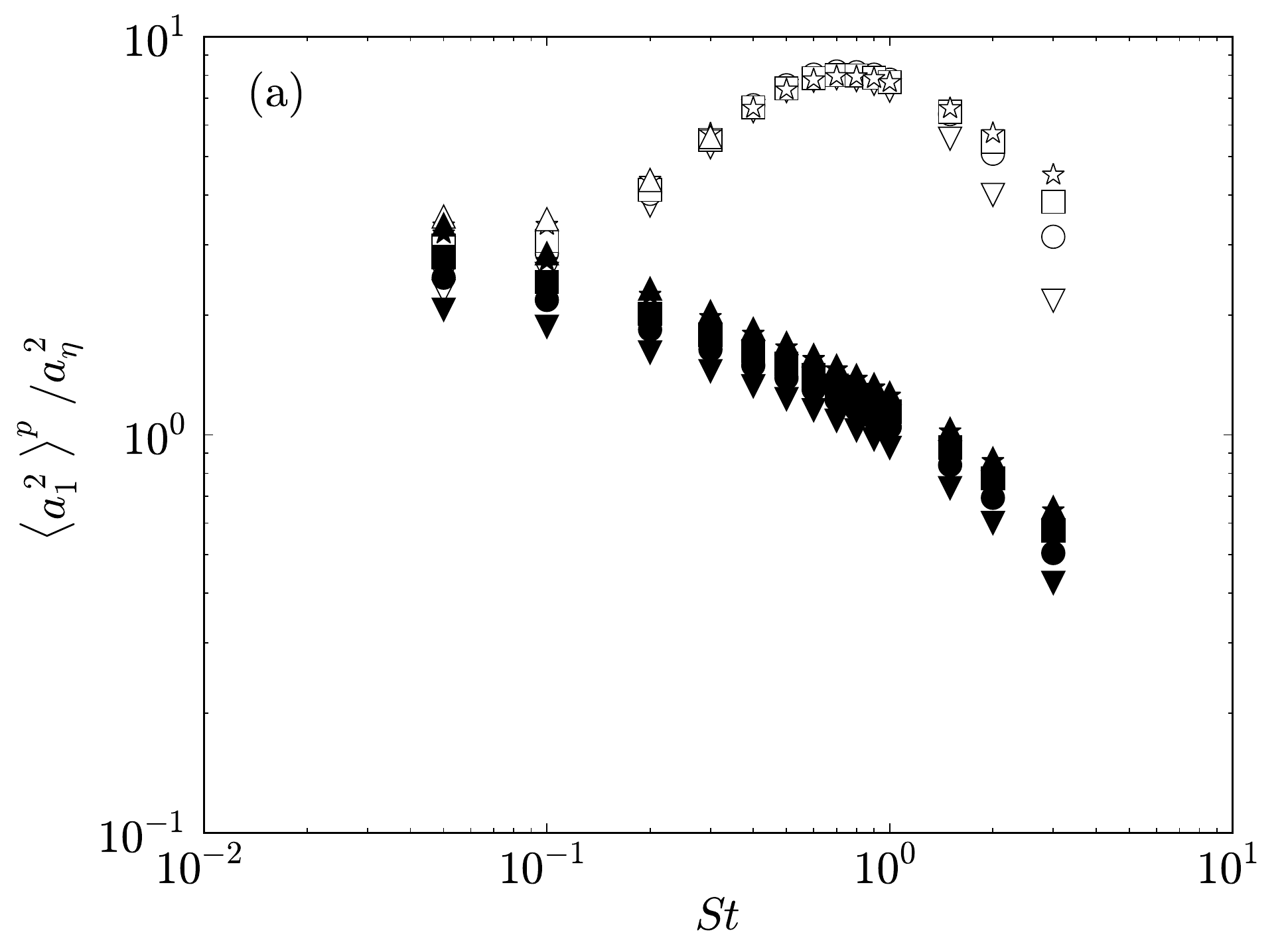}
 \includegraphics[width=2.6in]{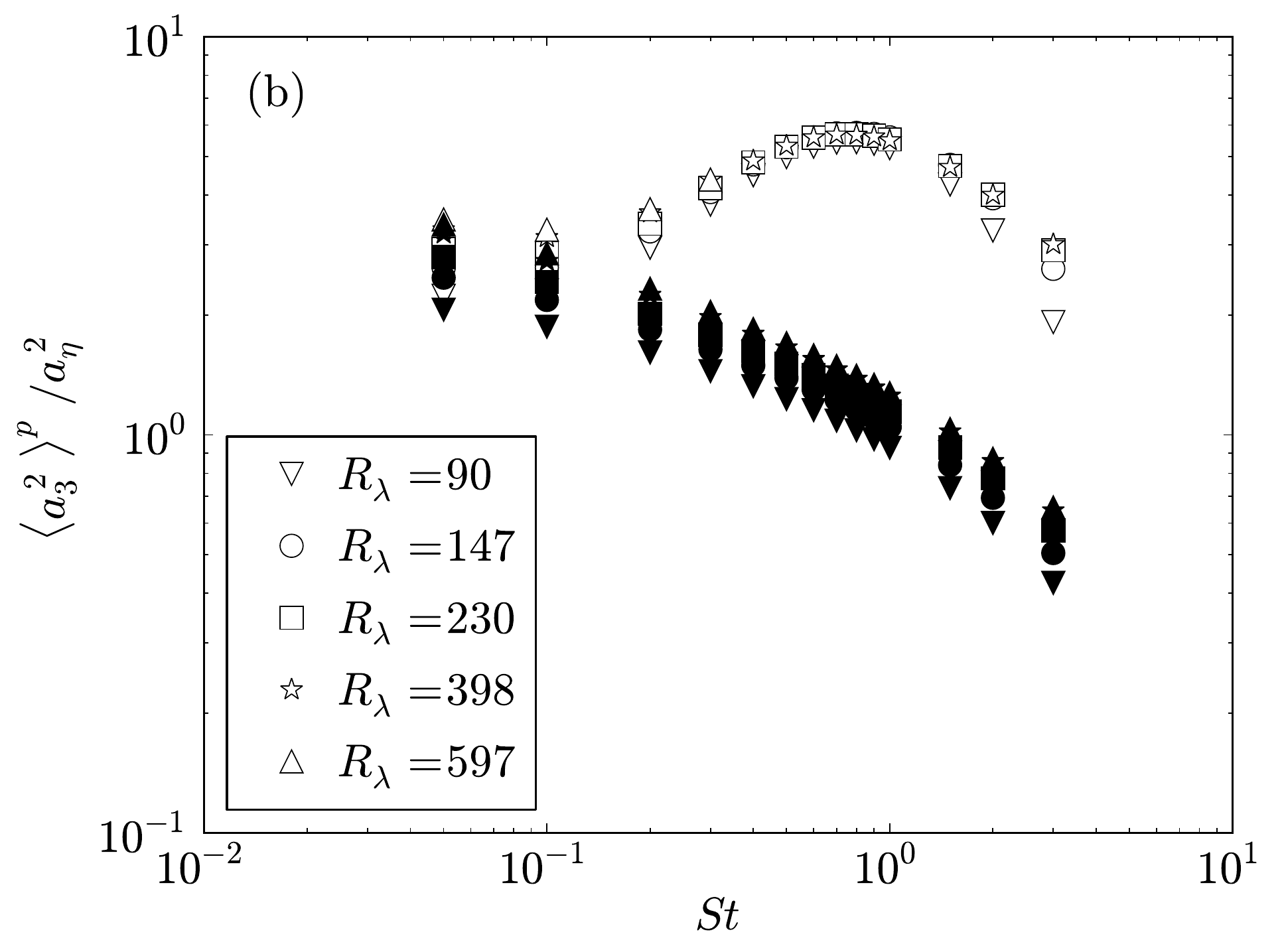}
 \caption{The variance of the particle accelerations in the horizontal (a) and
 vertical (b) directions. Open symbols denote data with gravity ($Fr = 0.052$),
 and filled symbols denote data without gravity ($Fr = \infty$).}
 \label{fig:acceleration_variance_particle}
\end{figure}

We now consider the acceleration statistics of the particles. 
As noted in \cite{chun05}, the relative velocity of different-sized particles 
(and thus their collision rate) is related to their accelerations. 
Figure~\ref{fig:acceleration_variance_particle} shows the particle acceleration variances 
$\langle a_1^2 \rangle^p \equiv \langle (dv^p_1(t)/dt)^2 \rangle$ and 
$\langle a_3^2 \rangle^p \equiv \langle (dv^p_3(t)/dt)^2 \rangle$, with and without gravity.  
Notice that the acceleration variances with gravity always exceed those without gravity, 
in some cases by as much as an order of magnitude.
Similar trends were observed in computational and experimental studies of channel flows \citep{gerashchenko08,lavezzo10},
and more recently in the computational study of \cite{parishani15} in HIT.

We also see that gravity reverses the trends in the particle accelerations at low $St$, 
causing the accelerations to increase with increasing $St$. 
This trend was recently reported in \cite{parishani15} at lower
values of $R_\lambda$ ($R_\lambda = 119-143$). Our results in Appendix~\ref{sec:periodicity}, however,
suggest that the acceleration statistics in \cite{parishani15}, while qualitatively correct,
have quantitative errors due the influence of artificial periodicity
effects on the small domain sizes in their simulations.
Experimental studies of inertial particle accelerations in HIT \citep{sathya06a} and 
Von K\'arm\'an flows \citep{volk08,volk08a} have only observed a monotonic 
decrease of the particle accelerations with increasing $St$, 
presumably because the Froude numbers in those studies 
($Fr \sim 0.3$ for HIT and $Fr \sim 30$ for the Von K\'arm\'an flow) 
were too large to observe this trend reversal. (Recall that in our study, $Fr = 0.052$.)

We now provide a physical explanation for these large accelerations.
We begin by considering the limit $St\ll 1$. If gravitational forces are negligible, 
to leading order the particle acceleration $\bm{a}^p(t)$ is equivalent to the 
fluid acceleration at the particle location \citep[e.g., see][]{bec06a}
\begin{equation}
 \bm{a}^p(t) \approx \frac{\partial \bm{u}(\bm{x}^p(t),t)}{\partial t} 
 + \bm{u}(\bm{x}^p(t),t) \cdot \nabla \bm{u}(\bm{x}^p(t),t) \mathrm{.}
 \label{eq:accel_lowSt_lowSv}
\end{equation}
As the gravitational force increases (i.e., as $Sv$ increases), 
it is the interplay between gravity and turbulence that determines the particle acceleration. 
For example, the downward drift due to the gravitational settling will cause the particle to 
experience different regions of the turbulence. We refer to this phenomenon as the 
`gravitational trajectory effect.' 
(Note that \cite{yudine59} coined `crossing trajectories' to refer to the same mechanism.)

\begin{figure}
 \centering
 \includegraphics[height=1.9in]{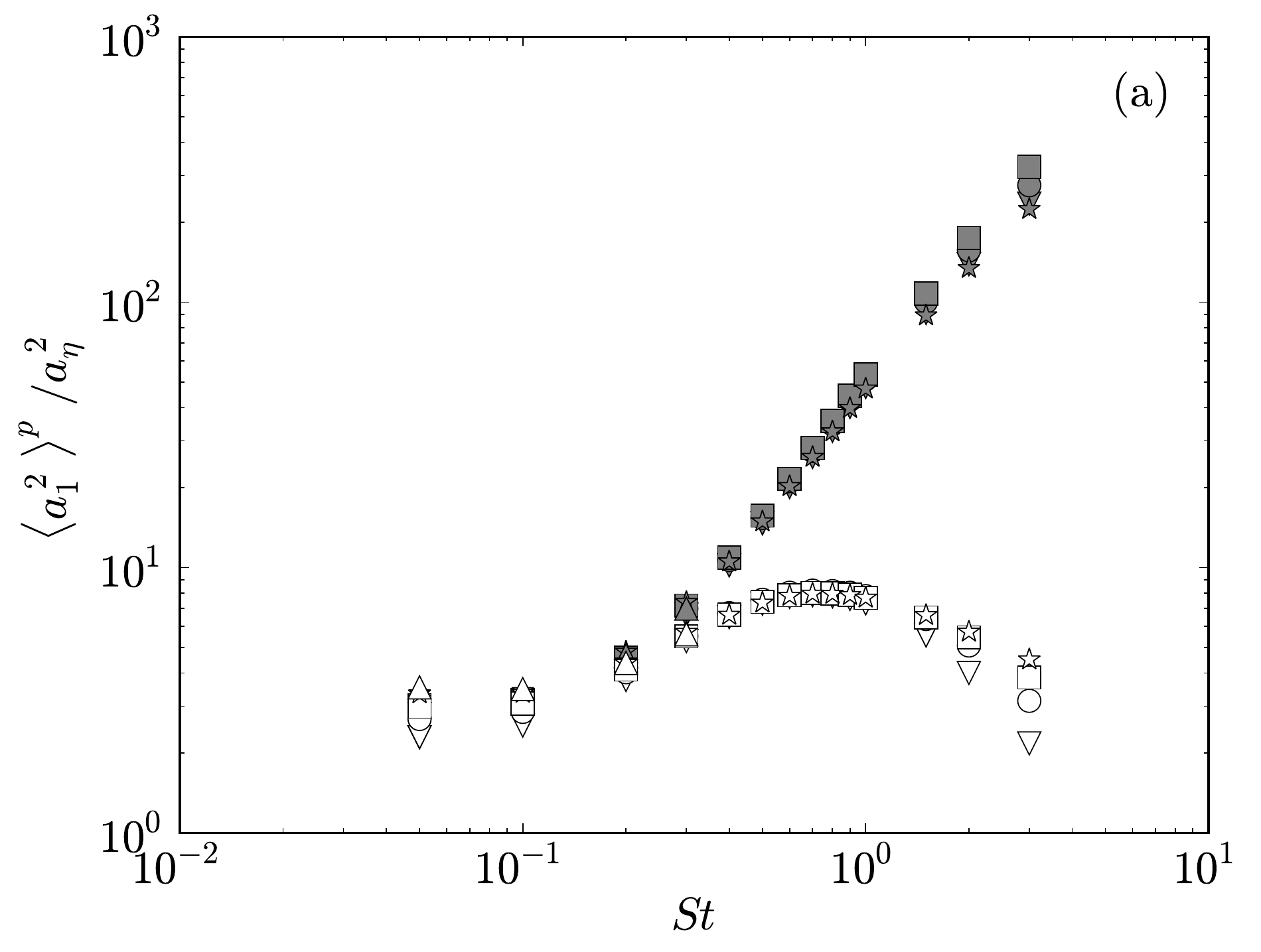}
 \includegraphics[height=1.9in]{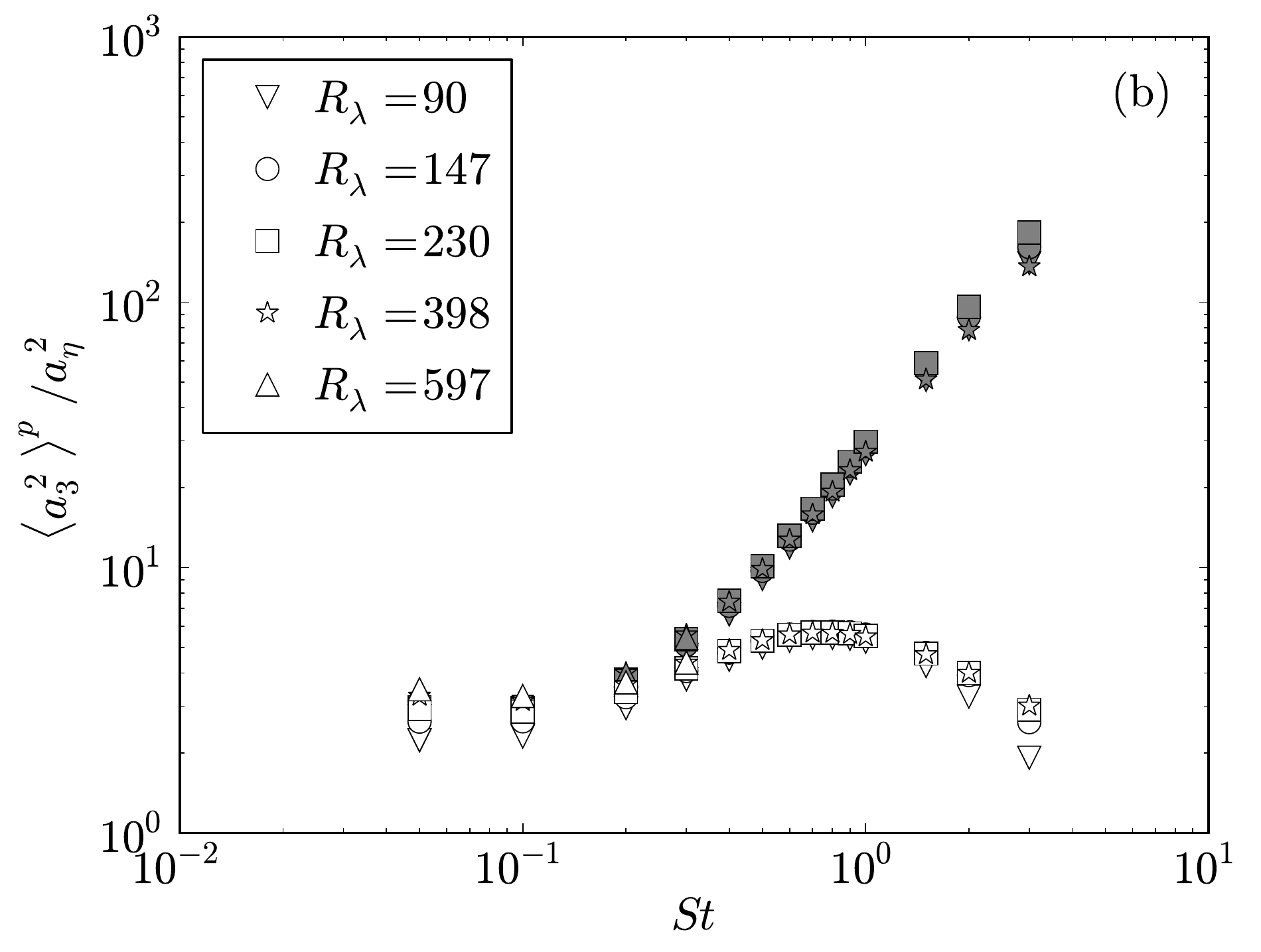}
 \caption{Inertial particle acceleration variances
 in the horizontal (a) and vertical (b) directions for different values of 
 $St$ and $R_\lambda$. All data are with gravity included ($Fr = 0.052$).
 Open symbols denote $\langle a^2_{1} \rangle^p$ (a) and $\langle a^2_{3} \rangle^p$ (b)
 and filled symbols denote $\langle (du_{1}/dt)^2 \rangle^p$ (a)
 and $\langle (du_{3}/dt)^2 \rangle^p$ (b).}
 \label{fig:acceleration_variance_particle_fluid}
\end{figure}

The particle acceleration in the limit of low $St$ and high $Sv$ (i.e., the limit $Fr\rightarrow0$) can be
 approximated by the derivative of the fluid velocity \textit{along the inertial particle trajectory} 
 (denoted here as $d\bm{u}^{p}/dt$), yielding \citep[see][]{bec06a}
\begin{equation}
 \bm{a}^p(t) \approx \frac{d \bm{u}^p}{dt} =\frac{\partial \bm{u}(\bm{x}^p(t),t)}{\partial t} 
 + \bm{v}^p(t) \cdot \nabla \bm{u}(\bm{x}^p(t),t) \mathrm{.}
 \label{eq:accel_lowSt_highSv}
\end{equation}
We compare the variances of the horizontal and vertical components of 
$d \bm{u}^p(t)/dt$ (denoted as $\langle (du_1/dt)^2 \rangle^{p}$ and $\langle (du_3/dt)^2 \rangle^{p}$, respectively) 
with $\langle a_1^2 \rangle^{p}$ and $\langle a_3^2 \rangle^{p}$ in figure~\ref{fig:acceleration_variance_particle_fluid}. 
The variances of $\bm{a}^p(t)$ and $d \bm{u}^p(t)/dt$ are almost identical at low $St$, as expected. 
As $St$ increases, the particles have larger settling velocities, 
causing the fluid seen by the particles to change more rapidly in time. 
As a result, the variance of $d\bm{u}^{p}/dt$ increases monotonically without bound with increasing $St$. 
In contrast, the inertial particle acceleration peaks around $St \sim 1$ and then decreases, 
causing the variance of $d\bm{u}^{p}/dt$ to exceed that of $\mathbf{a}^p(t)$ for $St \gtrsim 0.3$. 
The trend for the acceleration variance occurs because the particle's inertia also modulates its 
response to the underlying fluid through `inertial filtering' \citep{bec06a,salazar12b}. 
The particle is unable to respond to fluid fluctuations with frequencies greater than 
$O(1/\tau_p)$, causing a reduction in the acceleration variance. 
Beyond $St \sim 1$, inertial filtering dominates the enhancement due to gravitational settling, 
causing the particle accelerations to decrease with increasing $St$.

\cite{parishani15} also observed a peak in the particle acceleration variance
for $St \sim 1$ with gravity, and argued that the preferential sweeping mechanism identified
in \cite{wang93} is responsible for producing these large accelerations. 
While we observe a similar peak in the acceleration variances,
we find preferential sweeping to be negligible at $St \sim 1$ 
for $R_\lambda \lesssim 230$ (see \S \ref{sec:settling}).
Our data also indicate that the preferential sweeping of $St \sim 1$ particles
varies significantly with $R_\lambda$, while the acceleration variances are nearly
independent of $R_\lambda$. These findings suggest that preferential sweeping
cannot fully explain the trends in the acceleration
variances with gravity. In contrast, the approach we have taken is able to predict the
dependence of the accelerations on $St$ and $Sv$ without appealing to preferential-sweeping arguments.

By comparing figure~\ref{fig:acceleration_variance_particle_fluid}(a) 
and figure~\ref{fig:acceleration_variance_particle_fluid}(b),
we notice that $\langle a_1^2 \rangle^{p}$ generally exceeds $\langle a_3^2 \rangle^{p}$.
We use the acceleration model derived in \textsection\ref{sec:high_g_acceleration} 
to explain this result in the limit $Sv \gg u'/u_\eta$.
From \textsection\ref{sec:high_g_acceleration}, we have the following models
for the particle acceleration variances in the high-$Sv$ limit:
\begin{equation}
 \label{eq:accel_variance_perp_body}
 \frac{\langle a^2_1 \rangle^p}{a_\eta^2} = \left[ \frac{u'}{u_\eta} \right]^2
 \left[ \frac{Sv}{St} \right]
  \left[ \frac{2 St Sv + 3 \left( \frac{\ell}{\eta} \right)}
 {2 \left( St Sv + \frac{\ell}{\eta} \right)^2} \right] \mathrm{,}
\end{equation}
and
\begin{equation}
\label{eq:accel_variance_parallel_body}
 \frac{\langle a^2_3 \rangle^p}{a_\eta^2} =
 \left[ \frac{u'}{u_\eta} \right]^2 \left[ \frac{Sv}{St} \right]
 \left[ \frac{1}{St Sv + \frac{\ell}{\eta}} \right] \mathrm{.}
\end{equation}
By taking the ratio of (\ref{eq:accel_variance_perp_body}) and (\ref{eq:accel_variance_parallel_body}), we see that
\begin{equation}
 \frac{\langle {a^2_1} \rangle^p}{\langle {a^2_3} \rangle^p} = 
 \frac{2 StSv + 3 \left( \frac{\ell}{\eta} \right)}
      {2 StSv + 2 \left( \frac{\ell}{\eta} \right)}
       \mathrm{,}
\end{equation}
and thus the model also predicts that $\langle a_1^2 \rangle^{p}$ exceeds $\langle a_3^2 \rangle^{p}$ by an amount that increases with $R_\lambda$.

In the derivation of this model, we assumed that the particles' primary motion is in the vertical direction, 
from which it followed that the particle accelerations are related to the correlation lengthscales
of the fluid velocities sampled along the particle trajectories. 
Since in isotropic turbulence the longitudinal integral lengthscale
is twice the transverse integral lengthscale, the particles (as they fall vertically) 
will experience vertical fluid velocities that
are more strongly correlated over a time $\tau_p$ than are the horizontal fluid velocities.
(This phenomenon was referred to as the `continuity effect' in \cite{csanady63},
and is also explained in \cite{yudine59} and \cite{good14}.)
As a result, the horizontal fluid velocities sampled by the particles will change more rapidly, leading to
larger particle accelerations in those directions.

In \cite{parishani15}, the authors argued that the differences in the horizontal
and vertical particle accelerations are linked to the preferential sweeping mechanism
of \cite{wang93}. That is, a falling particle will accelerate toward the downward-moving
part of an eddy, where the vertical fluid velocity changes slowly, but the horizontal fluid
velocity changes rapidly. While this explanation is plausible for $St \lesssim 0.3$
(where preferential sweeping is strong, as shown in \S \ref{sec:settling}),
it fails to explain the strong difference between the horizontal and vertical accelerations
at higher values of $St$, where we observe negligible preferential sweeping
for $R_\lambda \lesssim 230$.
The models developed by \cite{parishani15} are for the low-$St$
limit, whereas our approach is based on the characteristics of the Eulerian velocity
field and provides a quantitative prediction for the accelerations at larger values of $St$.

\begin{figure}
 \centering
 \includegraphics[width=2.6in]{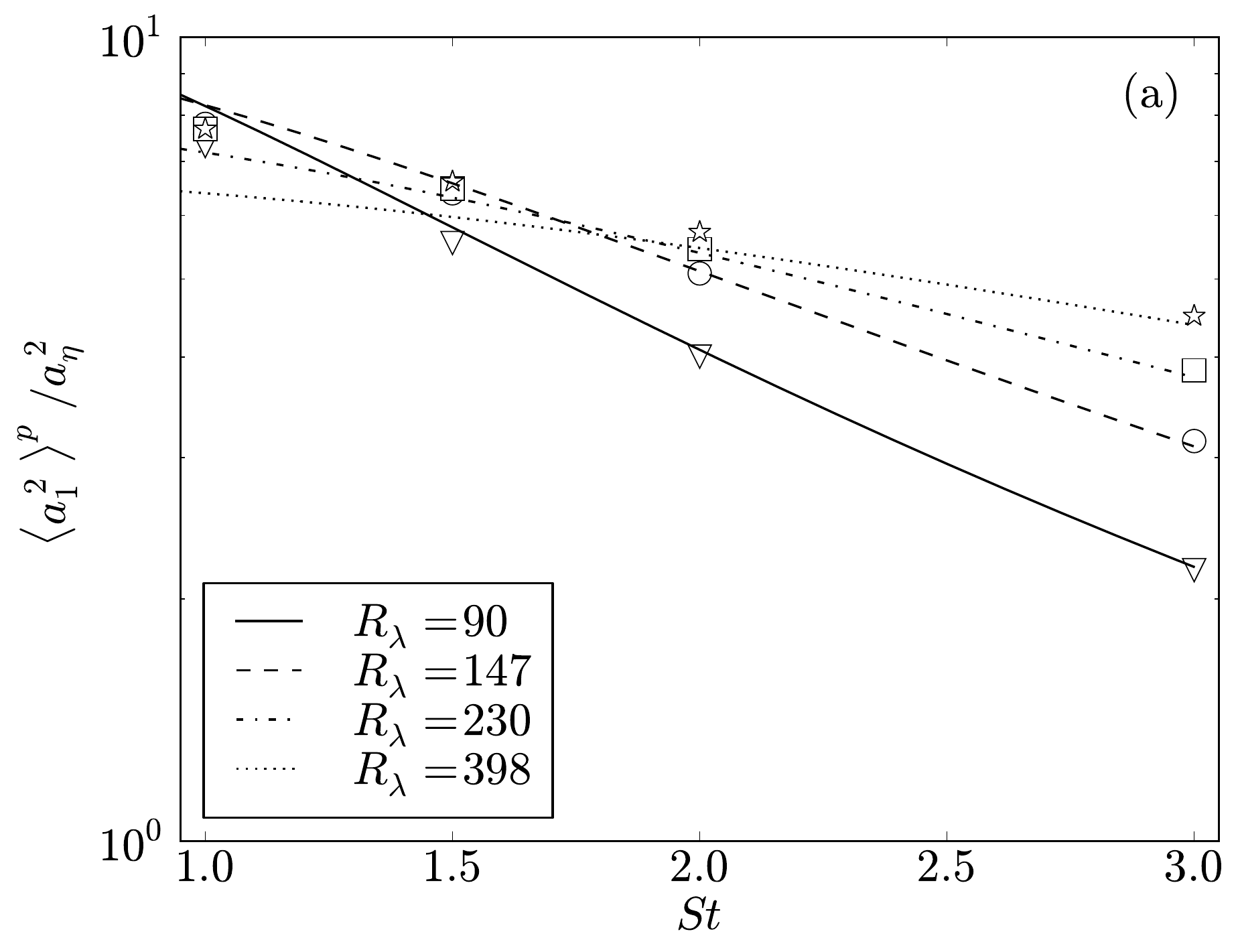}
 \includegraphics[width=2.6in]{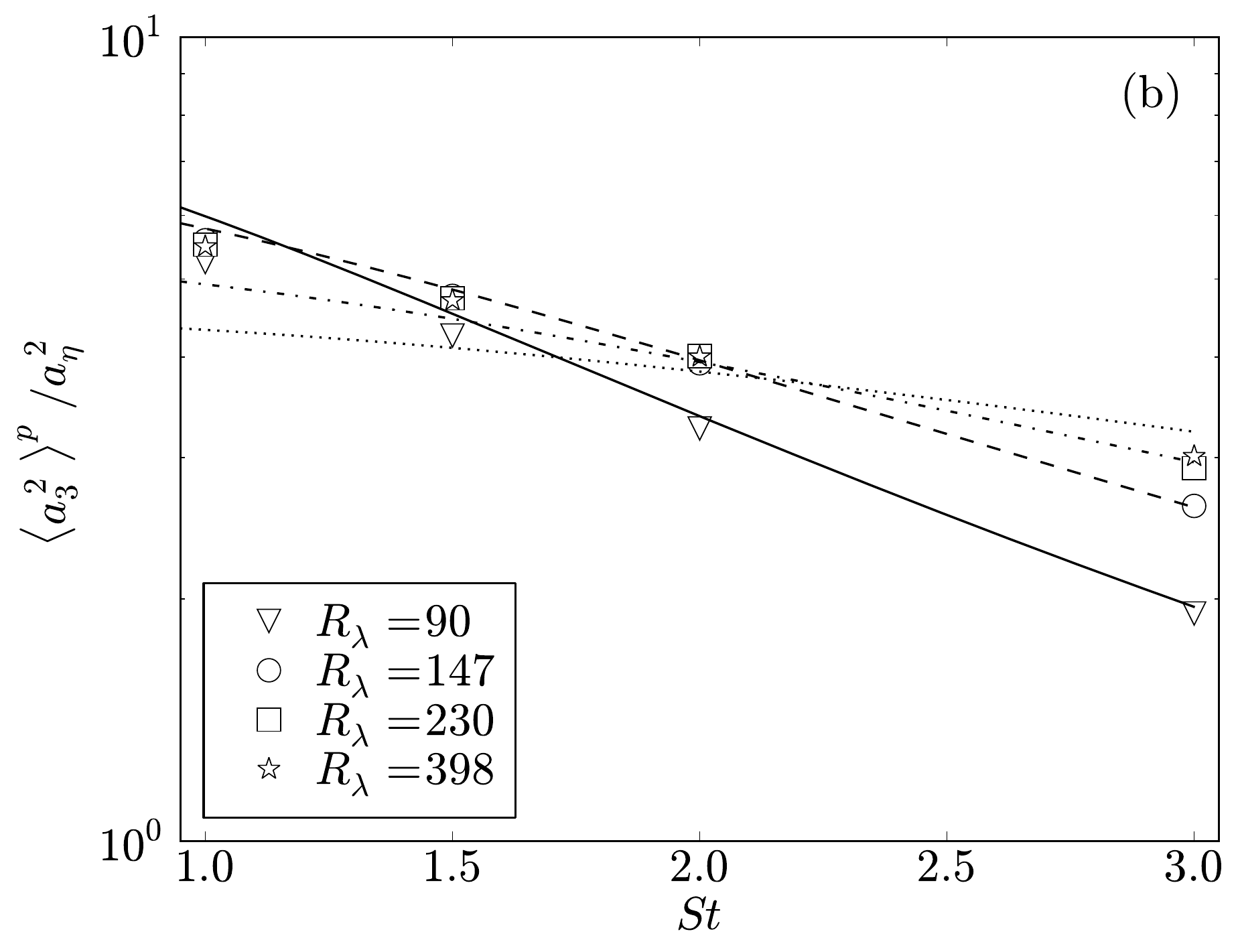}
 \caption{Comparision between DNS data and theoretical predictions
 for the variance of the particle accelerations in the horizontal (a) and
 vertical (b) directions at high $St$. Open symbols denote DNS data with gravity ($Fr = 0.052$), 
 and the lines in (a) and (b) indicate the predictions from 
 (\ref{eq:accel_variance_perp_body}) and (\ref{eq:accel_variance_parallel_body}), respectively,
 for the different Reynolds numbers simulated.}
 \label{fig:acceleration_variance_theory}
\end{figure}

In figure~\ref{fig:acceleration_variance_theory}, we compare the DNS data for
$\langle a_1^2 \rangle^{p}$ and $\langle a_3^2 \rangle^{p}$ to the model predictions
for $St \geq 1$ ($Sv \geq 19.3$).
We find that the results are in good agreement at the largest values of $St$, with the model being able
to predict both the trends with $R_\lambda$ and the decrease in the variances with increasing $St$.

\begin{figure}
 \centering
 \includegraphics[height=2.2in]{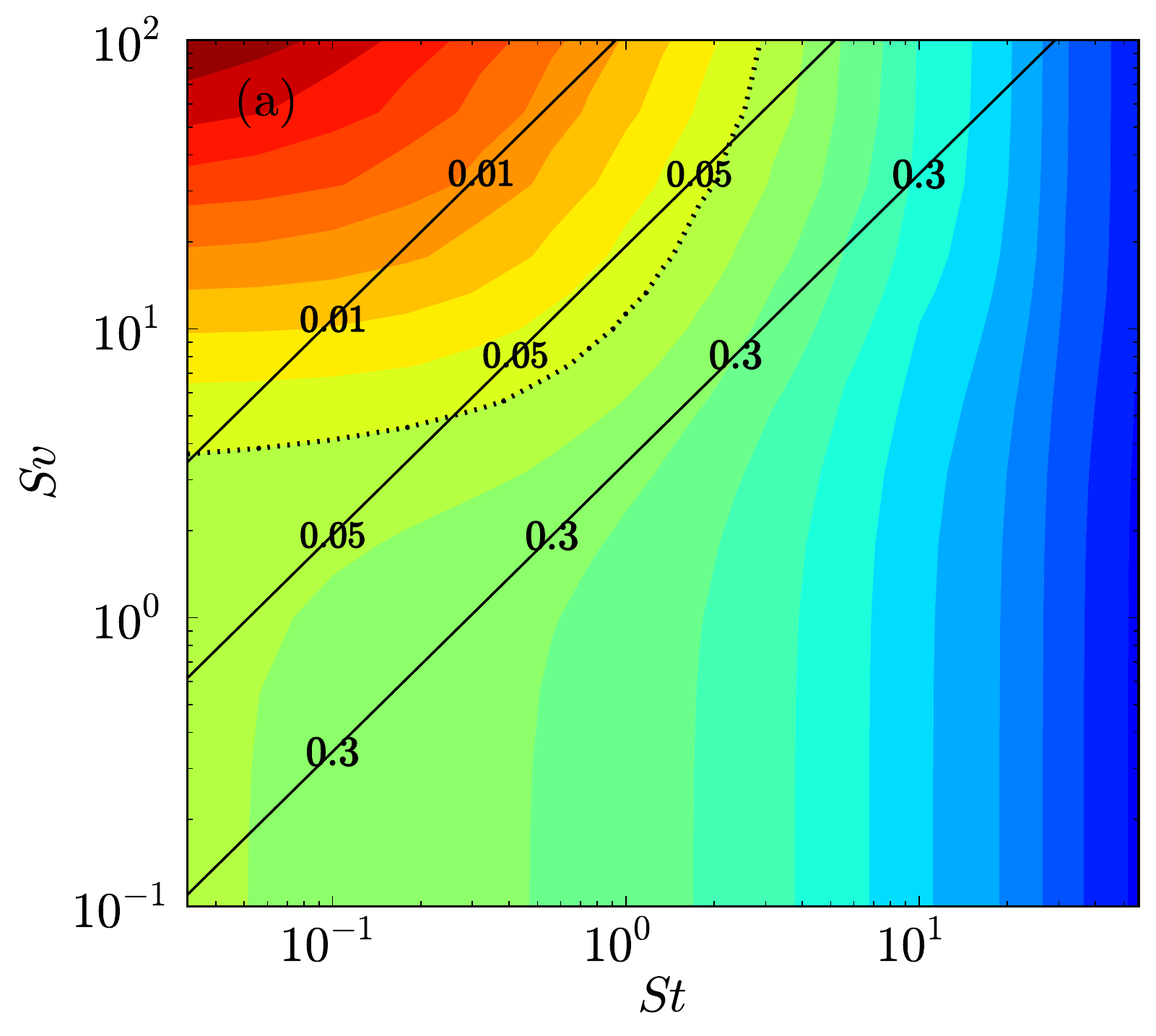}
 \includegraphics[height=2.2in]{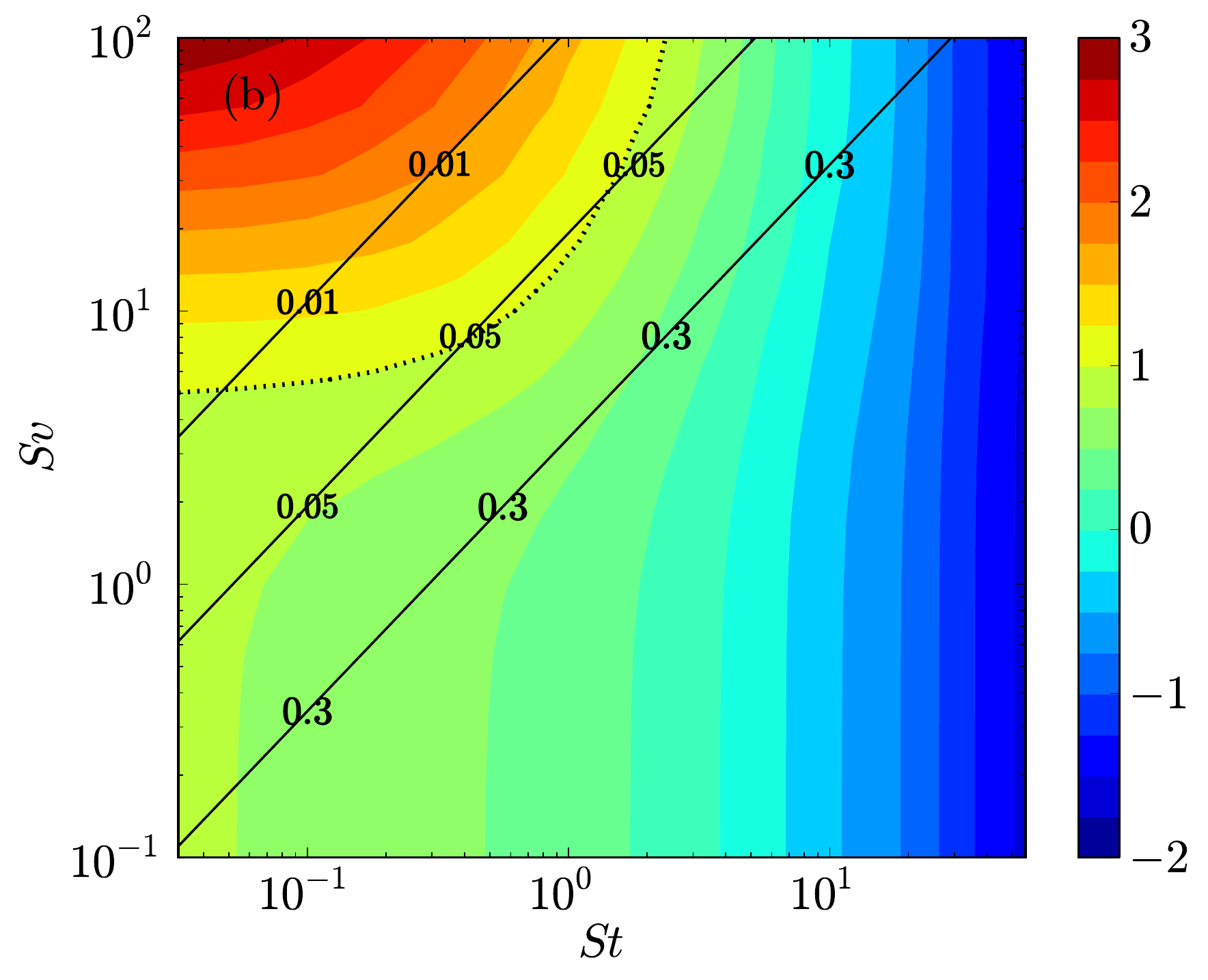}
 \caption{Filled contours of the particle acceleration variances, 
 $\langle {a_1^2} \rangle^p$ (a) and $\langle a_3^2 \rangle^p$ (b),
 normalized by the Kolmogorov acceleration variance $a_\eta^2$, for $R_\lambda = 227$.
 The contours are logarithmically scaled, and the labels on the colorbar
 denote exponents of the decade. The diagonal lines denote three different
 values of $Fr$, corresponding to conditions representative of stratiform clouds
 ($Fr = 0.01$), cumulus clouds ($Fr = 0.05$), and cumulonimbus clouds ($Fr = 0.3$).
 The dotted lines corresponds to $\langle a_{1}^2 \rangle^p = a^2_\eta$ in (a)
 and $\langle a_{3}^2 \rangle^p = a^2_\eta$ in (b).}
 \label{fig:accelerations_St_Sv}
\end{figure}

To further explore the trends in the accelerations for particles with varying levels
of inertia and gravity, we plot in figure~\ref{fig:accelerations_St_Sv} the inertial particle
acceleration variances for $0 < St \leq 56.2$, $0 < Sv \leq 100$, and $R_\lambda = 227$.
(Recall that statistics at the largest values of $Sv$ have errors due to unphysical effects of the periodic
boundary conditions. Refer to Appendix \ref{sec:periodicity} for a more detailed explanation.)
As expected, the particle acceleration variances are the largest for $St \ll 1$ and $Sv \gg 1$,
due to the gravitational trajectory effect, as predicted by
(\ref{eq:accel_variance_perp_body}) and (\ref{eq:accel_variance_parallel_body}).
For sufficiently large $St$, particles filter out
nearly all of the large-scale turbulence, and so acceleration variances are small.
For intermediate $St$ and $Sv$, the particle accelerations are determined by a combination
of preferential-sampling, gravitational-trajectory, and inertial-filtering effects.
We also observe from figure~\ref{fig:accelerations_St_Sv} that the particle acceleration variances
are the largest under conditions representative of stratiform clouds
($Fr\sim0.01$) and the smallest under conditions representative of cumulonimbus clouds ($Fr\sim0.3$).

\section{Two-particle statistics}
\label{sec:two_particle}
We now consider particle pair statistics, 
in particular, particle relative velocities (\textsection \ref{sec:relative_velocity}), 
clustering (\textsection \ref{sec:clustering}), 
and the collision kernels (\textsection \ref{sec:collision_kernel}). 
In each case, we compare our results to those without gravity (from Part I) to highlight the role of gravity.

\subsection{Particle relative velocities}
\label{sec:relative_velocity}

In this section, we examine the effect of gravity on the relative velocities of two identical inertial particles.
We first discuss the expected effect of gravity on the relative velocities 
from a theoretical framework (\textsection \ref{sec:relative_velocity_theory}), 
and then analyze the DNS results (\textsection \ref{sec:relative_velocity_dissipation}).

\subsubsection{Theoretical framework for particle relative velocities}
\label{sec:relative_velocity_theory}

As in Part I, we define the relative position and velocity as $\bm{r}^p(t)$ and $\bm{w}^p(t)$, respectively, 
and $\bm{\Delta u}(\bm{r}^p(t),t)$ as the difference in fluid velocities at the particle locations. 
By subtracting (\ref{eq:maxey_riley}) for two particles at a given instant in time, 
we get the following equation for the relative position and velocity of the two particles
\begin{equation}
 \frac{d^2 \bm{r}^p}{dt^2} = \frac{d \bm{w}^p}{dt} =
 \frac{\bm{\Delta u}\left(\bm{r}^p(t),t\right)-\bm{w}^p(t)}{\tau_p}\ .
 \label{eq:maxey_riley_relative}
\end{equation}
Notice that the gravity vector precisely cancels out of this equation, 
making it appear as though gravity does not influence the relative motion of the particle pair; 
however, this is not true, as gravity has an implicit effect on the statistics of 
$\bm{\Delta u}\left(\bm{r}^p(t),t\right)$, 
and this substantially modifies the relative position and velocity statistics of the particles 
(as will be shown below).

Following the nomenclature in Part I, we define the second-order particle relative velocity
structure function tensor as
\begin{equation*}
 \bm{S}^p_2(\bm{r}) = \Big \langle \bm{w}^p(t) \bm{w}^p(t) \Big \rangle_{\bm{r}} \mathrm{,}
\end{equation*}
where $\langle \cdot \rangle_{\bm{r}}$ denotes an ensemble average conditioned on 
$\bm{r}^p(t) = \bm{r}$.
We now use the formal solution of (\ref{eq:maxey_riley_relative}) to 
construct the exact expression for $\bm{S}^p_2(\bm{r})$. This will allow us to 
explain the expected trends in the structure functions with inertia and gravity.
The exact expression for $\bm{S}^p_2(\bm{r})$ is given by \citep[e.g., see][]{pan10}
\begin{equation}
\label{eq:Sp2_general}
 \bm{\hat{S}}^p_2(\bm{\hat{r}}) = 
 \frac{1}{St^2} \int_{-\infty}^0 \int_{-\infty}^0 
 \Big \langle \bm{\Delta \hat{u}}(\bm{\hat{r}}^p(\hat{s}),\hat{s}) 
              \bm{\Delta \hat{u}}(\bm{\hat{r}}^p(\hat{S}),\hat{S}) \Big \rangle_{\bm{\hat{r}}}
 \exp \left[ St^{-1} (\hat{s}+\hat{S}) \right] d\hat{s} d\hat{S} \mathrm{,}
\end{equation}
where $\hat{Z}$ denotes a variable $Z$ normalized by Kolmogorov scales.

We first consider (\ref{eq:Sp2_general}) in the absence of gravity.
From (\ref{eq:Sp2_general}), we observe that particle relative velocities
are affected by the fluid velocity difference
$\bm{\Delta u}(\bm{r}^p(t),t)$ along their trajectories. 
For small values of $St$, the 
high damping of the exponential term in (\ref{eq:Sp2_general}) 
causes contributions to the integral from the particle history to be negligible, 
and hence the particle relative velocities in the equation
become equivalent to the instantaneous fluid velocity difference at the current particle position,
reflecting the fact that preferential sampling is the dominant mechanism affecting relative particle motion.
As shown in Part I, preferential sampling
causes inertial particles to experience larger (smaller) fluid velocity differences than those of fluid
particles in directions parallel (perpendicular) to the particle separation vector.
As $St$ increases, in the absence of gravity,
the exponential damping diminishes and the influence of the particle history
becomes increasingly important. At higher $St$,
particles retain a memory of their interactions with the turbulence, 
and it is the fluid velocities at larger separations
along their path histories that dominate the contribution to the particle velocity dynamics, 
leading to particle relative velocities that are larger than the local fluid velocity difference.
This phenomenon has been referred
to as `caustics' \citep[e.g., see][]{wilkinson05,wilkinson06}.
Refer to \cite{bragg14b} for a more complete theoretical description of this
phenomenon, and to Part I for supporting DNS evidence.

We now introduce gravity and discuss its effect on the relative velocities. 
Recall in \textsection \ref{sec:topology} that gravity generally 
reduces the degree of preferential sampling. 
With gravity, we therefore expect the relative velocities of small-$St$ particles to be affected 
less by preferential sampling, and therefore to have dynamics closer to those of fluid particles. 
At higher values of $St$, gravity alters the inertial particle relative velocities 
through its influence on the path-history mechanism. 
We observed in \textsection \ref{sec:topology} that gravity reduces the 
correlation timescales of the flow along particle trajectories. 
This implies that gravity causes the fluid velocity differences in (\ref{eq:Sp2_general}) to 
become decorrelated more rapidly, and thus it reduces the 
correlation radius over which the relative velocities 
are influenced by their path-history interactions with the fluid. 
We therefore expect the particle relative velocities to be reduced by gravity.

In addition to the magnitude changes caused by gravity, 
its directional nature causes the symmetry of the particle field to become anisotropic. 
We define the mean inward, parallel and perpendicular relative velocity statistics as
\begin{eqnarray}
S^p_{-\parallel}(\bm{r}) & \equiv & -\int_{-\infty}^0 w_\parallel p(w_\parallel | \bm{r}) d w_\parallel\ ,\\
S^p_{2\parallel}(\bm{r}) & \equiv & -\int_{-\infty}^\infty \left[w_\parallel\right]^2 p(w_\parallel | \bm{r}) d w_\parallel\ ,\\
S^p_{2\perp}(\bm{r}) & \equiv & -\int_{-\infty}^\infty \left[w_\perp\right]^2 p(w_\perp | \bm{r}) d w_\perp\ ,
\end{eqnarray}
where $w_\parallel$ and $w_\perp$ are the parallel and perpendicular components 
of the relative velocity, respectively, 
and $p(w_\parallel | \bm{r})$ and $p(w_\perp | \bm{r})$ are 
their respective probability density functions, 
conditioned on the separation vector $\bm{r}$. 
In the absence of gravity, $S^p_{2\parallel}$, $S^p_{2\perp}$, and $S^p_{-\parallel}$ 
are functions of the separation distance $r\equiv|\bm{r}|$ only. 
With gravity, these statistics become functions of the angle between the separation vector 
$\bm{r}$ and the gravity vector $\boldsymbol{\mathfrak{g}}$, here denoted as $\theta$. 
A convenient way to express this dependence is to expand these functions 
in a spherical harmonic series 
\begin{eqnarray}
S^p_{-\parallel}(\bm{r}) & = & \sum_{\ell = 0}^{\infty} C_{2 \ell}^0 (r) Y_{2 \ell}^0 (\theta)\ ,\\
S^p_{2\parallel}(\bm{r}) & = & \sum_{\ell = 0}^{\infty} D_{2 \ell}^0 (r) Y_{2 \ell}^0 (\theta)\ ,\\
S^p_{2\perp}(\bm{r}) & = & \sum_{\ell = 0}^{\infty} E_{2 \ell}^0 (r) Y_{2 \ell}^0 (\theta)\ ,
\end{eqnarray}
where $C_{2 \ell}^0(r)$, $D_{2 \ell}^0(r)$, and $E_{2 \ell}^0(r)$, 
are the spherical harmonic coefficients, 
and $Y_{2 \ell}^0$ are the spherical harmonic functions. 
In discussing the results below, we will quantify the degree of anisotropy introduced by gravity 
by analyzing the relative magnitudes of the spherical harmonic coefficients. 
Similar expansions will be used to describe the anisotropic radial distribution function 
that results from gravity.

\subsubsection{Relative velocity results}
\label{sec:relative_velocity_dissipation}

We now analyze the results from the DNS using the decompositions described in 
\S\ref{sec:relative_velocity_theory}. 
To study how gravity affects the magnitude of the relative velocity, 
we first consider the zeroth-order coefficients, $C_0^0(r) \equiv S^p_{-\parallel}(r)$, 
$D_0^0(r) \equiv S^p_{2\parallel}(r)$, and $E_0^0(r) \equiv S^p_{2\perp}(r)$, 
where we have taken $Y_0^0 = 1$. 
In effect, these coefficients represent the spherical average over the anisotropic particle field.

\begin{figure}
 \centering
 \includegraphics[width=2.6in]{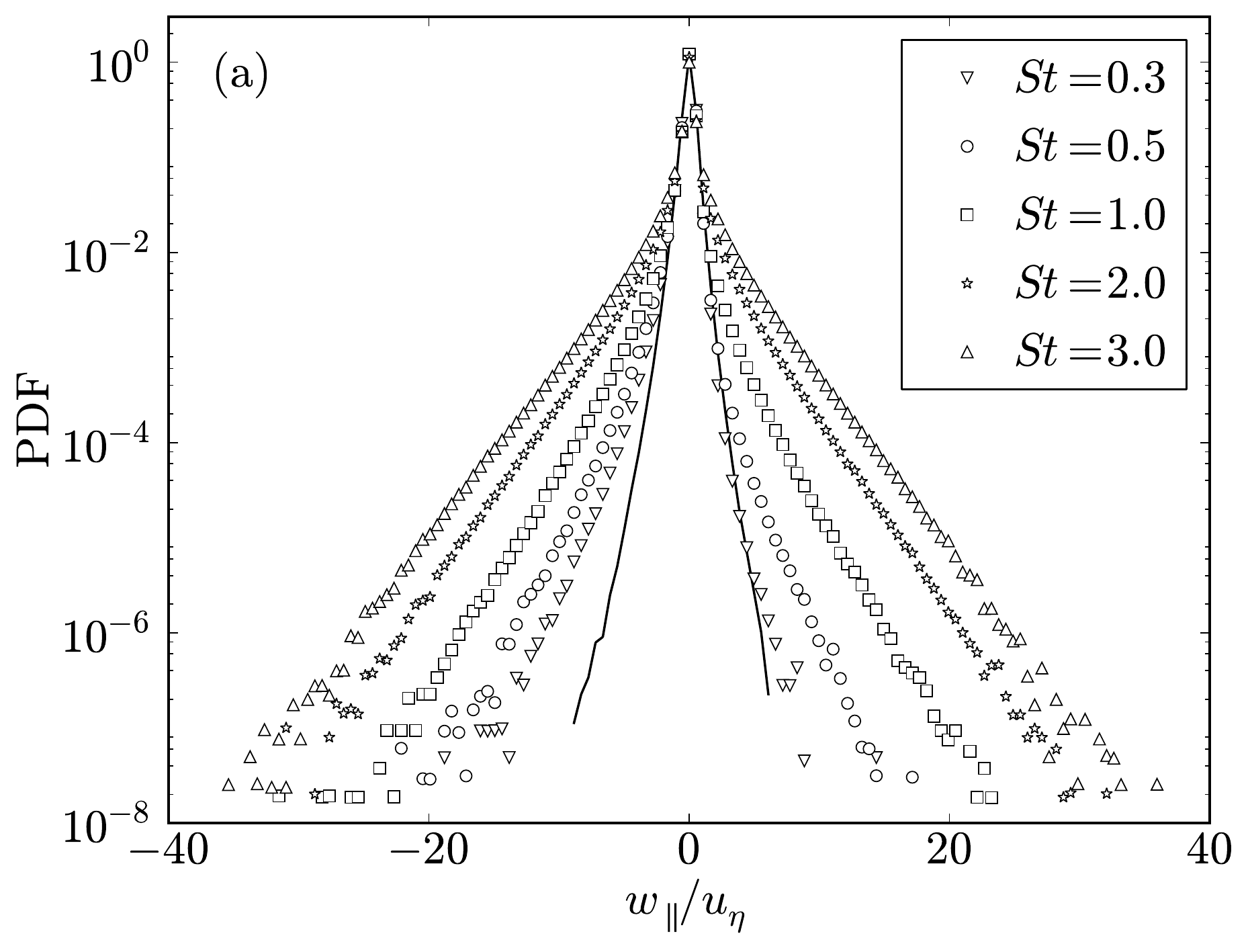}
 \includegraphics[width=2.6in]{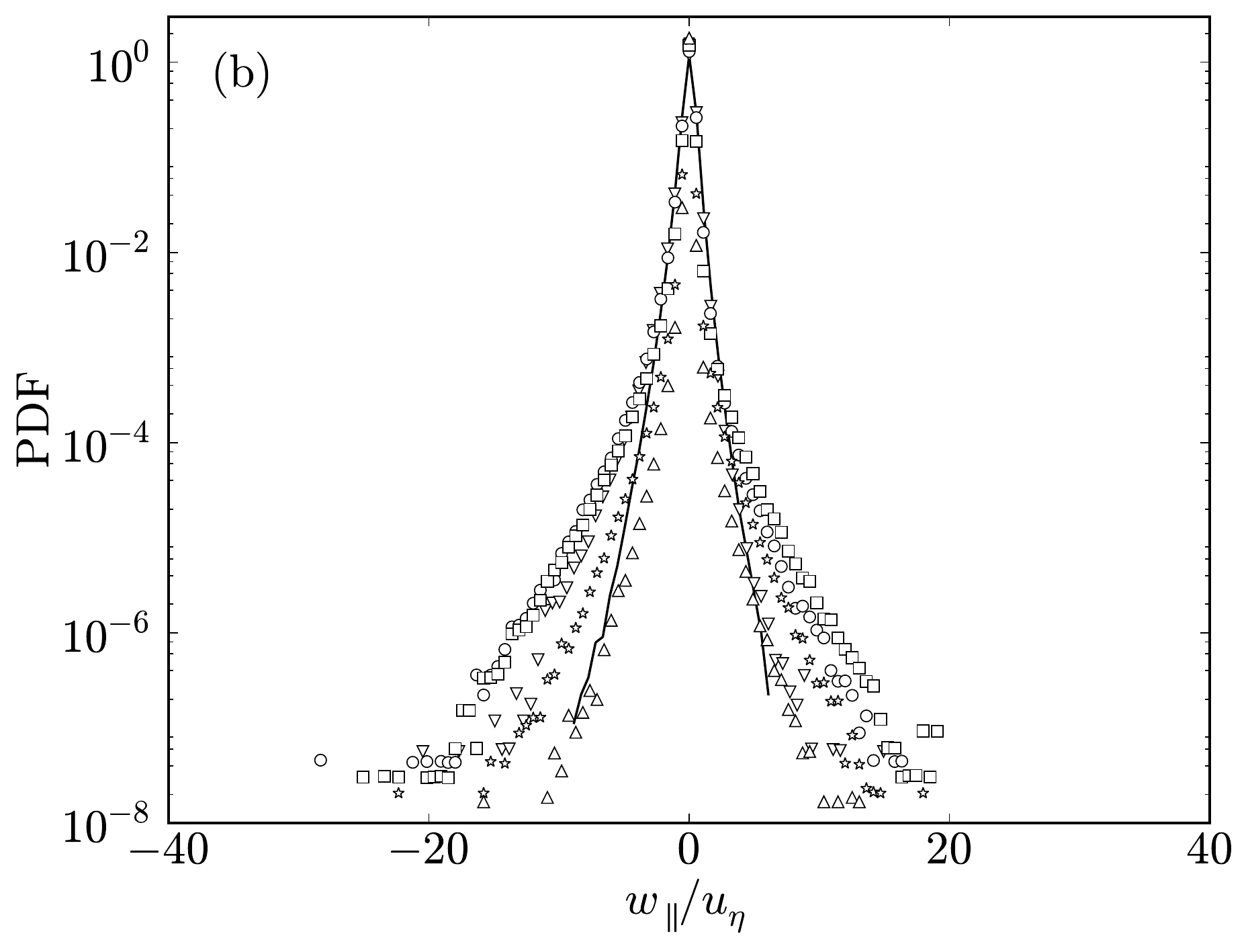}
 \caption{PDFs of the radial relative velocities without gravity (a)
 and with gravity ($Fr = 0.052$) (b) at $R_\lambda = 398$ and $0 \leq r/\eta \leq 2$. 
 The solid line denotes data for fluid ($St = 0$) particles.}
 \label{fig:relative_velocity_pdfs}
\end{figure}

Figure~\ref{fig:relative_velocity_pdfs} shows the probability density function (PDF) 
for the parallel component of the relative velocity for $0\le r/\eta\le 2$, 
at different Stokes numbers and a fixed Reynolds number of 398. 
The panel on the left shows the case of no gravity and the panel on the right the case of $Fr=0.052$. 
It is immediately apparent that gravity has a profound effect on the PDFs, 
causing a dramatic reduction in the tails of the distribution, particularly for particles with larger 
Stokes numbers. This is consistent with the explanation given in \S\ref{sec:relative_velocity_theory}, 
which argued that gravity suppresses path-history effects, thereby reducing the relative velocities.

\begin{figure}
 \centering
 \includegraphics[width=2.6in]{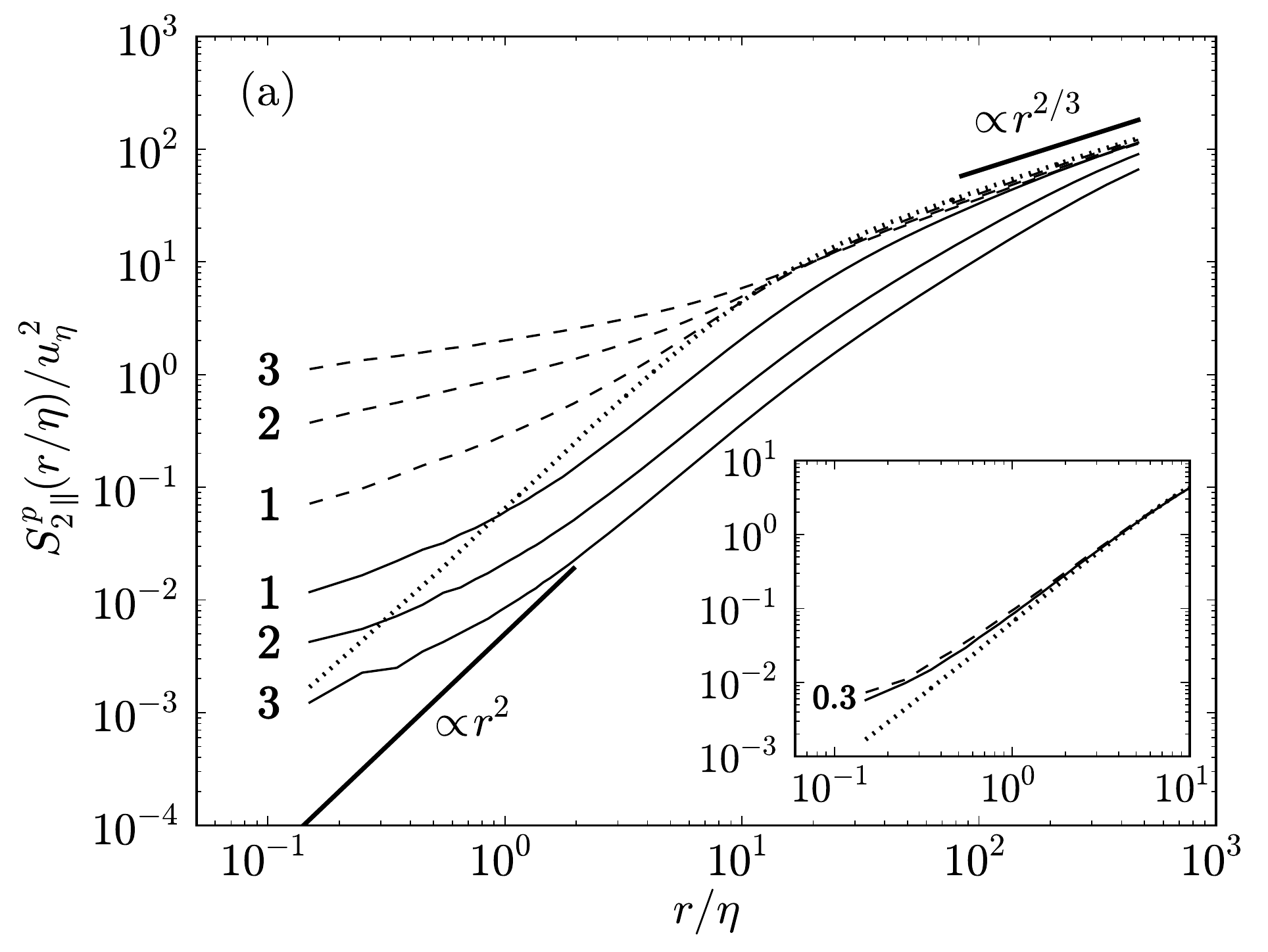}
 \includegraphics[width=2.6in]{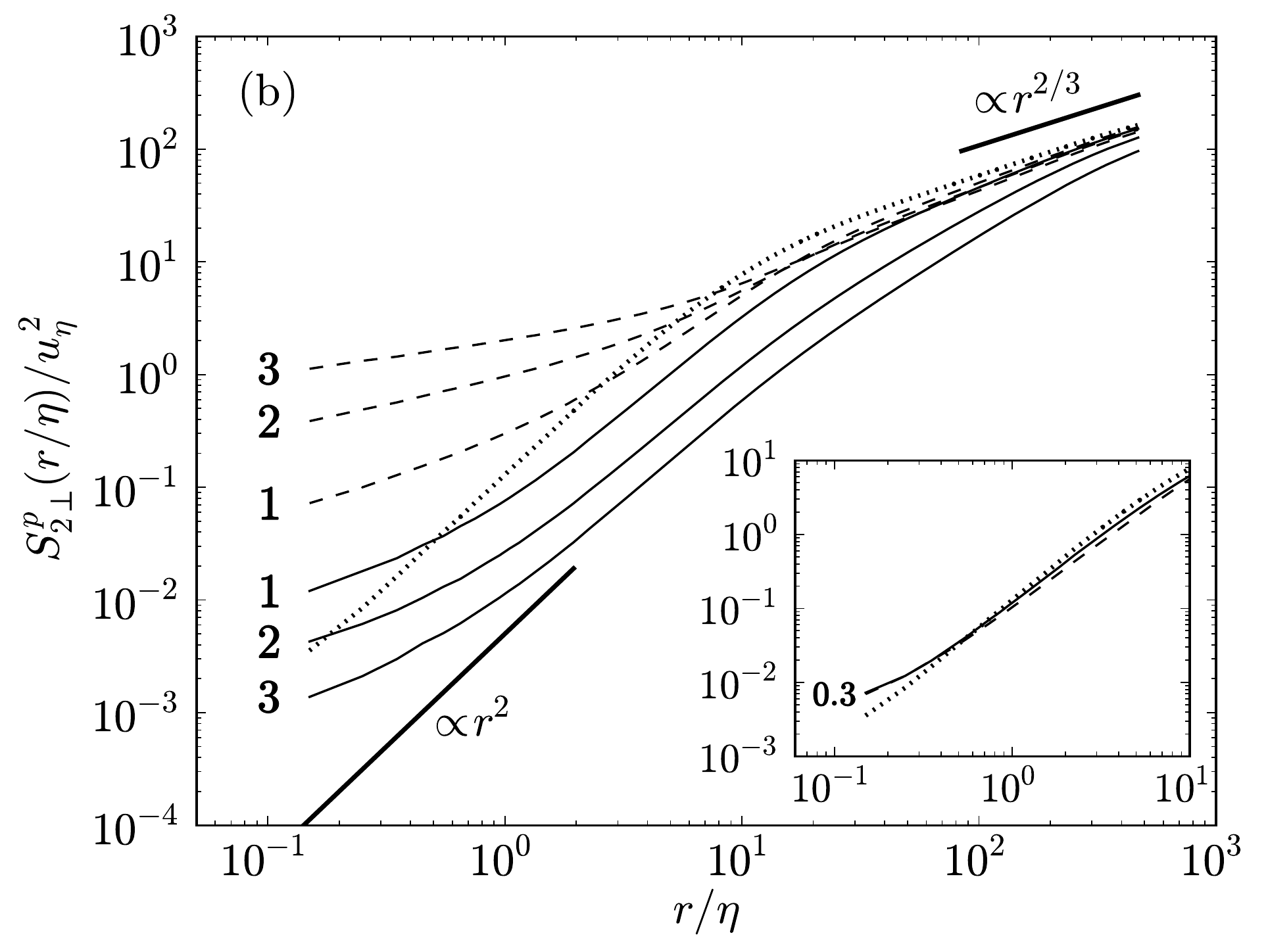}
 \caption{The variance of the particle relative velocities 
 parallel to the separation vector (a) and perpendicular to the separation vector (b) for $R_\lambda = 398$, 
 plotted as a function of  $r/\eta$ for different $St$. 
 The thin solid lines indicate data with gravity ($Fr = 0.052$), the dashed lines
 indicate data without gravity, and the thick dotted line indicates fluid ($St = 0$) particles.
 The Stokes numbers are indicated by the line labels, and
 the fluid velocity scalings are indicated by thick solid lines. 
 The main plots show $St = 0$ and $St \geq 1$,
 while the insets show $St=0$ and $St = 0.3$.}
 \label{fig:wr_lom_1024}
\end{figure}

We next consider the relative velocity variances as a function of the separation distance.
Figure~\ref{fig:wr_lom_1024} shows the spherically averaged quantities $S^p_{2\parallel}(r)$ and $S^p_{2\perp}(r)$.
For $St \geq 1$ (corresponding to $Sv \geq 19.3$), 
we observe that gravity strongly decreases the relative velocities, by orders of magnitude in some cases.
The explanation is that gravity reduces the effect of the path-history interactions, causing a reduction
in the relative velocities (cf. \S\ref{sec:relative_velocity_theory}).

Data at $St=0.3$ is shown in the insets in figure~\ref{fig:wr_lom_1024}.
For $St=0.3$ and very small separations ($r/\eta \lesssim 1$), the relative velocities
show evidence of path-history effects (see also Part I), causing the 
relative velocities to exceed those of $St=0$ particles in both the longitudinal and transverse directions.
However, at larger separations ($1 \lesssim r/\eta \lesssim 10$), we observe a crossover to the expected 
increase (decrease) in the longitudinal (transverse) relative velocities
as a result of preferential sampling. We also see that gravity causes the relative
velocities to approach those of the underlying fluid, since it decreases
preferential sampling effects. These trends are consistent with the discussion
in \textsection \ref{sec:relative_velocity_theory}. [Note that since particles
with $St < 0.3$ have only weak gravitational forces ($Sv < 5.79$), the 
effect of gravity on the relative velocities is less apparent,
and data from these particle classes are therefore not shown.]

The decreased influence
of path-history interactions is also apparent in the scaling of the relative
velocity variances in figure~\ref{fig:wr_lom_1024}.
In the dissipation range, the fluid relative velocity variances are proportional to $r^2$. 
In the absence of gravity, the relative velocities
for $St \geq 1$ do not exhibit $r^2$-scaling over any portion of the dissipation range, as was noted in Part I.
However, with gravity, we see a clearer $r^2$-scaling extending much deeper into the dissipation range
for $1 \lesssim r/\eta \lesssim 10$.  Compared to the case without gravity, 
one has to go to smaller separations to observe deviations from $r^2$ scaling, as gravity
reduces the path-history effect responsible for the formation of caustics.

\begin{figure}
  \centering
 \includegraphics[height=2.5in]{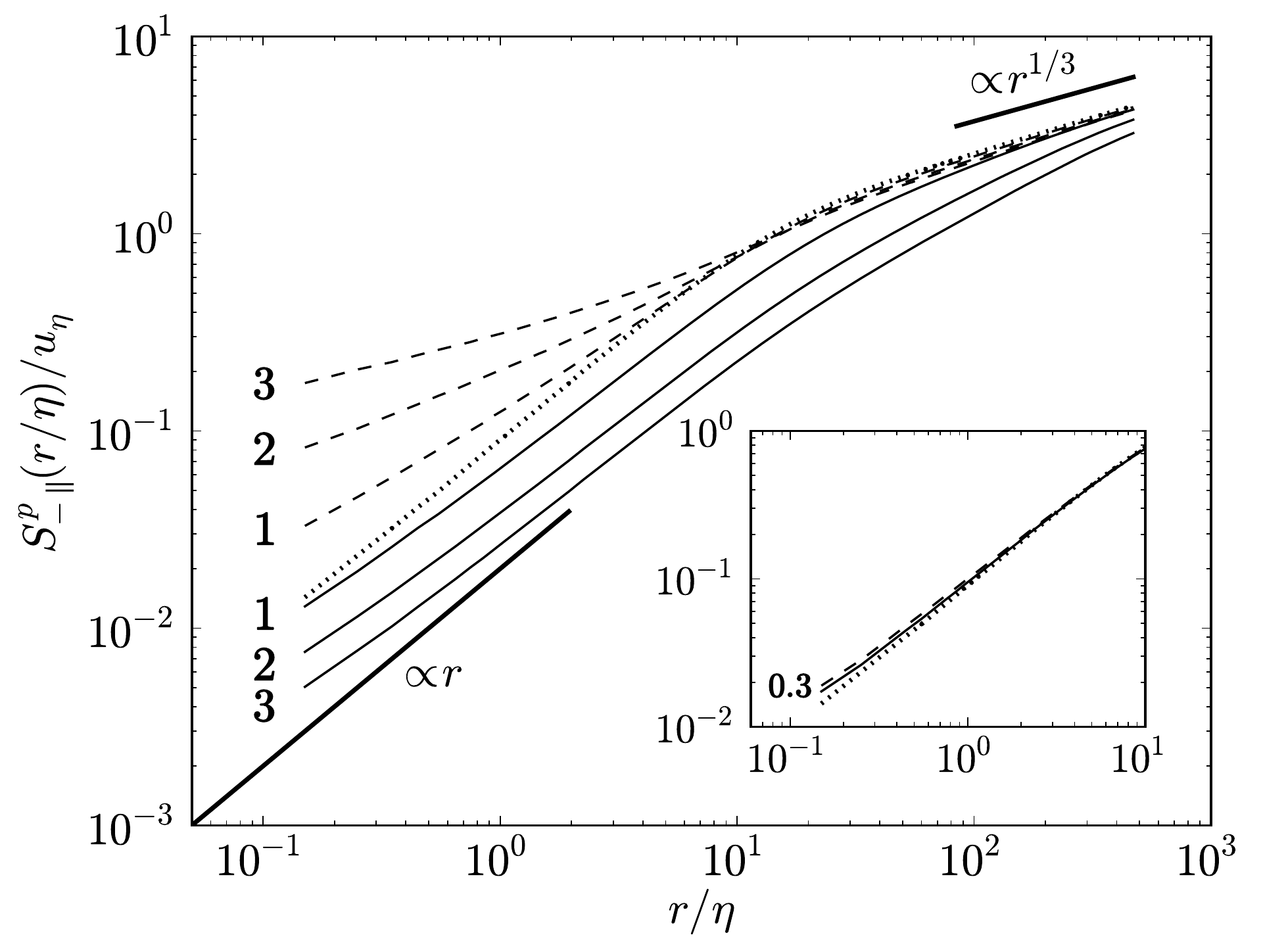}
 \caption{The mean inward particle relative velocity for $R_\lambda = 398$, 
 plotted as a function of  $r/\eta$ for different $St$.
 The thin solid lines indicate data with gravity ($Fr = 0.052$), the dashed lines
 indicate data without gravity, and the thick dotted line indicates fluid ($St = 0$) particles.
 The Stokes numbers are indicated by the line labels, and
 the fluid velocity scalings are shown by thick solid lines.
 The main plot shows $St = 0$ and $St \geq 1$,
 while the inset shows $St=0$ and $St = 0.3$.}
 \label{fig:wr_lom_1024_mean}
\end{figure}

Of particular interest to the collision kernel is the mean inward velocity (cf. \S\ref{sec:collision_kernel}). 
In general, the mean inward velocity is more difficult to analyze theoretically than the relative velocity 
variance, but qualitatively we expect both statistics to follow the same trends. 
In figure~\ref{fig:wr_lom_1024_mean}, we show $S^p_{-\parallel}(r)$ at $R_\lambda=398$, 
both with and without gravity. We see that the mean inward velocities, like the variances, 
decrease with the addition of gravity at large (small) $St$ due to the reduced influence of 
the path-history (preferential-sampling) mechanisms. 
Interestingly, with gravity,  the mean inward relative velocities 
(in contrast to the relative velocity variances) follow the local fluid scaling almost perfectly throughout 
the entire dissipation range. This result was also noted in \cite{bec14}. 
One plausible explanation is that since the mean inward velocity is a lower-order statistic than the velocity variance, 
it is less affected by path-history interactions and is more affected by the local fluid turbulence, 
causing the scaling exponents to increase (see Part I).

\begin{figure}
 \centering
 \includegraphics[height=2.2in]{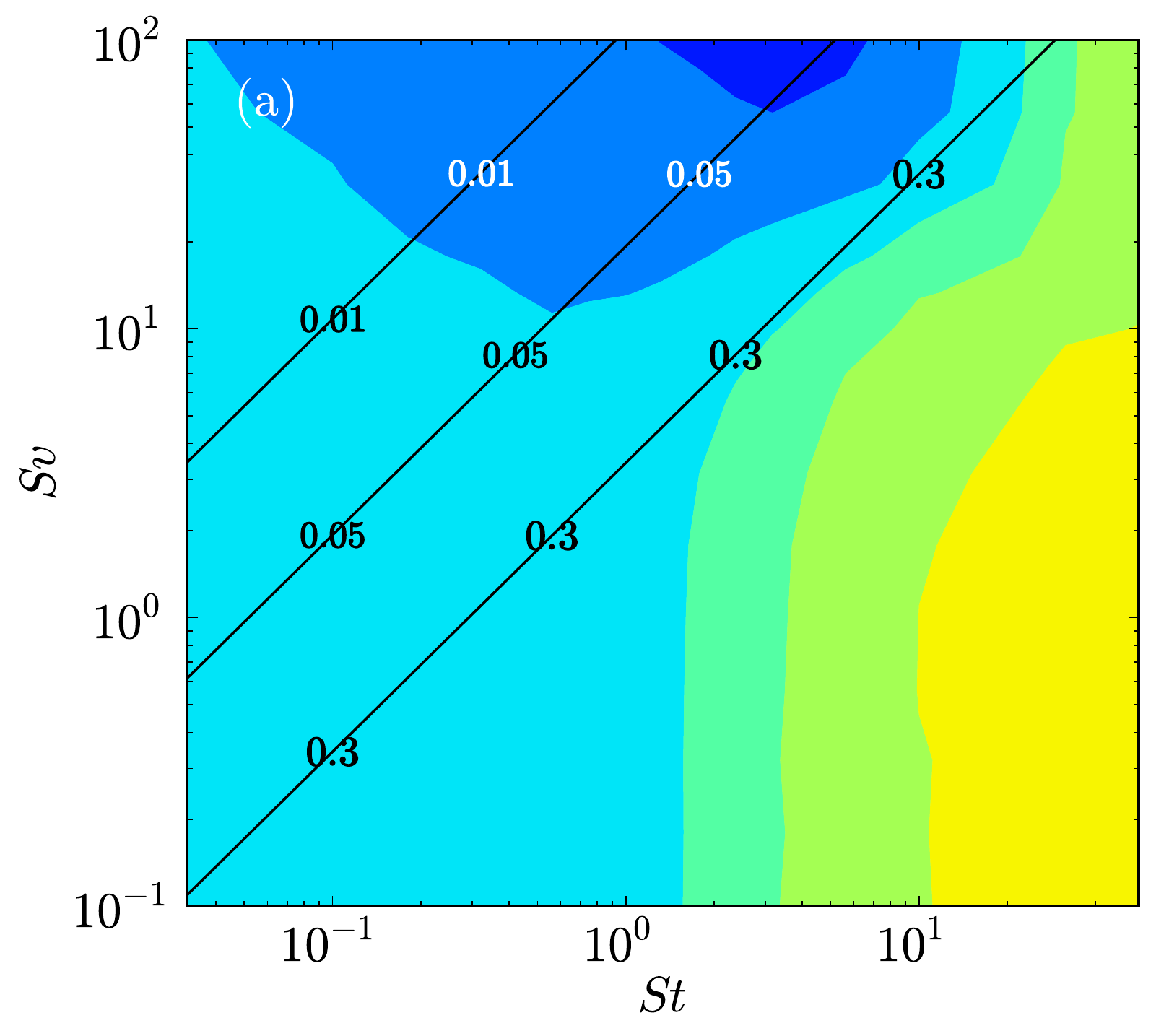}
 \includegraphics[height=2.2in]{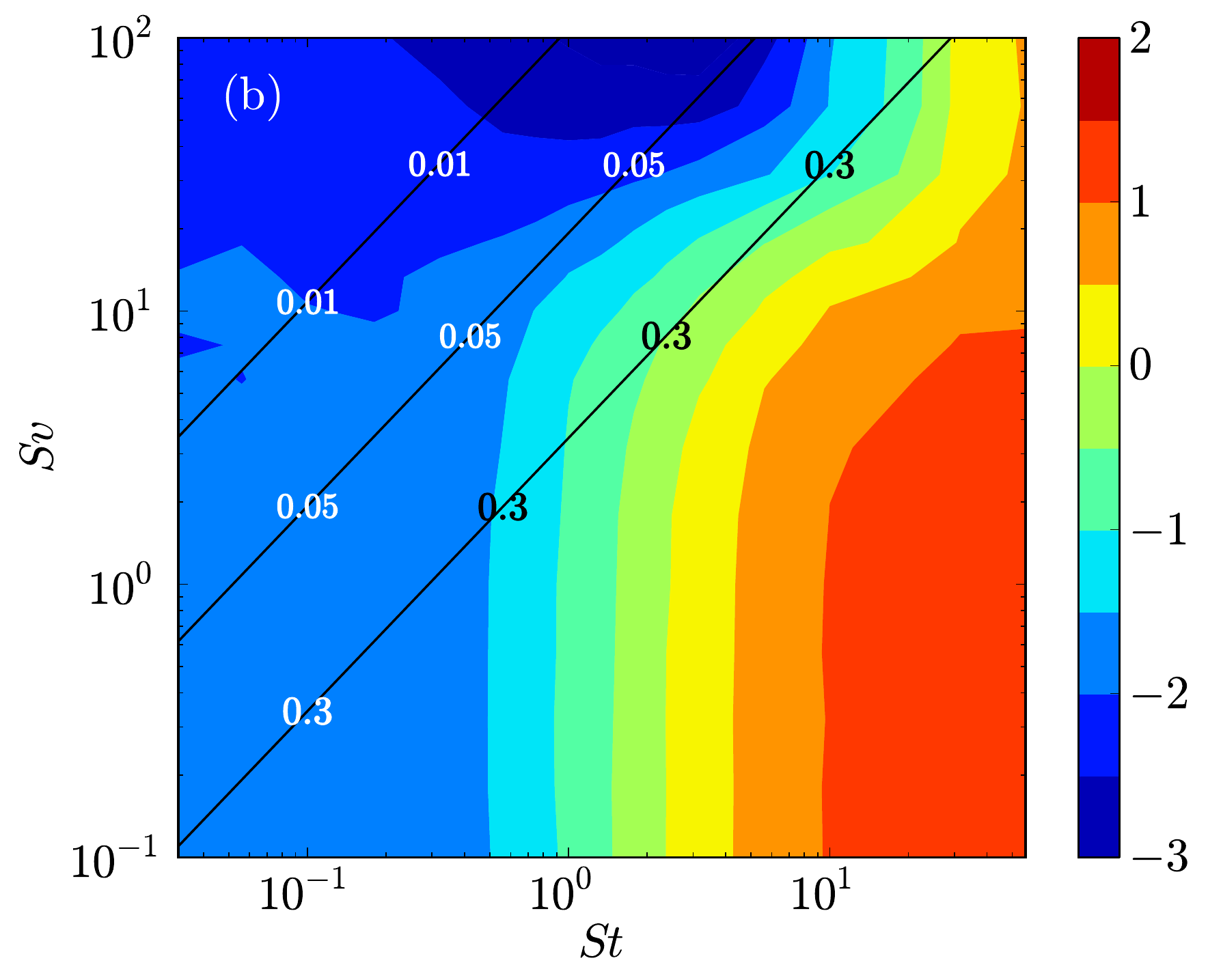}
 \caption{Filled contours of $S^p_{-\parallel}(r/\eta)/u_\eta$ (a) and $S^p_{2\parallel}(r/\eta)/u_\eta^2$ (b)
 evaluated at $r/\eta = 0.25$ and $R_\lambda=227$ for different values of $St$ and $Sv$.
 The contours are logarithmically scaled, and the colorbar labels indicate the exponents of the decade.
 The diagonal lines denote three different
 values of $Fr$, corresponding to conditions representative of stratiform clouds
 ($Fr = 0.01$), cumulus clouds ($Fr = 0.05$), and cumulonimbus clouds ($Fr = 0.3$).}
 \label{fig:wr_St_Sv}
\end{figure}

To further verify our arguments in \textsection \ref{sec:relative_velocity_theory},
it is helpful to decouple the effects of gravity and inertia by varying each 
independently. We do so in figure~\ref{fig:wr_St_Sv}, where we show $S^p_{-\parallel}(r/\eta = 0.25)$
and $S^p_{2\parallel}(r/\eta = 0.25)$ for $0 < St \leq 56.2$, $0 < Sv \leq 100$, and $R_\lambda = 227$.
While the results at high $Sv$ are likely artificially affected by the periodic boundary
conditions (see Appendix~\ref{sec:periodicity}), 
we can still use these data to discuss the qualitative trends in the
relative velocities at different values of $St$ and $Sv$.
We see that $S^p_{-\parallel}$ and $S^p_{2\parallel}$ have similar qualitative trends.
For $Sv \lesssim 10$, both quantities increase with increasing $St$,
either due to preferential-sampling effects (at low $St$) or path-history effects (at high $St$).
The relative velocities also decrease with increasing $Sv$, since gravity causes
both effects to be less significant, as discussed in \textsection \ref{sec:relative_velocity_theory}.
Finally, we observe that the relative velocities are the smallest for the largest values of $Sv$
and $St \sim 1$. At smaller (larger) $St$, preferential-concentration (path-history) effects are more significant,
leading to an increase in the relative velocities.
An implication of these results for clouds is that for a given value of $St$, droplets in stratiform clouds ($Fr = 0.01$) 
will generally have smaller relative velocities than droplets in cumulonimbus clouds ($Fr = 0.3$).

\begin{figure}
 \centering
 \includegraphics[width=2.6in]{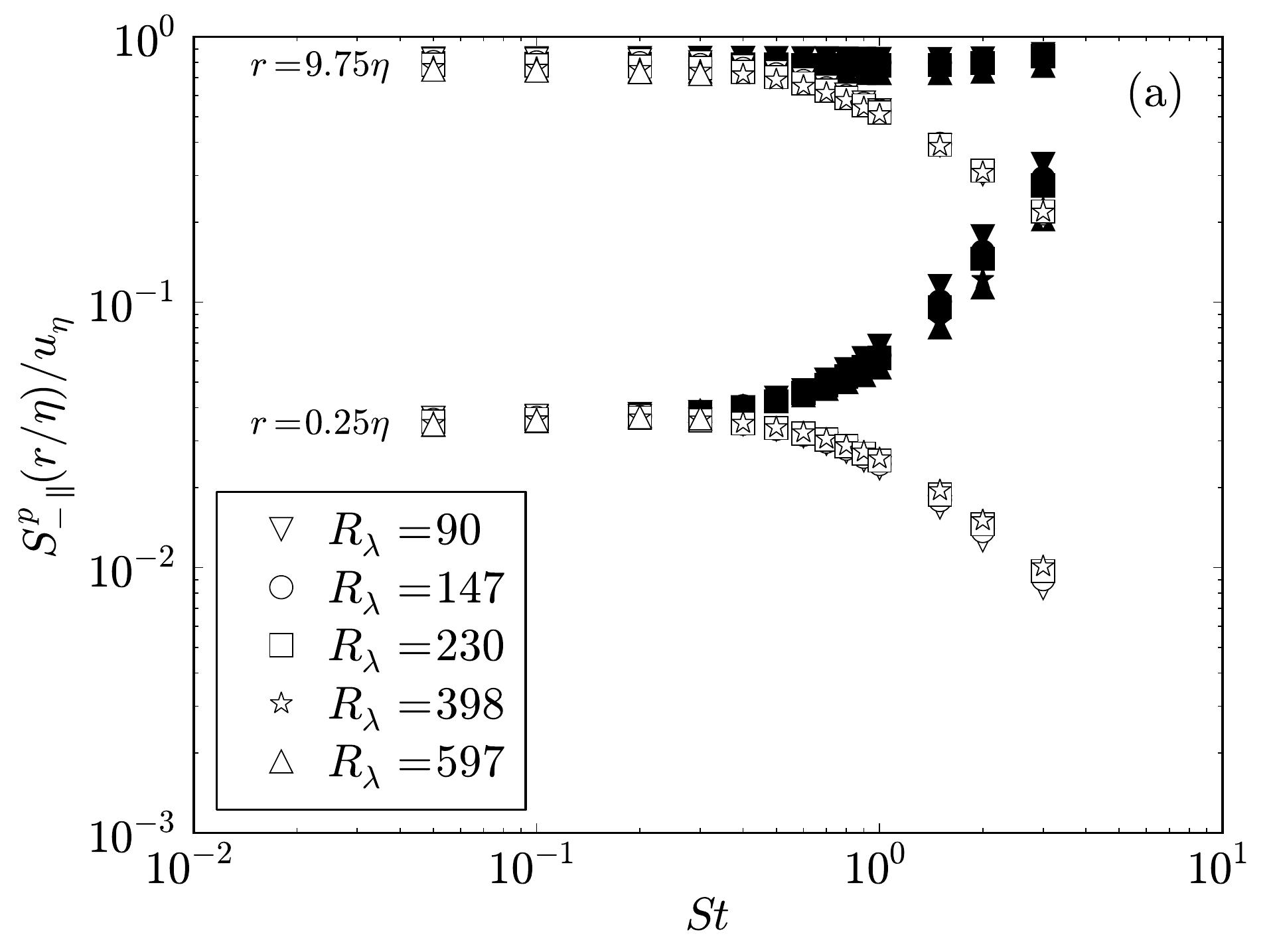}
 \includegraphics[width=2.6in]{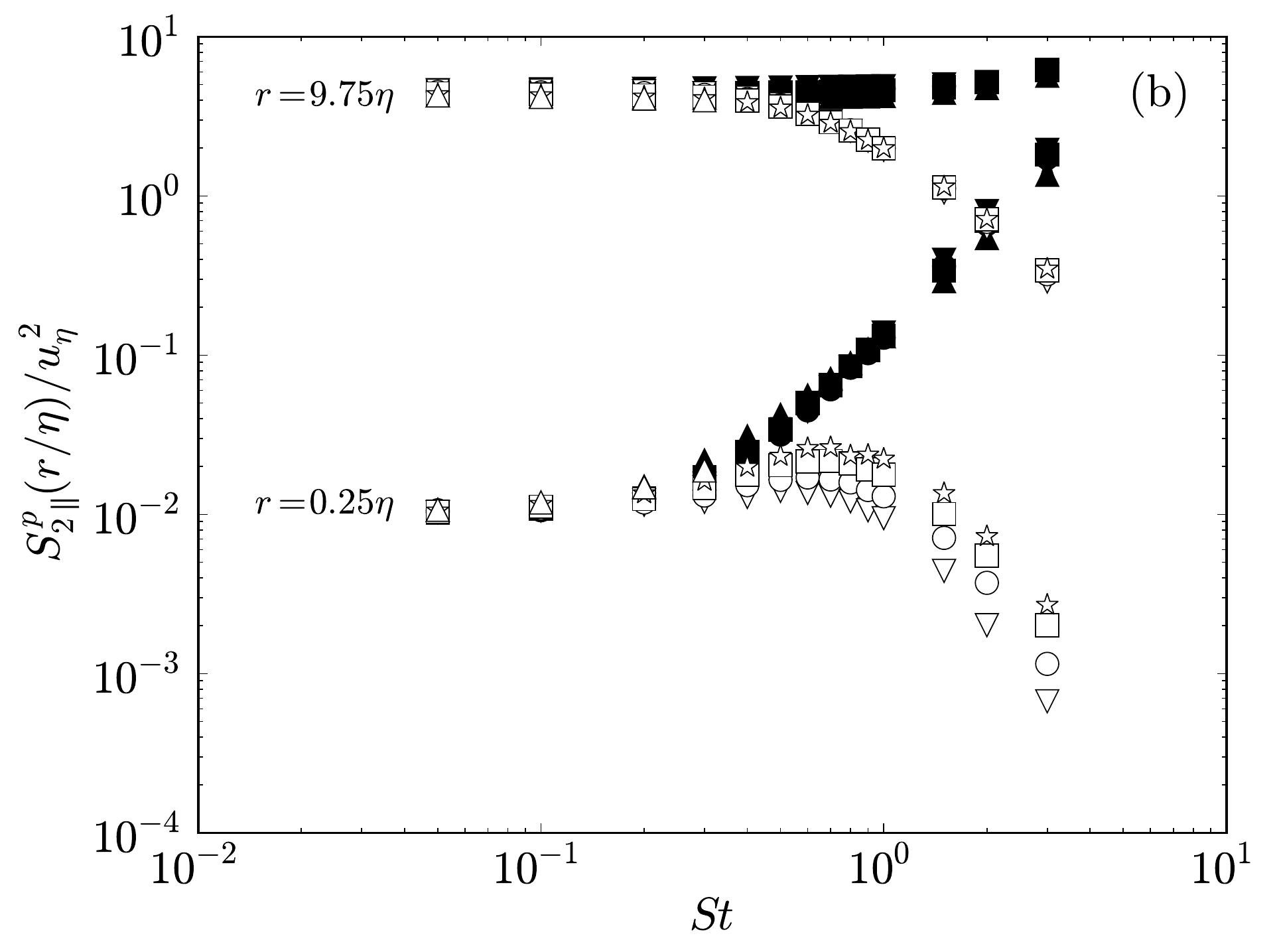}
 \caption{The normalized mean inward radial relative velocities (a) and  
 relative velocity variances (b) plotted as a function
 of $St$ for different values of $R_\lambda$. The open symbols denote the case with
 gravity ($Fr = 0.052$), and the filled symbols denote the case without gravity.
 Data are shown for particles with separations $r = 0.25 \eta$ and $r = 9.75 \eta$.}
 \label{fig:wr_re}
\end{figure}

We have thus far examined and explained relative velocity statistics for fixed Reynolds numbers.
We now consider how these statistics are affected by changes in $R_\lambda$.
In figure~\ref{fig:wr_re}, we show both $S^p_{-\parallel}(r/\eta)/u_\eta$ and $S^p_{2\parallel}(r/\eta)/u_\eta^2$.
For $0 \leq St \leq 3$, the longitudinal relative velocities have a very weak dependence on $R_\lambda$,
both with and without gravity. 
In Part I, we noted that $S^p_{2\parallel}(r/\eta)/u_\eta^2$ increases weakly with increasing $R_\lambda$
in the absence of gravity for $0.3 \lesssim St \lesssim 1$, 
and attributed this trend to the increased scale separation
with increasing $R_\lambda$, which causes a few inertial particles to be 
affected by their memory of increasingly energetic turbulence.
(We also suggested that increased intermittency effects at higher Reynolds numbers
could contribute to the observed trends.)
While we expected the increased scale separation of the turbulence to also increase the
relative velocities for higher-$St$ particles, we instead
found that the relative velocities decreased with increasing $R_\lambda$ for $1 \lesssim St \lesssim 3$.
We argued that this trend was caused by a corresponding decrease in the rotation
timescales $T^p_{\mathcal{R} \mathcal{R}}/\tau_\eta$, which
in turn reduced the influence of path-history effects and decreased the relative velocities.
However, with gravity, we find that $T^p_{\mathcal{R} \mathcal{R}}/\tau_\eta$ is generally
independent of $R_\lambda$ (see \textsection \ref{sec:topology}), and thus 
we expect that this allows the increased scale separation of the turbulence to cause $S^p_{2\parallel}(r/\eta)/u_\eta^2$
to uniformly increase with increasing $R_\lambda$ for $St \gtrsim 0.3$.
Our results in figure~\ref{fig:wr_re} confirm this expectation.
We also see that $S^p_{-\parallel}(r/\eta)/u_\eta$ is generally unaffected by changes in $R_\lambda$,
since it is less influenced by the relatively infrequent occurrences of the effects discussed above.

\begin{figure}
 \centering
 \includegraphics[width=2.6in]{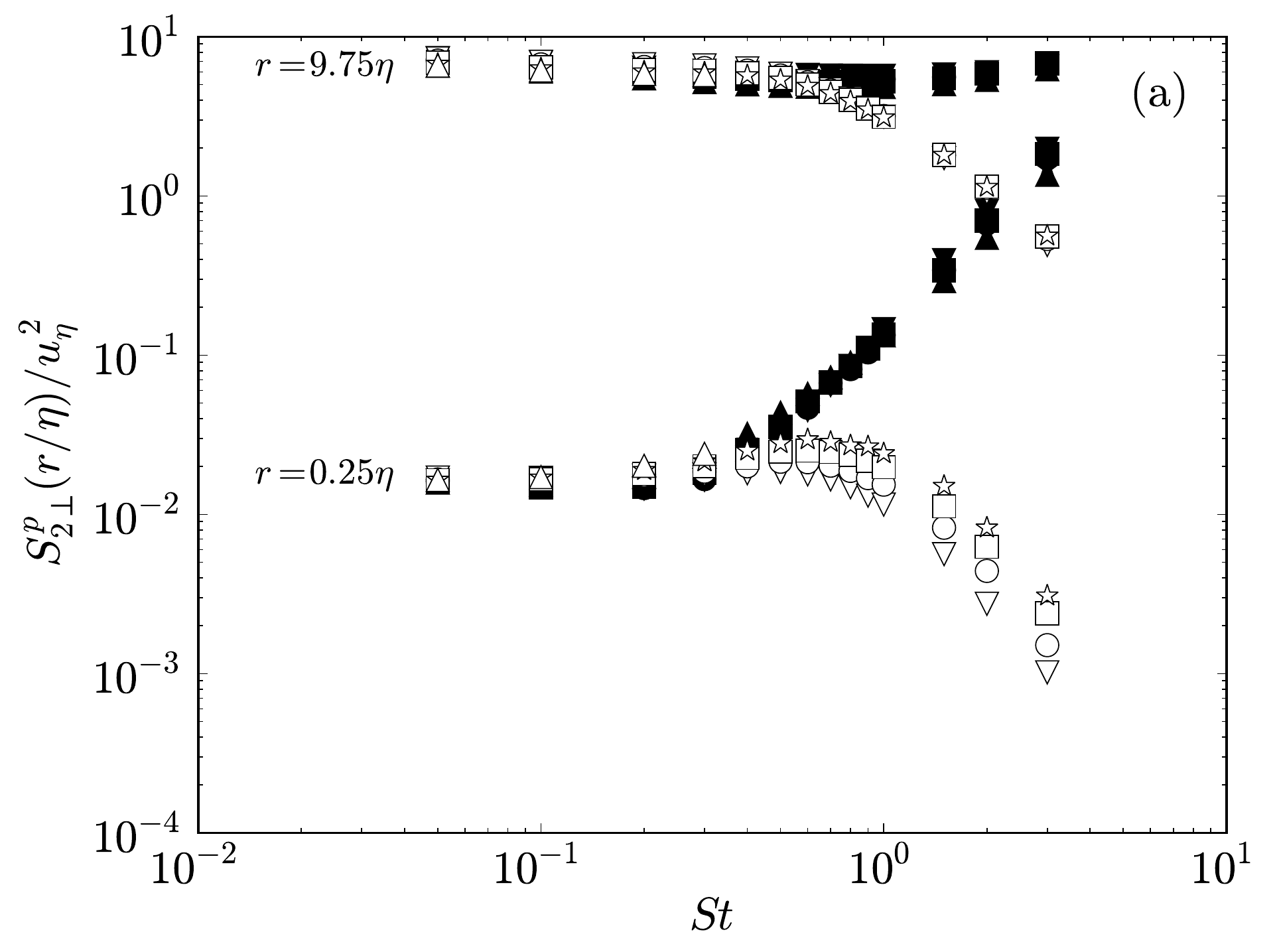}
 \includegraphics[width=2.6in]{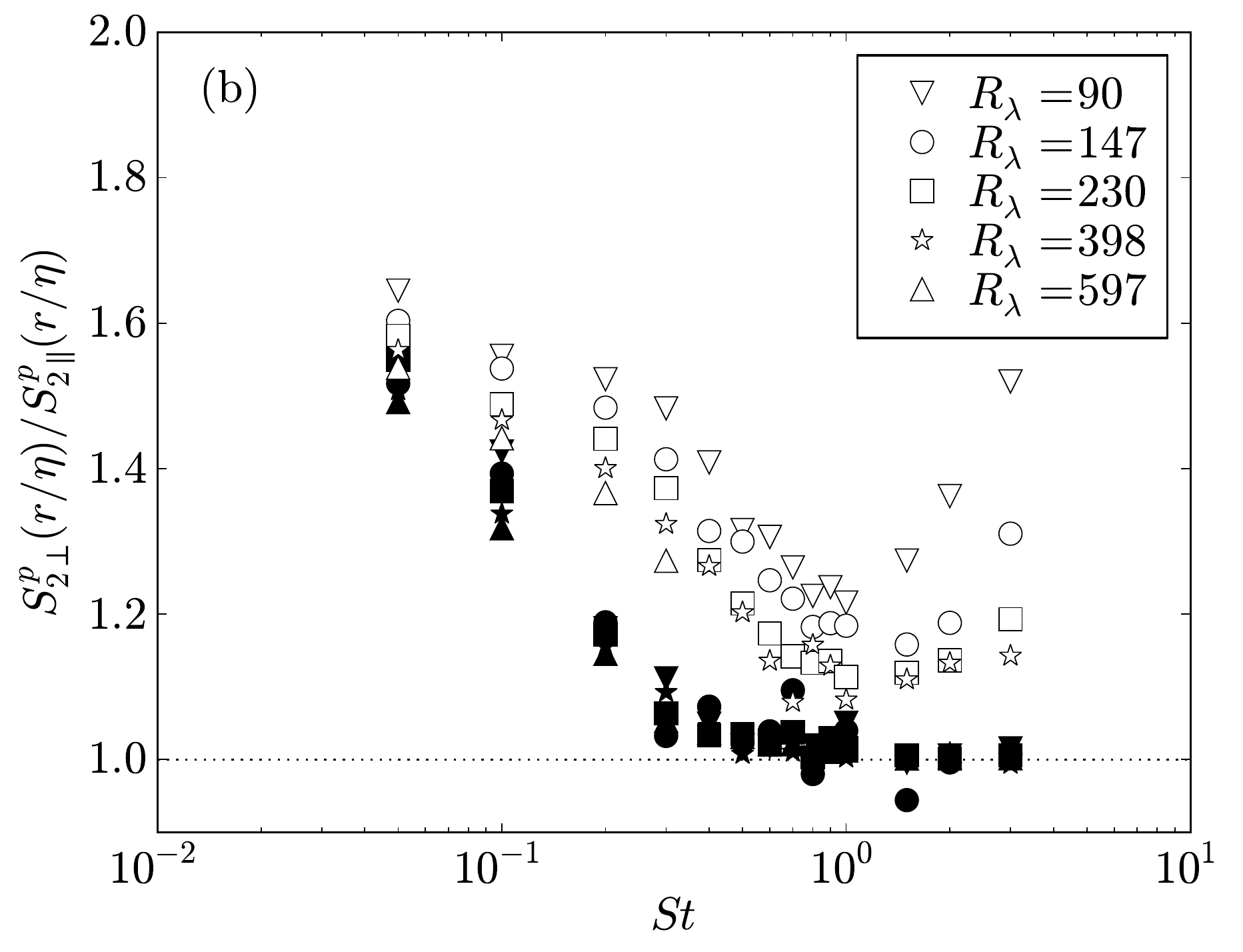}
 \caption{(a) The transverse relative velocity variances,
 plotted as a function of $St$ for different values of $R_\lambda$.
 Data are shown for particles with separations $r = 0.25 \eta$ and $r = 9.75 \eta$.
 (b) The ratio between the transverse and longitudinal relative velocity variances
 $S^p_{2\perp}(r/\eta)/S^p_{2\parallel}(r/\eta)$, evaluated at $r/\eta = 0.25$.
 In both plots, the open symbols denote the case with
 gravity ($Fr = 0.052$), and the filled symbols denote the case without gravity.}
 \label{fig:wr_re_per}
\end{figure}

We are also interested in the effect of the Reynolds number on the transverse 
relative velocities, since these statistics will also affect the degree
of particle clustering (see \textsection \ref{sec:clustering}). 
Figure~\ref{fig:wr_re_per}(a) shows the transverse relative velocity
variances $S^p_{2\perp}(r/\eta)$ for different values of $St$ and $R_\lambda$.
We see that the trends with $R_\lambda$ in the transverse direction
are identical to those in the longitudinal direction.

In Part I, we found that without gravity, the longitudinal
and transverse relative velocities became equivalent at small separations for $St \gtrsim 0.3$.
The physical explanation is that the path-history contribution to their relative velocities decreases the 
coherence of the pair motion,
and in the ballistic limit where the pairs move independently of each other, the 
longitudinal and transverse components are equal. However, with gravity, path-history effects 
are weaker, and thus the longitudinal and transverse velocities may not be the same in this regime.

We compare these two quantities in figure~\ref{fig:wr_re_per}(b) by plotting $S^p_{2\perp}(r/\eta)/S^p_{2\parallel}(r/\eta)$ at $r/\eta = 0.25$.
We see that both with and without gravity,
the ratio $S^p_{2\perp}(r/\eta)/S^p_{2\parallel}(r/\eta)$ approaches 2 at low Stokes numbers, the value for fluid particles \citep[e.g., see][]{pope}.
For $St \gtrsim 0.3$ without gravity, 
$S^p_{2\perp}(r/\eta)$ and $S^p_{2\parallel}(r/\eta)$ are equivalent, as expected.
At high values of $St$ with gravity, however, the longitudinal
and transverse components are not equivalent, since path-history effects are weakened by gravity. 
As $R_\lambda$ increases, the particle relative velocities are affected
by increasingly energetic turbulence along their path histories.
As a result, the relative velocities are larger and the particles move more ballistically,
causing the ratio $S^p_{2\perp}(r/\eta)/S^p_{2\parallel}(r/\eta)$ to decrease with increasing $R_\lambda$.

We now examine the effects of Reynolds number on the scaling of the relative velocities. 
As in Part I, we compute the scaling exponents of the mean inward relative velocity
($\zeta^-_\parallel$) and the relative velocity variance ($\zeta^2_\parallel$)
by performing linear least-squares power-law fits over 
separations $0.75 \leq r/\eta \leq 2.75$. While the results in figure~\ref{fig:wr_lom_1024} 
suggest that the power-law exponent may vary considerably over this range, 
we are forced to use this relatively large range to gather sufficient statistics
at the smaller separations. Thus, while these scaling
exponents allow us to assess the trends with $R_\lambda$,
they only provide a qualitative understanding, as the scaling varies considerably
throughout the dissipation range, and this variation is not completely captured
by this analysis.

\begin{figure}
 \centering
 \includegraphics[width=2.6in]{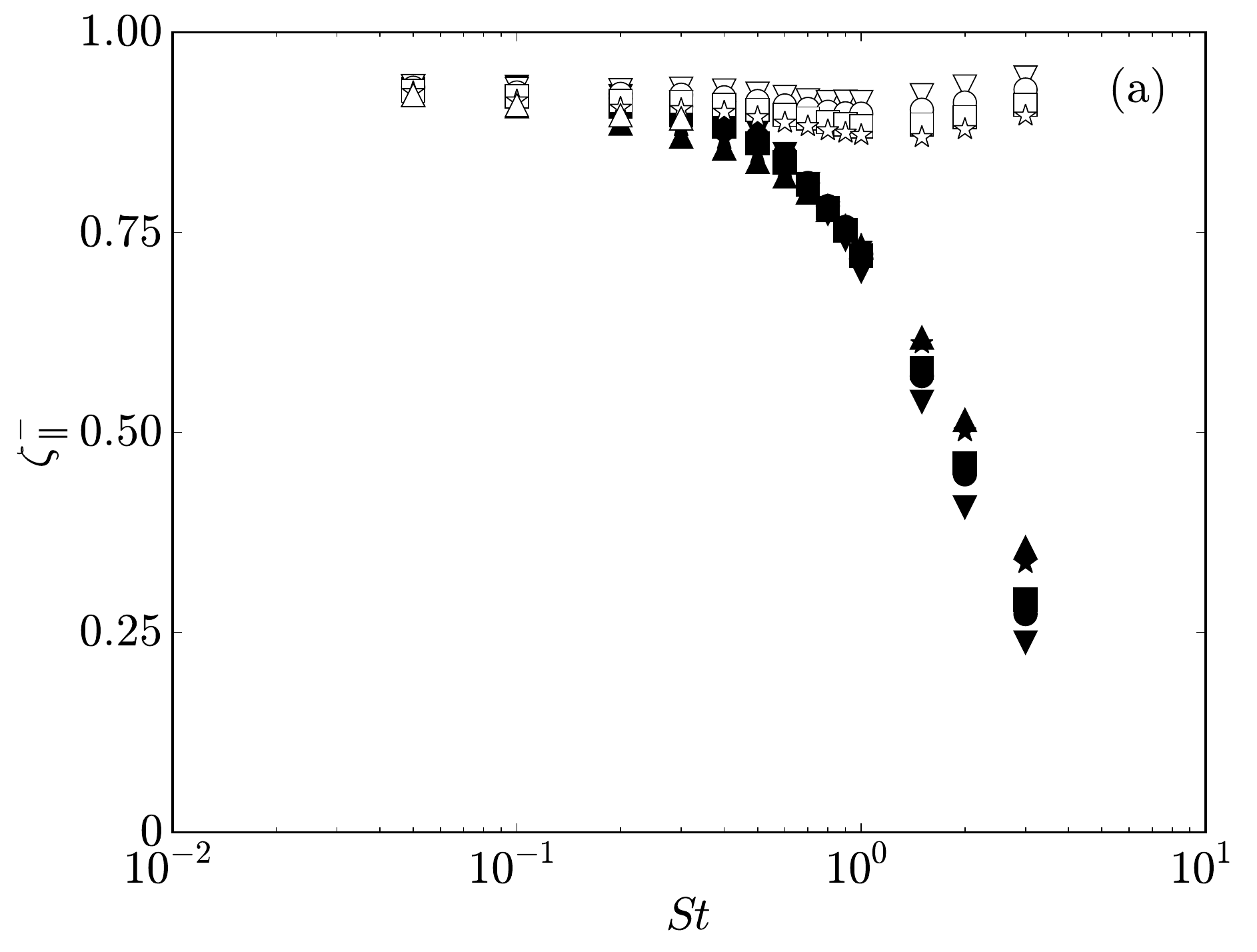}
 \includegraphics[width=2.6in]{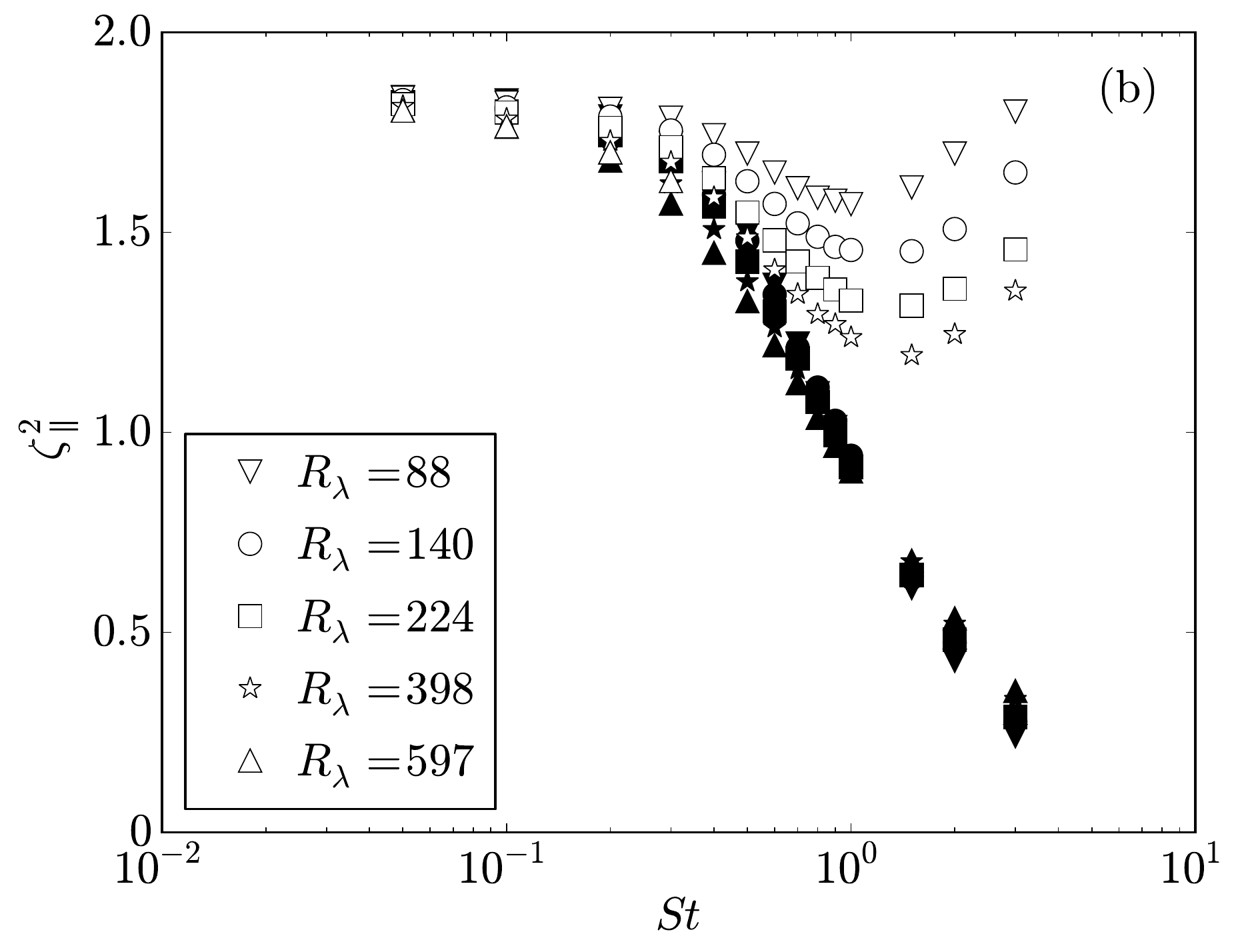}
 \caption{The scaling exponents of the longitudinal mean inward relative velocities (a) and  
 relative velocity variances (b) plotted as a function
 of $St$ for different values of $R_\lambda$. The open symbols denote the case with
 gravity ($Fr = 0.052$), and the filled symbols denote the case without gravity.
 The scaling exponents are computed using linear least-squares regression
 over the range $0.75 \leq r/\eta \leq 2.75$.}
 \label{fig:wr_scaling_exponent}
\end{figure}

The scaling exponents are shown in figure~\ref{fig:wr_scaling_exponent}. 
The trends at a given value of $R_\lambda$ are in agreement with our observations from 
figure~\ref{fig:wr_lom_1024} and figure~\ref{fig:wr_lom_1024_mean}, which are explained above.
We also see that with gravity, the scaling exponents tend to decrease with increasing $R_\lambda$,
since the particles are more influenced by their memory of increasingly energetic turbulence,
and therefore tend to move ballistically, as noted above.

\begin{figure}
 \centering
 \includegraphics[height=1.8in]{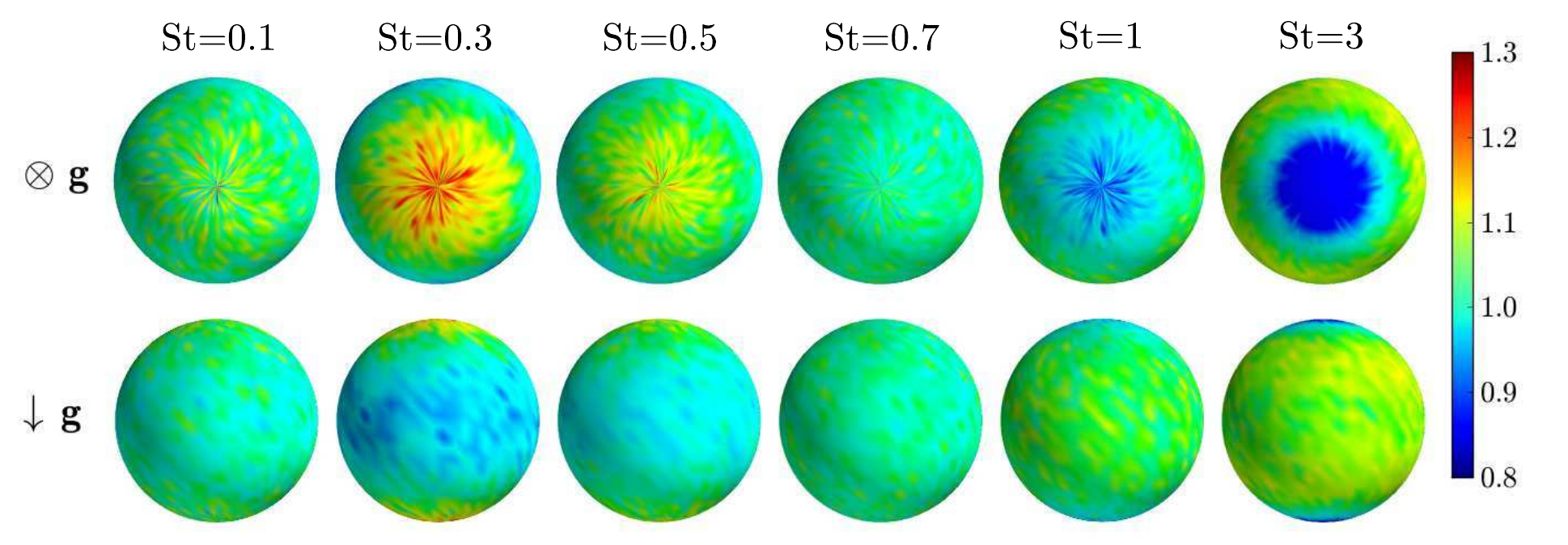}
 \caption{The directionally dependent mean inward relative velocity $S^p_{-\parallel}(\bm{r})$,
 normalized by the spherically averaged mean inward relative velocity $S^p_{-\parallel}(r)$, shown on a unit sphere
 for $R_\lambda = 398$ and $r < \eta$ with gravity ($Fr = 0.052$). The different columns
 correspond to different values of $St$.  The top row shows the projection where gravity is directed
 into the page, and the bottom row shows the projection where gravity is directed downward.}
 \label{fig:wr_3d}
\end{figure}

Angular distributions of $S^p_{-\parallel}(\bm{r})/S^p_{-\parallel}(r)$ are presented in 
figure~\ref{fig:wr_3d} for a range of Stokes numbers, $R_\lambda = 398$, and $r < \eta$. 
The anisotropy of the distribution is reflected in the color variation over the surface of the sphere. 
We see immediately that the asymmetry in the relative velocities follows opposite trends at small and large $St$.
That is, at small $St$, the relative velocities are largest for particles separated in the vertical direction, 
whereas for $St\gtrsim 1$, the opposite is true.
The explanation is as follows.
At small separations and $St \lesssim 0.3$, particle pairs that are separated
vertically will have longitudinal relative velocities
that are proportional to $\mathcal{S}_{33}(\bm{x}^p(t),t)$,
the longitudinal velocity gradient in the vertical 
direction as sampled by inertial particles (see \textsection \ref{sec:topology}).
Particle pairs that are separated along the horizontal direction,
however, will have longitudinal relative velocities that are proportional to 
$\mathcal{S}_{11}(\bm{x}^p(t),t)$, the longitudinal velocity gradient in the horizontal direction.
In \textsection \ref{sec:topology}, we observed that particles
tend to preferentially sample flow where the vertical (horizontal) velocity gradients
are larger (smaller). As a result, the relative velocities
are expected to be larger (smaller) for particles which are separated in the vertical (horizontal) directions.
At large $St$, however, the relative velocities are smallest for particles that are separated
vertically.  The physical explanation is
that the correlation timescales of the fluid velocity gradients along this direction are the smallest 
(see \textsection \ref{sec:topology}), 
and thus the particles have a weaker response to vertical fluid velocity fluctuations and correspondingly
lower particle relative velocities along this direction. 

\begin{figure}
 \centering
 \includegraphics[width=2.6in]{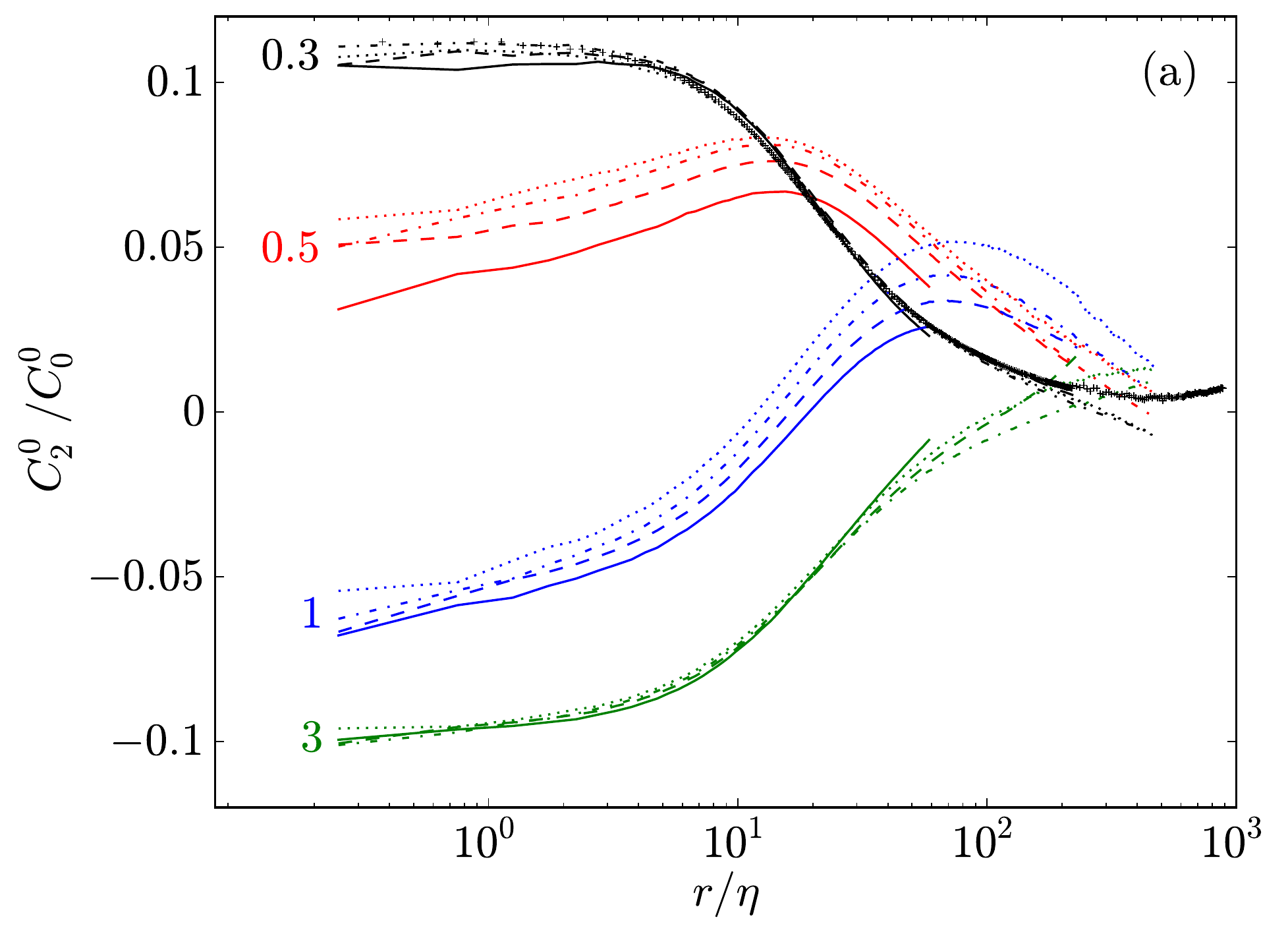}
 \includegraphics[width=2.6in]{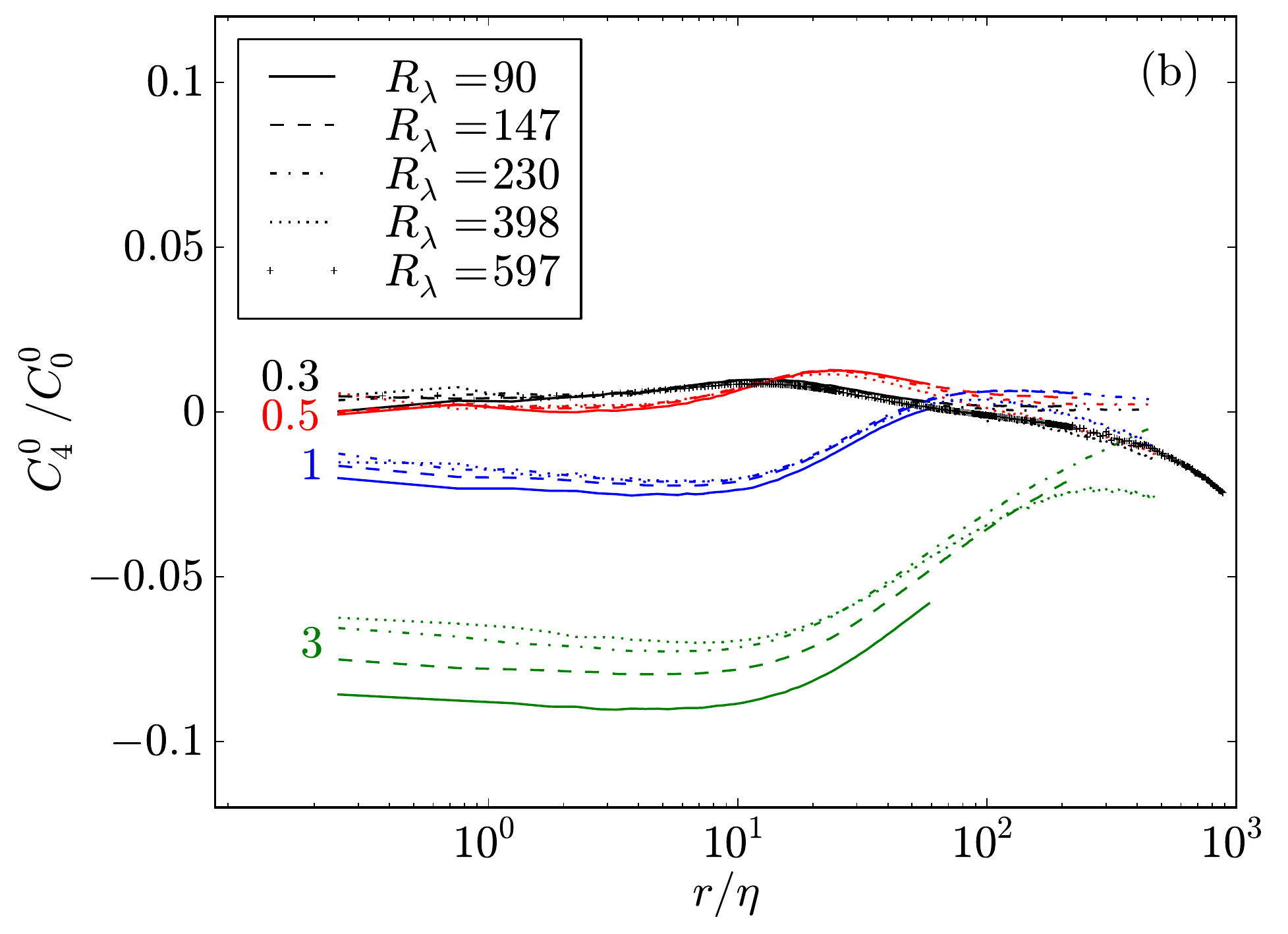}
 \caption{The second (a) and fourth (b) spherical harmonic coefficients
 of $S^p_{-\parallel}(\bm{r})$, 
 normalized by the zeroth spherical harmonic coefficient, plotted as a function of $r/\eta$
 for different $St$ and $R_\lambda$ with gravity ($Fr = 0.052$). 
 The different values of $St$ considered ($0.3$, $0.5$, $1$, $3$)
 are shown in black, red, blue and green, respectively, and 
 the Stokes numbers are indicated by the line labels.}
 \label{fig:spher_harm_wrel}
\end{figure}

Next, we consider the dependence of the anisotropy on the separation distance
by plotting the spherical harmonic coefficients of order 2 and 4 as a function of $r$ in
figure~\ref{fig:spher_harm_wrel}. (Coefficients above order 4 are too small to measure accurately and thus are not shown.)
We see that the anisotropy in the relative velocities generally decreases with increasing separation. 
The physical explanation is that with increasing separation distance, 
the particle motions are increasingly dependent on larger eddies, 
and the relative velocities induced by these isotropic eddies are increasingly energetic relative 
to the anisotropic velocities induced by gravity.

For small (large) values of $St$, $c_2^0$ and $c_4^0$ are positive (negative), 
indicating that the particle relative velocities
are strongest for particles that are separated vertically (horizontally). 
These observations are in agreement with the trends
shown in figure~\ref{fig:wr_3d}.
We also see that $c_2^0$ and $c_4^0$ tend to become more (less) isotropic
at high (low) $St$ as $R_\lambda$ increases. 
While the physical explanation for the trend at low $St$ is unclear,
at high $St$, we expect that this increase in isotropy is 
linked to the increase in the relative velocities with increasing $R_\lambda$.
That is, as the overall relative velocities increase, 
the anisotropic velocities induced by gravity will be comparatively weaker,
and thus the particle relative velocities will be more isotropic.

\subsection{Particle clustering}
\label{sec:clustering}
In this section, we consider how gravity affects the inertial particle clustering process.
We first provide a theoretical explanation for the clustering (\textsection \ref{sec:particle_clustering_theory})
and then compare the predictions to DNS results (\textsection \ref{sec:particle_clustering_dns}).

\subsubsection{Theoretical framework for particle clustering}
\label{sec:particle_clustering_theory}

We use the angular distribution function 
$g(\bm{r})$ \citep[hereafter `ADF,' see][]{gualtieri09} to quantify
the degree and orientation of particle clustering. We define $g(\bm{r})$ as
\begin{equation}
g(\bm{r}) \equiv \frac{N(r,\theta,\phi)/V(r,\theta,\phi)}{N/V} \mathrm{.}
\label{eq:adf}
\end{equation}
In this equation, $N(r,\theta,\phi)$ denotes the number of particle pairs in a truncated spherical cone with 
nominal radius $r$, polar angle $\theta$, and azimuthal angle $\phi$.
The volume of the truncated spherical cone $ V(r,\theta,\phi)$ is given by
\begin{equation*}
 V(r,\theta,\phi) \equiv \sin(\theta) \Delta \theta \Delta \phi \left[(r + \Delta r)^3 - (r-\Delta r)^3 \right]/3 \mathrm{,}
\end{equation*}
where $\Delta r$ is the radial width, $\Delta \theta$ is the extent of the polar angle,
and $\Delta \phi$ is the extent of the azimuthal angle.

As discussed in \textsection \ref{sec:relative_velocity_theory}, the equation governing 
the relative motion of like particles, (\ref{eq:maxey_riley_relative}), is the same with and without gravity.
Hence the theory of \cite{zaichik09}, developed without consideration of gravitational settling, 
is not explicitly changed by the presence of gravity; 
instead gravity modifies the equation through implicit changes to its coefficients. 
From \cite{zaichik09} \citep[see also][]{bragg14}, 
the equation describing $g(\bm{r})$ at steady-state is
\begin{equation}
\label{eq:ss_rdf}
\bm{0}= -St \tau_\eta \Big( \bm{S}^p_2
            + \boldsymbol{\lambda} \Big) \cdot \boldsymbol{\nabla}_{\bm{r}} g
            - St \tau_\eta g \boldsymbol{\nabla}_{\bm{r}} \cdot \bm{S}^p_2
             \mathrm{,}
\end{equation}
where $\boldsymbol{\lambda}$ is the dispersion tensor describing the influence of the fluid velocity
difference field on the dispersion of the particles \citep[see][]{bragg14}.

Since $\bm{S}^p_2$ is anisotropic with gravity (cf. \S\ref{sec:relative_velocity_dissipation}), 
so too is the ADF $g(\bm{r})$, as is evident in (\ref{eq:ss_rdf}). 
We similarly adopt an expansion of the ADF in terms of spherical harmonic functions
\begin{equation}
 g(\bm{r}) = \sum_{\ell = 0}^{\infty} \mathcal{C}_{2 \ell}^0 (r) Y_{2 \ell}^0 (\theta) \mathrm{,}
\end{equation}
where $\mathcal{C}_{2 \ell}^0$ are the spherical harmonic coefficients for the ADF. We only attempt to model the zeroth-order term, which we denote as
\begin{equation}
 g(r) = \mathcal{C}_{0}^0 (r) \mathrm{.}
\end{equation}
where $Y_{0}^0\equiv1$. Note that $g(r)$ is just the spherical average of the ADF $g(\bm{r})$, or the equivalent of the radial distribution 
function (RDF) defined in Part I. 
While this approach captures the leading-order contribution of gravity to the ADF, 
it does not predict the gravity-induced anisotropy. 
(Note that a similar approach was adopted by \cite{alipchenkov13} to model particle clustering 
in a homogeneous turbulent shear flow.) 
We will assess the accuracy of this approach in \textsection \ref{sec:particle_clustering_dns}.

We therefore consider the isotropic form of (\ref{eq:ss_rdf}), which is given as
\begin{equation}
\label{eq:ss_rdf_isotropic}
 0 = \underbrace{-St \left( \hat{S}^p_{2\parallel} + \hat{\lambda}_\parallel \right)}_{\equiv \mathfrak{D}} \nabla_{\hat{r}} g
     \underbrace{-St\left(
     \nabla_{\hat{r}} \hat{S}^p_{2\parallel} + 2 \hat{r}^{-1} 
     \left[ \hat{S}^p_{2\parallel} - \hat{S}^p_{2\perp} \right] \right)}_{\equiv \mathfrak{d}}g \mathrm{,}
\end{equation}
where $\hat{Z}$ denotes a variable $Z$ normalized by Kolmogorov scales. 
$\lambda_\parallel$ denotes the projection of the dispersion tensor 
$\boldsymbol{\lambda}$ along a direction parallel to the particle separation vector.
We note that the first term on the right-hand-side is associated with outward diffusion
(which acts to reduce the RDF),
while the second is associated with an inward drift (which acts to increase the RDF).
The corresponding diffusion and drift coefficients are denoted as 
$\mathfrak{D}$ and $\mathfrak{d}$, respectively \citep[see][]{bragg14}.

As explained in \cite{bragg14}, the overall degree of clustering is determined
by the relative strength of the drift and diffusion coefficients, 
$\mathfrak{d}/\mathfrak{D}$.
Increases (decreases) in this ratio are associated with increased (decreased) clustering.
To understand how gravity affects the clustering, we therefore need to understand
its effect upon the ratio $\mathfrak{d}/\mathfrak{D}$.

\cite{bragg14} demonstrated that in the limit $St \ll 1$, 
$\mathfrak{D}$ and $\mathfrak{d}$ simplify to
\begin{equation}
\label{eq:diff_low_St}
 \mathfrak{D} \approx -\hat{r}^2 B_{nl} \mathrm{,}
\end{equation}
and
\begin{equation}
\label{eq:drift_low_St}
 \mathfrak{d} \approx -St \hat{r} \left( \langle \hat{\mathcal{S}}^2 \rangle^p - \langle \hat{R}^2 \rangle^p \right)/3 \mathrm{,}
\end{equation}
respectively, where $B_{nl}$ is the non-local coefficient defined in \cite{chun05}
that is independent of the particle Stokes number.
As is evident in (\ref{eq:diff_low_St}), the diffusion term is independent of the Stokes
number--hence $\mathfrak{D}$ is unaffected by gravity in this limit.
In contrast, $\mathfrak{d}$ is proportional to both $St$ and $\langle \hat{\mathcal{S}}^2 \rangle^p - \langle \hat{\mathcal{R}}^2 \rangle^p$.
Gravity reduces the correlation between the particles and the strain and rotation fields, and thereby reduces
$\mathfrak{d}$. The net effect of gravity in this limit is therefore to reduce the ratio $\mathfrak{d}/\mathfrak{D}$ and the
degree of clustering.
We also note that at low $St$, the trends in the strain and rotation rates with $R_\lambda$ 
are very weak. 
Consistent with Part I, these trends imply the clustering to be nearly independent of $R_\lambda$
in the limit $St \ll 1$, with or without gravity.

For inertial particles with intermediate Stokes numbers in the range $0.2 \lesssim St \lesssim 0.7$
(note that this range is approximate and is likely sensitive to changes in Reynolds number, as mentioned in Part I),
we no longer have simple relationships for the drift and diffusion terms. However, we can still
make qualitative predictions for the trends in the clustering based on our physical
understanding and the explanations in \cite{bragg14}.
Over this range of $St$, \cite{bragg14} argued that both preferential-sampling effects and path-history effects 
act to increase clustering, with the path-history effects becoming increasingly dominant as $St$ increases.
Since gravity decreases both preferential-sampling and path-history effects, we therefore expect it to reduce
clustering over the approximate range of $0.2 \lesssim St \lesssim 0.7$.

We next use (\ref{eq:ss_rdf_isotropic}) to understand clustering at larger values of $St$. 
In Part I, we argued that without gravity, $\hat{S}^p_{2\parallel} \approx \hat{S}^p_{2\perp}$ 
for $St \gtrsim 0.3$, and we were thus able to neglect the term 
$2 \hat{r}^{-1} ( \hat{S}^p_{2\parallel} - \hat{S}^p_{2\perp} )$ in (\ref{eq:ss_rdf_isotropic}) 
at high $St$. However, with gravity, the longitudinal 
and transverse relative velocity variances are generally not equal
(as discussed in \textsection \ref{sec:relative_velocity_dissipation}), 
and hence this term must be retained. 
We are still able to neglect $\hat{\lambda}_\parallel$ at small separations and high $St$, 
since this term is inversely proportional to $St$ and decreases as the timescales of the fluid velocity seen 
by the particles decrease \citep[see][]{bragg14}.

The simplified form of (\ref{eq:ss_rdf_isotropic}) at high $St$ is therefore
\begin{equation}
\label{eq:zaichik_rdf_high_St}
 0 = \underbrace{-St \hat{S}^p_{2\parallel}}_{\mathfrak{D}} \nabla_{\hat{r}} g
     \underbrace{-St \left(
     \nabla_{\hat{r}} \hat{S}^p_{2\parallel} + 2 \hat{r}^{-1} 
     \left[ \hat{S}^p_{2\parallel} - \hat{S}^p_{2\perp} \right] \right)}_{\mathfrak{d}} g \mathrm{.}
\end{equation}
Taking the ratio between the drift and diffusion coefficients gives us
\begin{equation}
\label{eq:zaichik_diffusion_drift}
\frac{\mathfrak{d}}{\mathfrak{D}} = \frac{\nabla_{\hat{r}} \hat{S}^p_{2\parallel} + 2 \hat{r}^{-1} 
     \left[ \hat{S}^p_{2\parallel} - \hat{S}^p_{2\perp} \right]}
 {\hat{S}^p_{2\parallel}} =
 \frac{1}{\hat{r}} \left( \zeta^2_\parallel + 2 - 2 \hat{S}^p_{2\perp}/\hat{S}^p_{2\parallel} \right) \mathrm{,}
\end{equation}
where $\zeta^2_\parallel$ is the scaling exponent of the longitudinal relative velocity variance 
(see \S\ref{sec:relative_velocity_dissipation}). 

From \S\ref{sec:relative_velocity_dissipation}, 
we see that with gravity, $\zeta^2_\parallel$ increases very strongly, 
while $2 \hat{S}^p_{2\perp}/\hat{S}^p_{2\parallel}$ increases weakly. 
We expect that $\zeta^2_\parallel - 2 \hat{S}^p_{2\perp}/\hat{S}^p_{2\parallel}$ will therefore increase 
with gravity, leading to an increase in the clustering at high $St$. 
We attribute these trends
to the fact that gravity reduces the influence of path-history effects 
on particle relative motion in the dissipation range at high $St$. 
Path-history effects primarily act to diminish
clustering for $St \gtrsim 1$ \citep[see][]{bragg14}, 
and thus gravity reduces their influence and therefore acts to increase $g(\hat{r})$.
We emphasize that this does not necessarily imply that gravity increases the inward drift
and decreases the outward diffusion. In fact, our data indicate that gravity
decreases both $\mathfrak{d}$ and $\mathfrak{D}$. The reduction in $\mathfrak{D}$ with gravity,
however, is stronger than the corresponding reduction in $\mathfrak{d}$, causing the ratio $\mathfrak{d}/\mathfrak{D}$,
and therefore the clustering strength, to increase.
We also see from \S\ref{sec:relative_velocity_dissipation} that increasing $R_\lambda$, 
with gravity, generally leads to a decrease in both $\zeta^2_\parallel$ and 
$\hat{S}^p_{2\perp}/\hat{S}^p_{2\parallel}$, 
due to the increased influence of path-history effects 
on particle relative motion at higher Reynolds numbers. 
While both quantities seem to be decreased by about the same amount with increasing $R_\lambda$, 
the latter quantity has a greater effect on the clustering, 
since it is multiplied by a factor of two in (\ref{eq:zaichik_diffusion_drift}). 
This explains why $\mathfrak{d}/\mathfrak{D}$ decreases
as $R_\lambda$ increases, leading to a decrease in the clustering.

In summary, gravity reduces clustering at low and intermediate values of $St$,
but increases clustering at high values of $St$.
Gravity reduces both preferential-sampling and path-history effects.
Preferential-sampling effects increase clustering at low values of $St$,
and thus gravity reduces clustering by weakening the preferential sampling
drift mechanism. At intermediate values of $St$, both preferential sampling
and path-history effects act to increase clustering,
and thus gravity, by weakening both effects, reduces clustering.
At higher values of $St$, path-history effects act to diminish clustering,
and gravity increases clustering by decreasing path-history effects.

This explanation is consistent with arguments put forth by 
\cite{franklin07,ayala08a,woittiez09,onishi09,rosa13} for low-Stokes-number particles. 
However, at higher Stokes numbers, several of these authors have argued that gravity facilitates 
interactions with large-scale turbulent eddies, which in turn leads to increased particle clustering 
\citep{franklin07,woittiez09,rosa13}. 
In contrast, we argue that gravity reduces
the temporal correlation radius over which the particles are affected by their
path-history interactions with the turbulence, and therefore causes the particle relative motion
in the dissipation range
to be \textit{less} affected by their interaction with larger-scale 
turbulence.

Another recent paper by \cite{park14} suggests the increase in particle clustering with gravity at high 
Stokes numbers is linked to the skewness of the vertical velocity gradients of the underlying fluid. 
This proposed explanation, however, seems to be flawed, and we
suggest that their finding reflects a certain correlation in the system, but not a causal relation.
This conclusion is supported by the recent work of \cite{gustavsson14}, who observed qualitatively 
similarly strong clustering of high-Stokes-number particles in the presence of gravity in their random 
Gaussian flow field, whose fluid velocity, by definition, has no skewness.

\subsubsection{Particle clustering results}
\label{sec:particle_clustering_dns}

\begin{figure}
 \centering
 \includegraphics[height=3.5in]{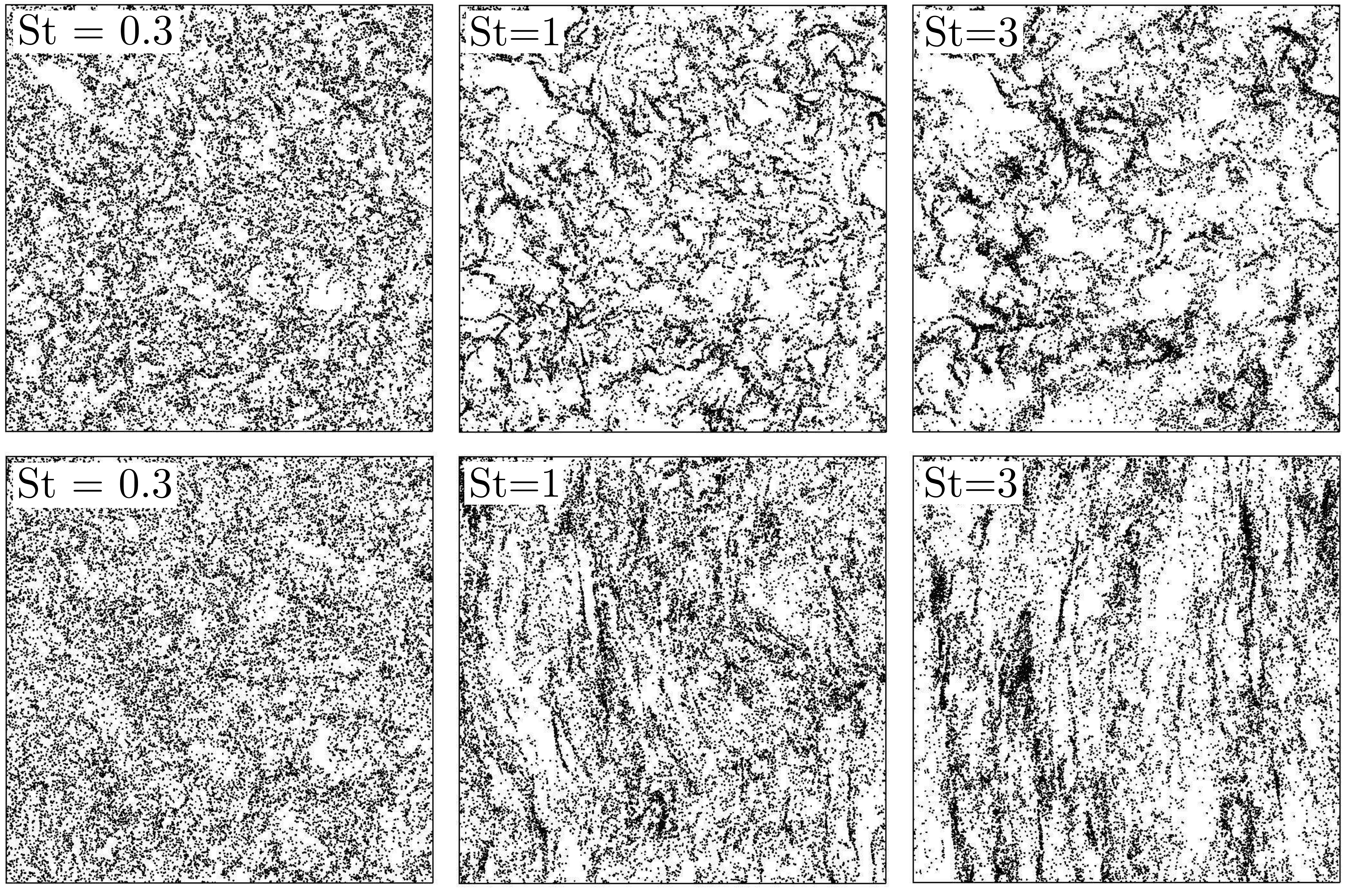}
 \caption{Instantaneous particle locations in $1000\eta \times 1000\eta \times 10\eta$ slices of the domain
 for different values of $St$ and $R_\lambda = 398$. The top row is for the case without gravity, and the bottom row 
 is for the case with gravity ($Fr = 0.052$), with the gravitational vector pointing downward.}
 \label{fig:particle_vis}
\end{figure}

We begin by first examining instantaneous snapshots of particle positions in the simulations in figure~\ref{fig:particle_vis}. 
From these visualizations, it is evident that gravity alters both the degree and orientation of the clusters, 
and generally causes them to be aligned with the gravity vector.
Recall that the clustering observed at $St\gtrsim1$ is not related to the Maxey centrifuge mechanism 
\citep{maxey87}, but rather is due to the history effect \citep{bragg14}. 
This explains why higher-Stokes-number particles show signs of clustering even though they nearly 
uniformly sample the strain and rotation fields (cf. figure~\ref{fig:S2_R2}). 
We conclude the vertical streaks in figure~\ref{fig:particle_vis} 
are a manifestation of the non-local history mechanism acting in the presence of gravity.

\begin{figure}
 \centering
 \includegraphics[width=2.6in]{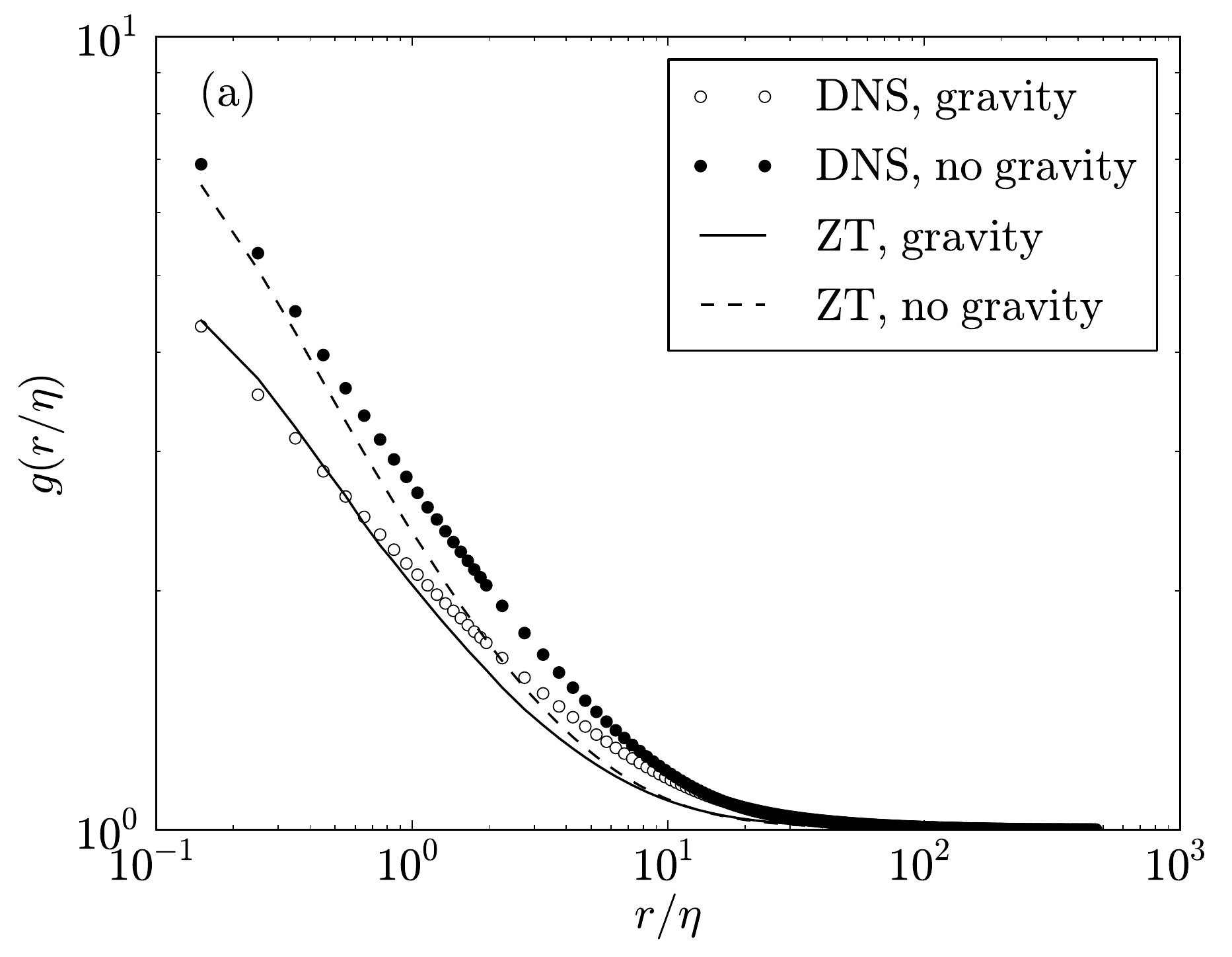}
 \includegraphics[width=2.6in]{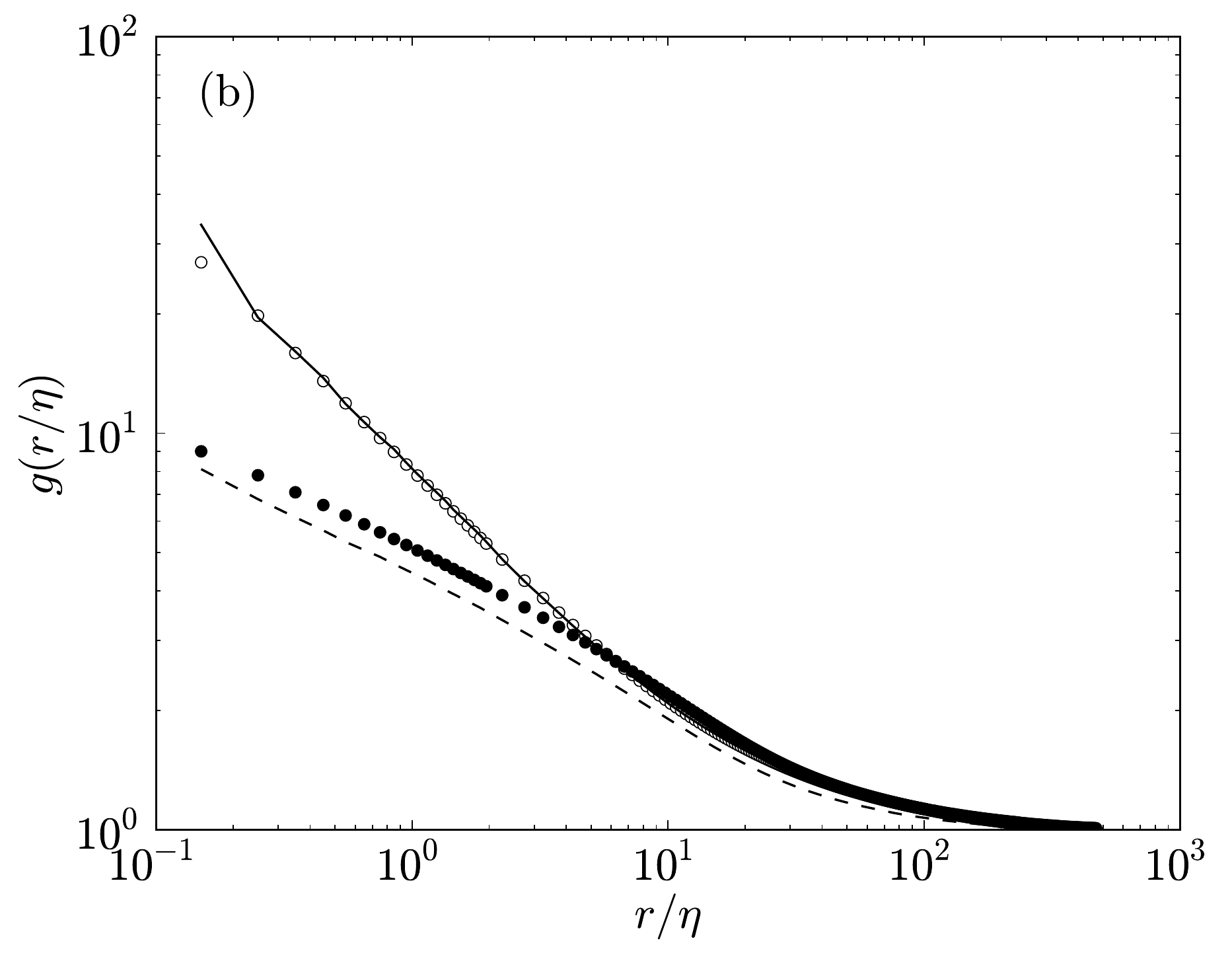}
 \caption{DNS (symbols) and theoretical (lines) data for the RDFs 
 $g$ plotted as a function of $r/\eta$ at $R_\lambda = 398$
 both with gravity (open symbols and solid lines) and without gravity (filled symbols and dashed lines). 
 (a) shows RDFs for $St=0.3$, and (b) for $St = 3$. 
 The theoretical predictions are calculated from the equations in \cite{zaichik09}
 with the non-local correction from \cite{bragg14}, using our DNS data
 to specify $T^p_{\mathcal{S}\mathcal{S}}$, $S^p_{2\parallel}$, and $S^p_{2\perp}$.}
 \label{fig:rdfs}
\end{figure}

To quantify the clustering, we first consider the spherically averaged RDFs $g(r)$ to analyze the overall effect of gravity on the degree of clustering. 
Figure~\ref{fig:rdfs} shows plots of the RDFs for $St=0.3$ and $St=3$ both with gravity ($Fr = 0.052$) 
and without gravity at $R_\lambda = 398$. 
The results show gravity causes the RDFs to decrease (increase) at low (high) values of $St$, 
in agreement with our arguments in \textsection\ref{sec:particle_clustering_theory} 
and the earlier findings \citep{franklin07,ayala08a,woittiez09,onishi09,rosa13,bec14,gustavsson14}. 
For sufficiently large separations, the RDFs with and without gravity approach each other.

To test the theory in \textsection \ref{sec:particle_clustering_theory}, 
we compute $g(r)$ from (\ref{eq:ss_rdf_isotropic}) 
using the relative velocity statistics $S^p_{2\parallel}$ and $S^p_{2\perp}$ obtained from DNS. 
In addition, we use the directionally averaged strain timescales $T^p_{\mathcal{S} \mathcal{S}}$ 
from the DNS (see \textsection \ref{sec:topology}) and the non-local closure proposed in 
\cite{bragg14} to compute the dispersion tensor. 
This allows us to test the formulation of (\ref{eq:ss_rdf_isotropic}) 
and the theoretical arguments in \textsection \ref{sec:particle_clustering_theory}. 
Figure~\ref{fig:rdfs} shows that the theory captures the quantitative results in the DNS well, 
indicating that (\ref{eq:ss_rdf_isotropic}) is an accurate model even for an anisotropic particle phase, 
and thus verifying the physical explanations presented in \textsection\ref{sec:particle_clustering_theory}.

\begin{figure}
 \centering
 \includegraphics[width=2.6in]{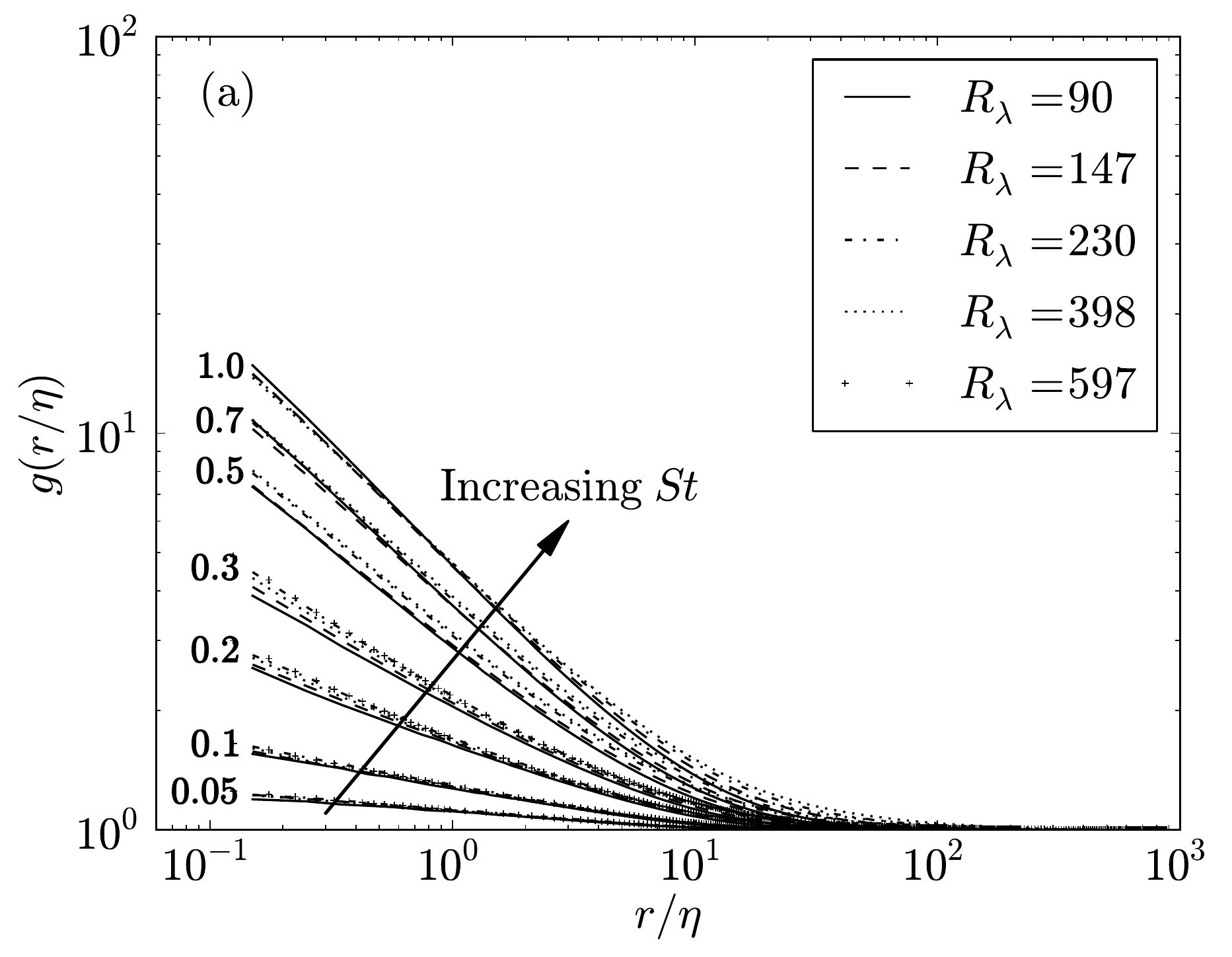}
 \includegraphics[width=2.6in]{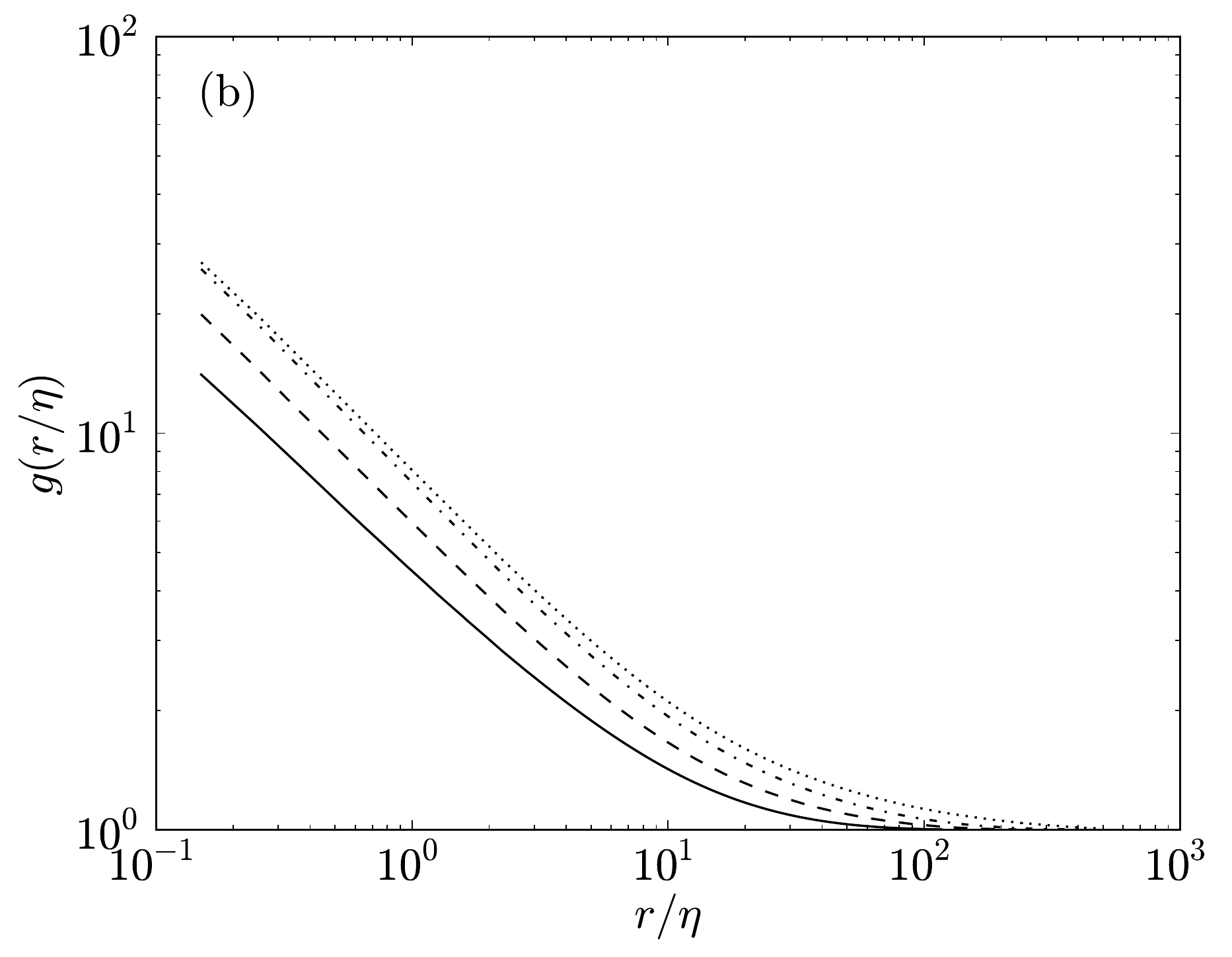}
 \caption{The RDFs $g(r/\eta)$ with gravity ($Fr=0.052$) for different $R_\lambda$ for
 $St \leq 1$ (a) and for $St=3$ (b). The Stokes numbers in (a) are indicated
 by the line labels.}
 \label{fig:rdfs_re_dependence}
\end{figure}

Next, we consider the dependence of the RDFs on $R_\lambda$ in figure~\ref{fig:rdfs_re_dependence}.
As was the case without gravity (see Part I), 
the RDFs are largely independent of $R_\lambda$
for $St \lesssim 1$ with gravity (figure~\ref{fig:rdfs_re_dependence}(a)).
This is in agreement with our arguments in \textsection \ref{sec:particle_clustering_theory}
and implies that small-$St$ clustering is a small-scale phenomenon (both with and without gravity)
that is generally unaffected by the intermittency of the turbulence.
At larger $St$ (figure~\ref{fig:rdfs_re_dependence}(b)) 
the RDFs increase monotonically with increasing $R_\lambda$, since 
the ratio between the drift and diffusion increases,
as discussed in \textsection \ref{sec:particle_clustering_theory}.

\begin{figure}
 \centering
 \includegraphics[width=2.6in]{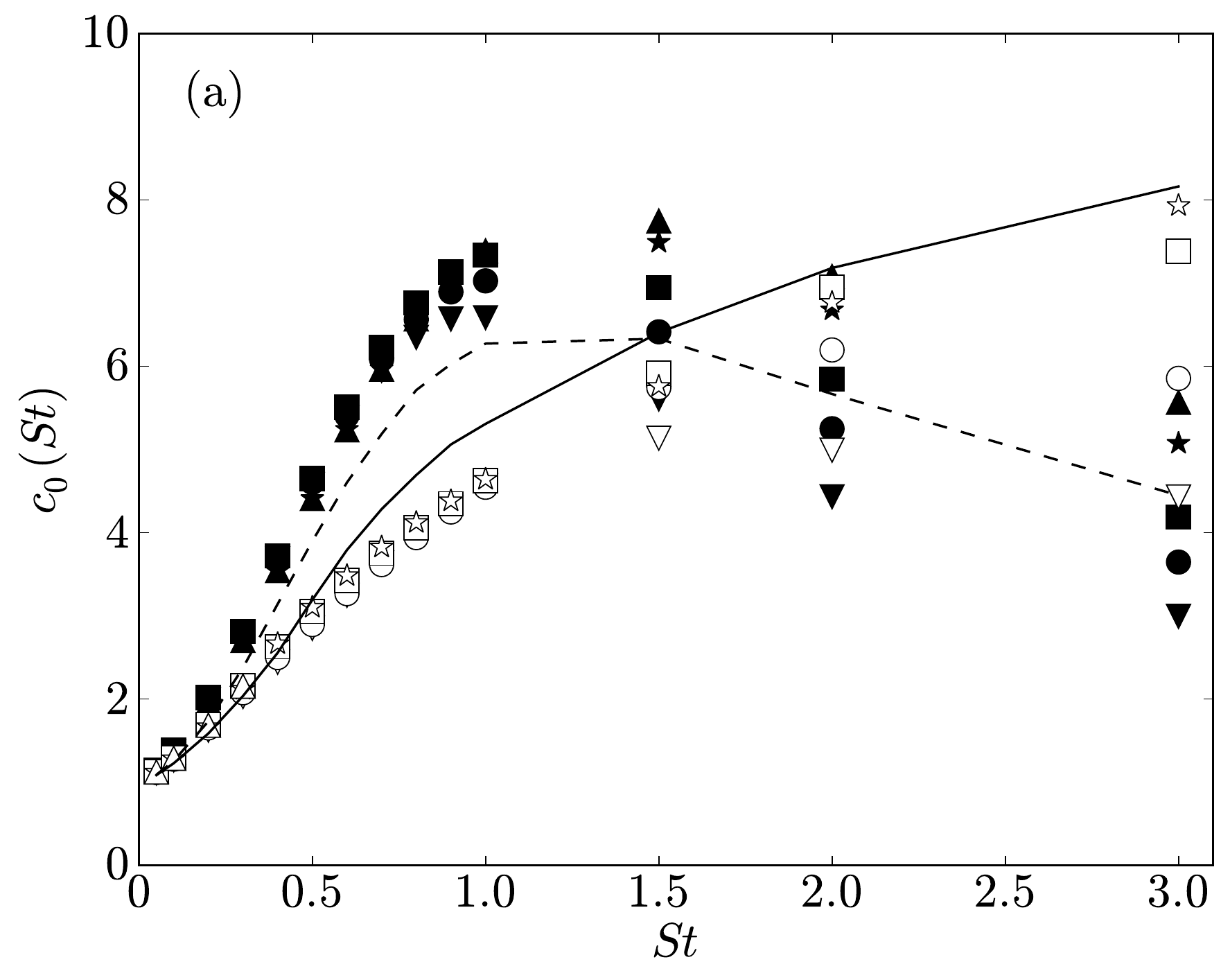}
 \includegraphics[width=2.6in]{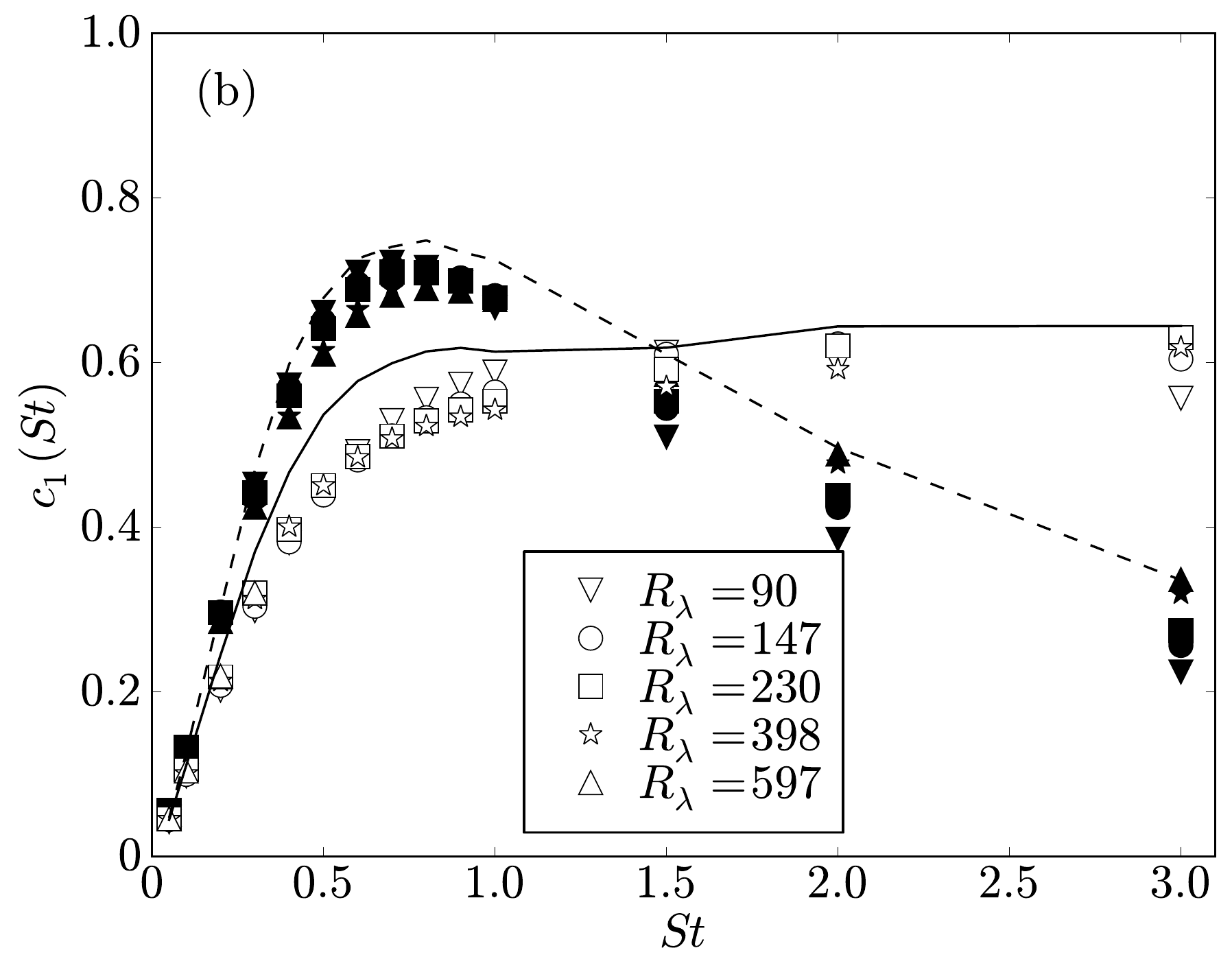}
 \caption{The prefactor $c_0$ (a) and the exponent $c_1$ (b) of the power-law fits
 of the RDFs using (\ref{eq:power_law_rdf}).
 Open symbols denote data with gravity ($Fr = 0.052$), and filled symbols denote data
 without gravity. The predictions from \cite{zaichik09}
 for $R_\lambda = 398$ (where DNS data are used to specify the relative velocities and the strain timescales here)
 are shown with solid lines (gravity) and dashed lines (no gravity).}
 \label{fig:c0_c1}
\end{figure}

We can also quantify the degree of small-scale clustering by performing a power-law fit of the RDFs
\citep{reade00} at small separations,
\begin{equation}
 g(r) \approx c_0 \left( \frac{\eta}{r} \right)^{c_1} \mathrm{,}
 \label{eq:power_law_rdf}
\end{equation}
as discussed in Part I. The power-law fits are performed over the range $0.75 \leq r/\eta \leq 2.75$,
and the calculated values of $c_0$ and $c_1$ are plotted in figure~\ref{fig:c0_c1}.
We observe that both the power-law coefficient $c_0$ and the exponent $c_1$ decrease
when gravity is introduced for $St \lesssim 1.5$ and increase 
when gravity is introduced for $St \gtrsim 1.5$,
consistent with our explanations in \textsection \ref{sec:particle_clustering_theory}. 
The theoretical model for $g(r/\eta)$
from (\ref{eq:ss_rdf_isotropic}) (with the relative velocities and strain timescales specified from the DNS)
is in good agreement with the DNS data, both with and without gravity.
We also note that our DNS results for the exponent $c_1$
agree well with those of \cite{rosa13} and \cite{bec14} (not shown).

\begin{figure}
 \centering
 \includegraphics[width=2.6in]{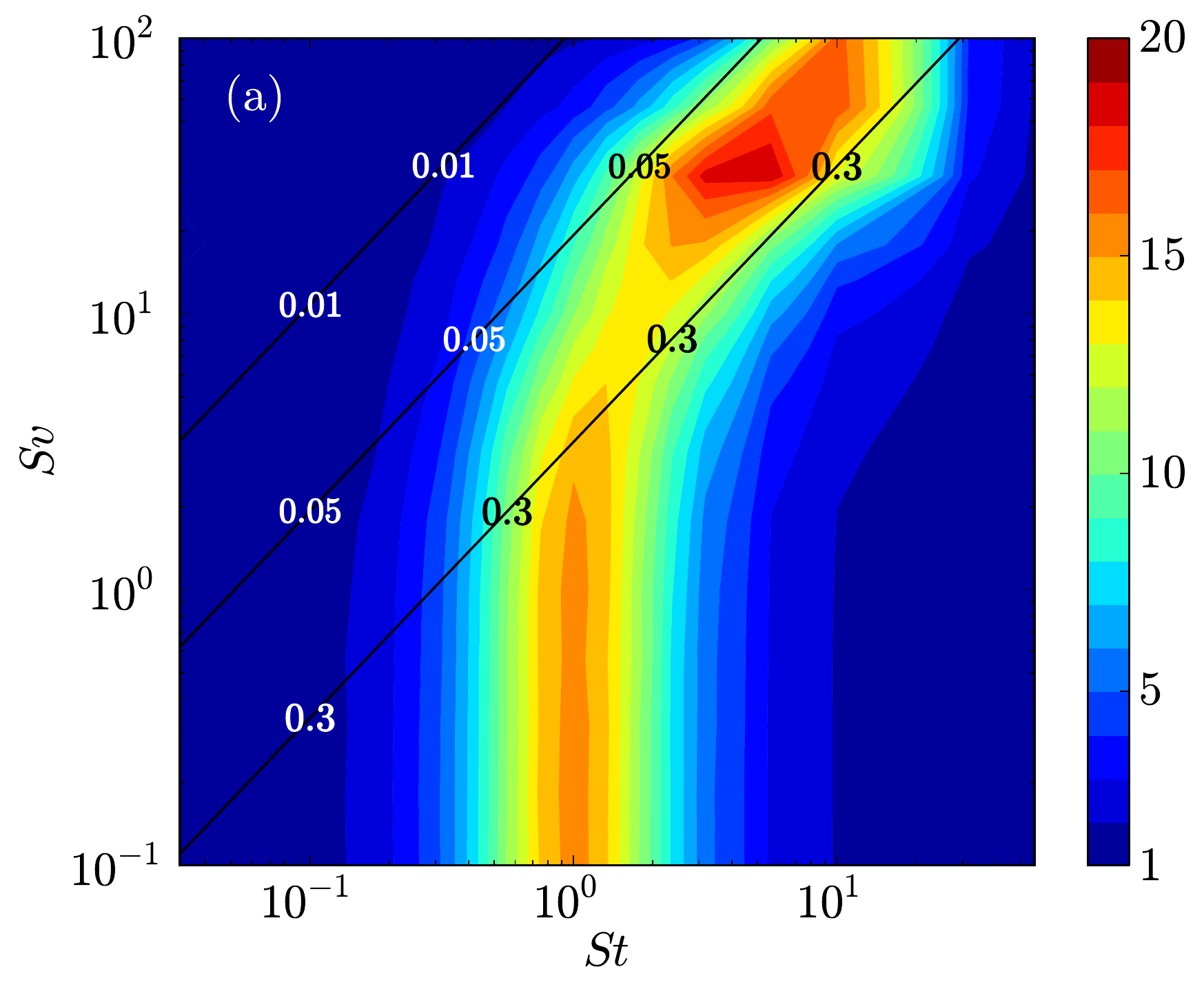}
 \includegraphics[width=2.6in]{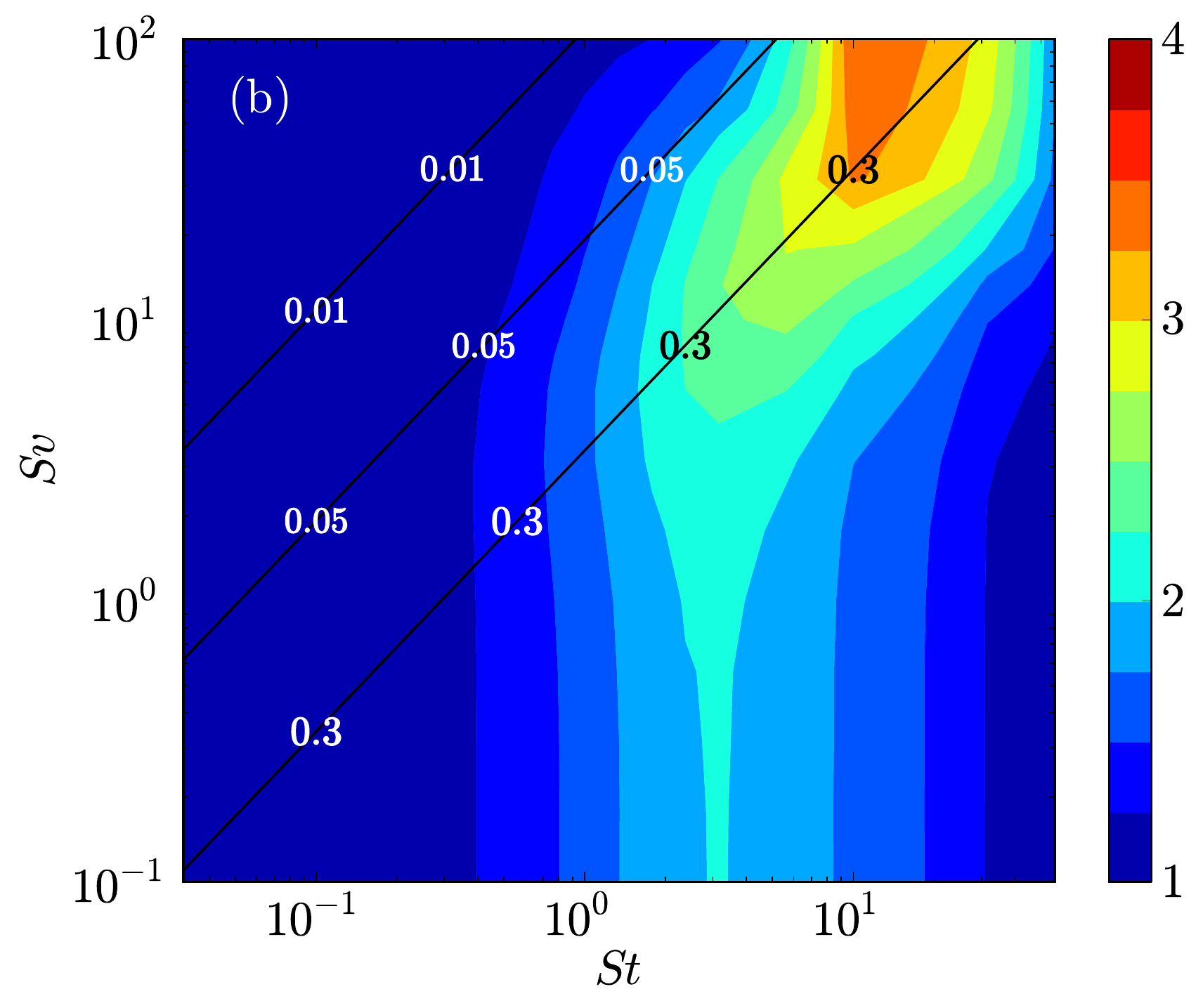}
 \caption{Filled contours of the RDF evaluated at $r/\eta = 0.25$ (a) and $r/\eta = 9.75$ (b)
 for different values of $St$ and $Sv$, for $R_\lambda = 227$.
 The diagonal lines denote three different
 values of $Fr$, corresponding to conditions representative of stratiform clouds
 ($Fr = 0.01$), cumulus clouds ($Fr = 0.05$), and cumulonimbus clouds ($Fr = 0.3$).}
 \label{fig:rdfs_St_Sv}
\end{figure}

To further investigate the two-parameter space of particle inertia and gravity,
we consider the RDFs at $R_\lambda = 227$ for $0 < St \leq 56.2$ and $0 < Sv \leq 100$
in figure~\ref{fig:rdfs_St_Sv}. 
While the particle clustering behavior here is generally complex and 
varies strongly with $St$ and $Sv$, we are able to provide
physical explanations for several of the observed trends.
For $St \gg 1$, the RDFs generally decrease
with increasing $St$, since the particles become unresponsive to almost
all of the underlying turbulence, as expected. Also, for $St \lesssim 1$, the RDFs
tend to decrease with increasing $Sv$, since the preferential-sampling 
and path-history mechanisms may act to increase clustering here, and they are both
reduced by gravity.
We also see that for $1 < St < 10$, the RDFs generally increase
with increasing $Sv$, as expected, since gravity increases
the ratio between the drift and the diffusion,
as discussed in \textsection \ref{sec:particle_clustering_theory}.

\begin{figure}
 \centering
 \includegraphics[height=1.8in]{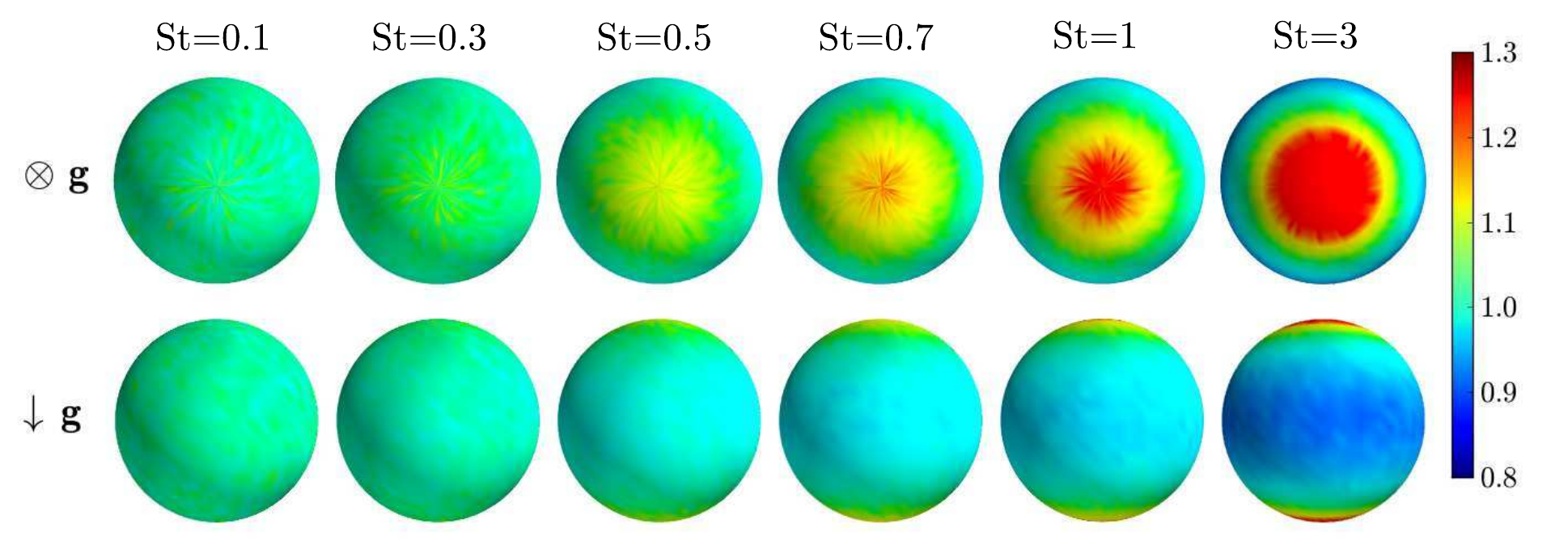}
 \caption{The ADF $g(\bm{r})$ (normalized by the RDF $g(r)$) shown on a unit sphere
 for $R_\lambda = 398$ and separations $r < \eta$ with gravity ($Fr = 0.052$). The different columns
 correspond to different values of $St$.  The top row shows the projection where gravity is directed
 into the page, and the bottom row shows the projection where gravity is directed downward.}
 \label{fig:adfs_3d}
\end{figure}

We now investigate the anisotropy of the particle field by introducing higher-order spherical harmonic functions,
\begin{equation}
 \frac{g(\bm{r})}{g(r)}
 = \sum_{\ell = 1}^{\infty} \frac{\mathcal{C}_{2 \ell}^0 (r)}{\mathcal{C}_0^0(r)} Y_{2 \ell}^0 (\theta)\ ,
\end{equation}
in figure~\ref{fig:adfs_3d} for various values of $St$ with gravity ($Fr = 0.052$) at $R_\lambda = 398$.

Consistent with the qualitative observations from figure~\ref{fig:particle_vis}, 
we see that particle clustering is strongest along the vertical direction, 
in agreement with earlier findings of \cite{dejoan13}, \cite{bec14} and \cite{park14}. 
At low $St$, the anisotropic clustering is caused by the fact that particles tend to preferentially 
sample downward-moving flow (see \textsection \ref{sec:settling}). \cite{bec14} 
showed that this preferential sampling causes particles to form 
vertical clusters when $St$ is small.

When $St$ is large, the effects of preferential sampling vanish, and the anisotropy 
is related to the way gravity influences the path-history effect. 
Since the analysis presented in \S\ref{sec:particle_clustering_theory} is restricted to spherically averaged clustering, 
the results cannot be used to predict the trends in anisotropy. However, we hope 
that the strain and rotation timescale results presented in \textsection \ref{sec:topology} 
will aid in the development of future models to predict this anisotropy.

We also note that the anisotropy increases with increasing $St$. 
The physical explanation is that as $St$ (and thus $Sv$) increases, 
gravitational forces become more significant, causing the particle motion to be 
more anisotropic. However, we expect that at some sufficiently large value of $St$ and $Sv$, 
the particles are unaffected by the fluid turbulence, 
and thus clustering (and the associated anisotropy) will vanish.

\begin{figure}
 \centering
 \includegraphics[width=2.6in]{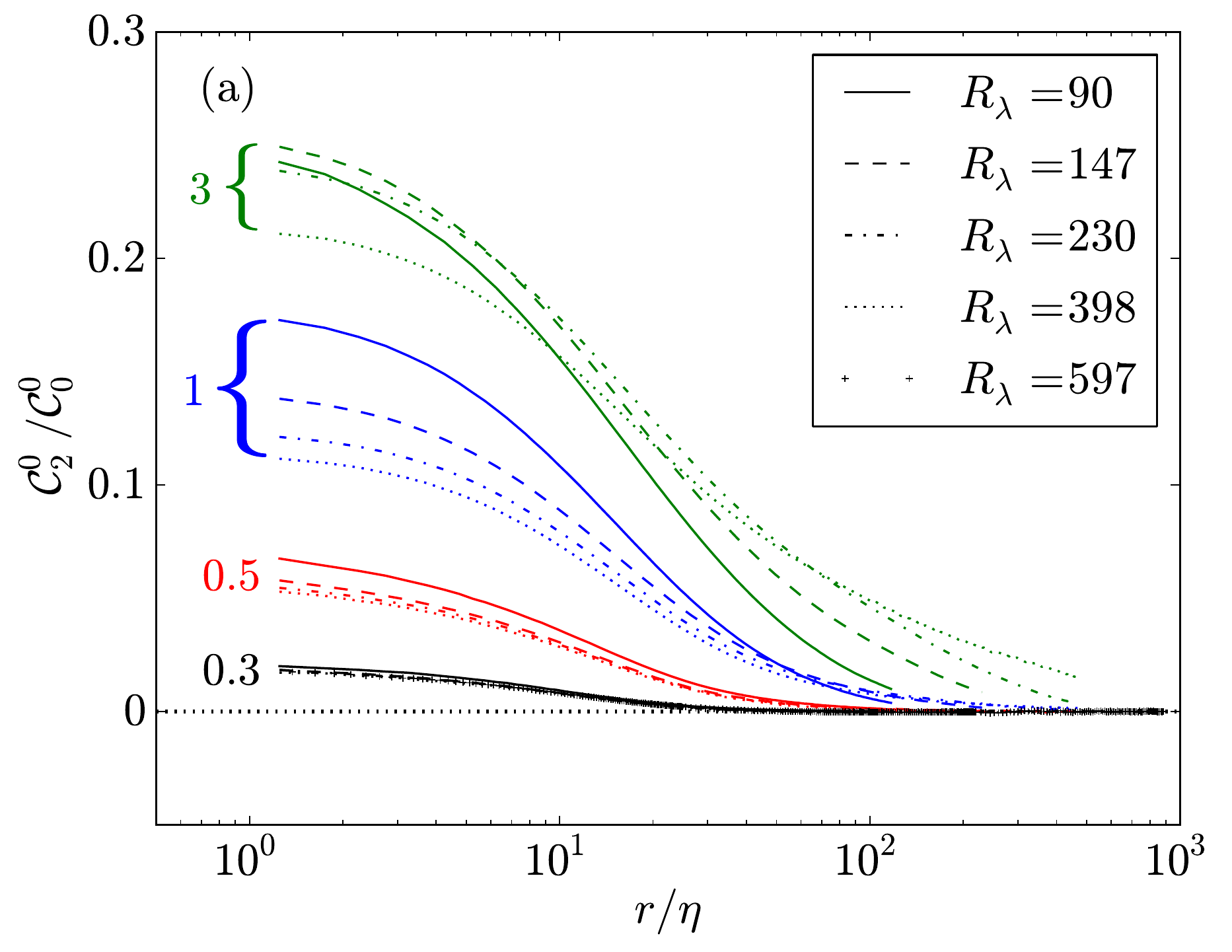}
 \includegraphics[width=2.6in]{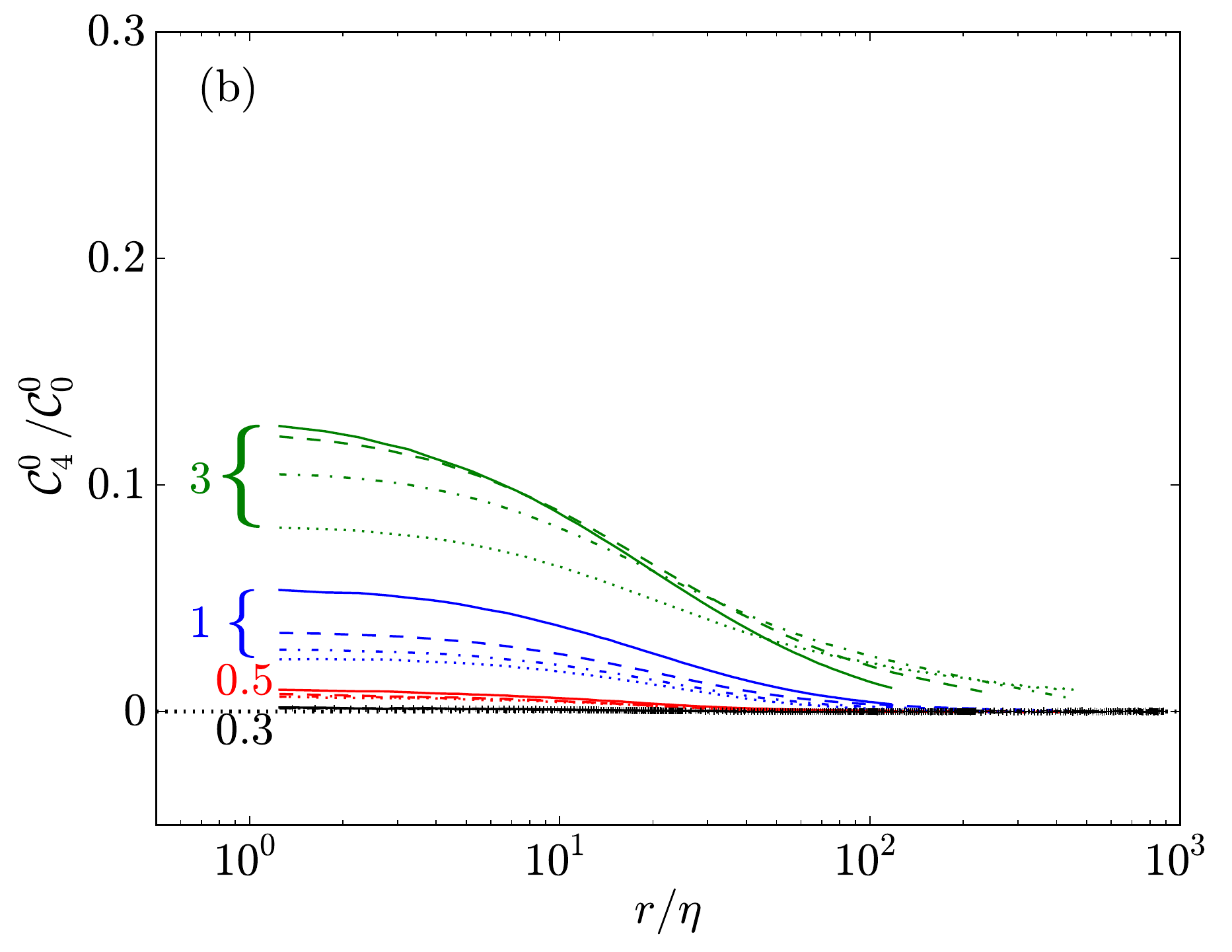}
 \caption{The second (a) and fourth (b) spherical harmonic coefficients
 of the angular distribution function, 
 normalized by the zeroth spherical harmonic coefficient, plotted as a function of $r/\eta$,
 for different $St$ and $R_\lambda$ with gravity ($Fr = 0.052$). 
 The different values of $St$ considered ($0.3$, $0.5$, $1$, $3$)
 are shown in black, red, blue and green, respectively, and 
 the Stokes numbers are indicated by the line labels.}
 \label{fig:spher_harm_adf}
\end{figure}

We explore the degree of anisotropy as a function of the separation distance $r$ and the Reynolds number 
by plotting the spherical harmonic coefficients $\mathcal{C}_2^0$ and $\mathcal{C}_4^0$ in figure~\ref{fig:spher_harm_adf}
(coefficients above order four are too small to be statistically significant, and hence are not shown). 
In agreement with figure~\ref{fig:particle_vis}, we see that the anisotropy increases with increasing $St$.
We also observe that as $r/\eta$ increases, the anisotropy approaches zero, 
since both the clustering and the clustering anisotropy vanish in the limit $r/\eta\rightarrow\infty$.

The degree of anisotropy is also Reynolds-number-dependent and decreases with increasing $R_\lambda$ 
for larger values of $St$. 
Since the relative velocities of high-$St$ particles generally also tend to become more isotropic as $R_\lambda$ 
increases (see \textsection \ref{sec:relative_velocity_dissipation}), 
it is likely that the reduction in the anisotropy of the relative 
velocities will cause a similar reduction in the anisotropy of the clustering.

\subsection{Particle collision kernels}
\label{sec:collision_kernel}

\begin{figure}
 \centering
 \includegraphics[height=2.5in]{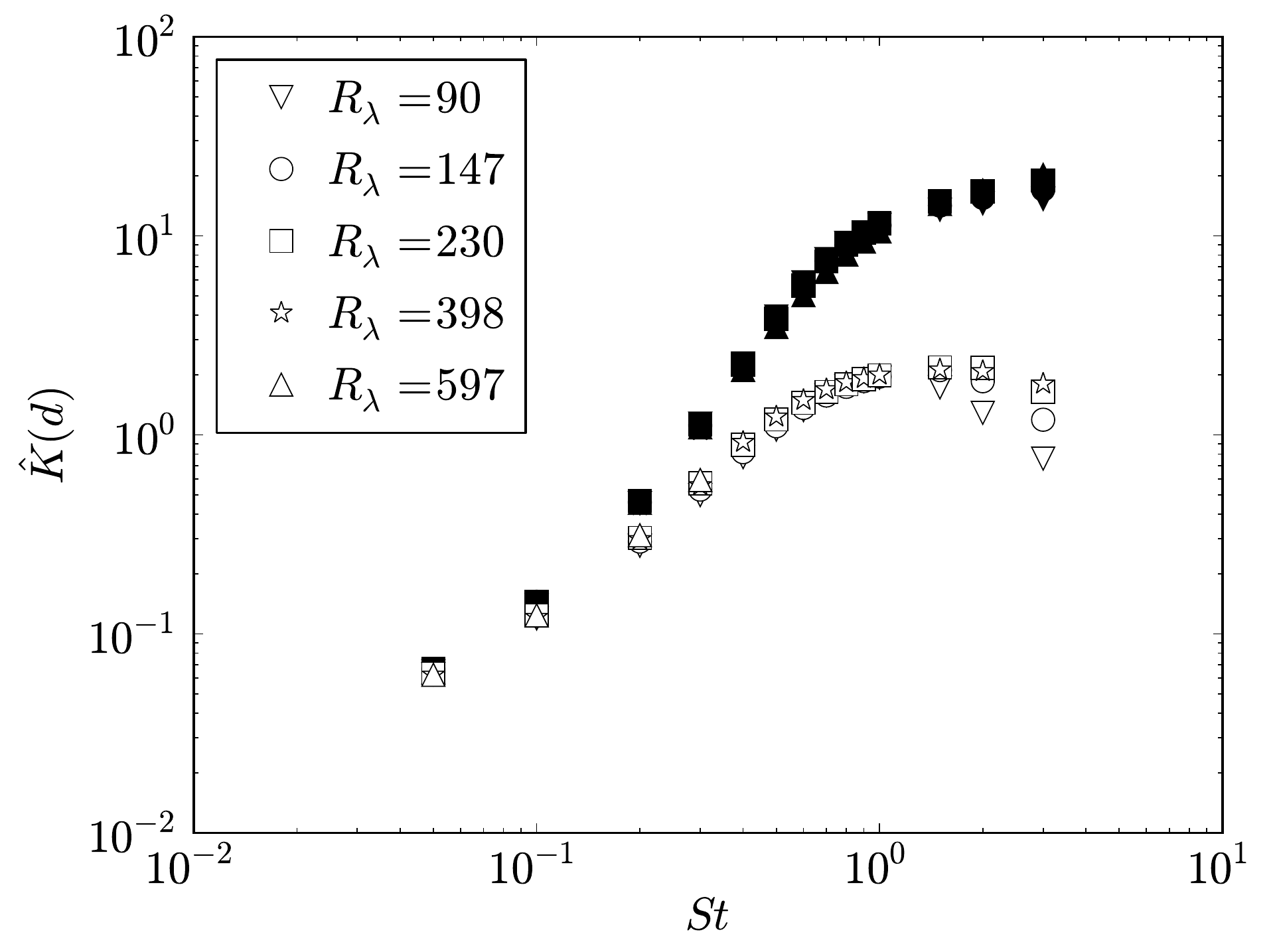}
  \caption{The non-dimensional collision kernel $\hat{K}(d)$ plotted as a function of $St$
  for $\rho_p/\rho_f = 1000$ and different values of $R_\lambda$. The open symbols denote data with gravity
  ($Fr = 0.052$), and the filled symbols denote data without gravity.}
 \label{fig:collision_kernel}
\end{figure}

The final two-particle statistic we consider is the kinematic collision kernel $K$ for inertial particles.
\cite{sundaram4} and \cite{wwz00} showed that for an isotropic particle field, $K$ is given by
\begin{equation}
 K(d) = 4 \pi d^2 g(d) S^p_{-\parallel}(d) \mathrm{,}
 \label{eq:collision_kernel}
\end{equation}
where $d$ is the particle diameter.
In Appendix~\ref{sec:collision_kernel_theory}, we show mathematically
that (\ref{eq:collision_kernel}) holds even for an anisotropic particle field.
As in Part I, we plot the non-dimensional collision kernel $\hat{K}(d) = K(d) / (d^2 u_\eta)$ 
in figure~\ref{fig:collision_kernel}
\citep[see also][]{vosskuhle14}.
Note that while we simulate only point particles, we define
$d$ from $St$ for this plot by prescribing the density ratio to be $\rho_p/\rho_f = 1000$.
Since our statistics are generally not sufficient to compute the RDFs and relative velocities
at separations on the order of the particle diameter, we fit both quantities
using linear least-squares power-law regression and extrapolate the resulting power-law fits
to $r=d$ \citep[refer to][]{rosa13}.

Figure~\ref{fig:collision_kernel} shows the dimensionless collision kernel, with and without gravity, 
as a function of the Stokes number. 
Notice that the collision kernel is reduced by gravity. 
At low $St$, the RDF and the relative velocity both decrease with gravity, 
thereby decreasing the collision kernel. At higher $St$, 
the reduction in the relative velocity with gravity (see \textsection \ref{sec:relative_velocity_dissipation}) 
is sufficient to compensate for the slight increase in the RDF with gravity 
(see \textsection \ref{sec:particle_clustering_dns}), 
sustaining the net reduction in the collision kernel. 
Finally, we note that at large $St$ with gravity, the collision rates increase with increasing 
$R_\lambda$ since both the relative velocity and RDF increase with increasing Reynolds number.

\begin{figure}
 \centering
 \includegraphics[height=2.5in]{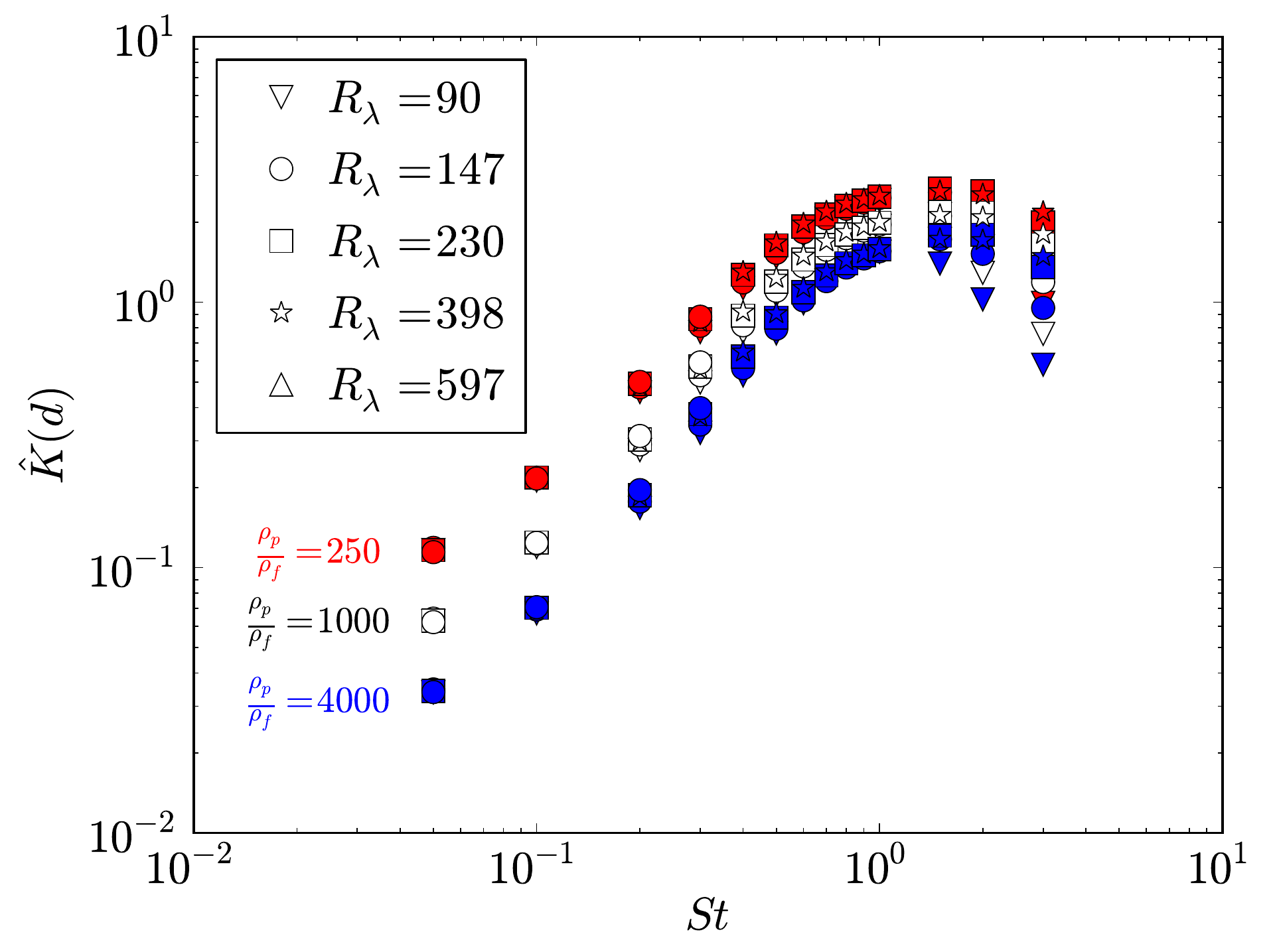}
  \caption{The non-dimensional collision kernel $\hat{K}(d)$ plotted as a function of $St$
  for different values of $\rho_p/\rho_f$ and $R_\lambda$. 
  All data are with gravity. Filled red symbols indicate $\rho_p/\rho_f = 250$,
  open symbols indicate $\rho_p/\rho_f = 1000$, and filled blue symbols indicate $\rho_p/\rho_f = 4000$.}
 \label{fig:collision_kernel_rhop_rhof}
\end{figure}

We found in Part I that $\hat{K}(d)$ without gravity is nearly independent of $\rho_p/\rho_f$ for $St \gtrsim 1$, 
in agreement with \cite{vosskuhle14}. The physical explanation is that $g(d)$ and $S^p_{-\parallel}(d)/u_\eta$ 
are either independent of $d$ (for $St \geq 10$) or have inverse power-law scalings with $d$ that precisely cancel (for $1 \lesssim St \leq 3$). 
In contrast, with gravity, path-history effects are suppressed, 
and these quantities have different scaling behaviors 
(see \textsection \ref{sec:relative_velocity_dissipation} and \textsection \ref{sec:particle_clustering_dns}). 
We therefore find that with gravity, $\hat{K}(d)$ decreases with increasing 
$\rho_p/\rho_f$ at all values of $St$ considered, as shown in figure~\ref{fig:collision_kernel_rhop_rhof}.

\begin{figure}
 \centering
 \includegraphics[height=2.5in]{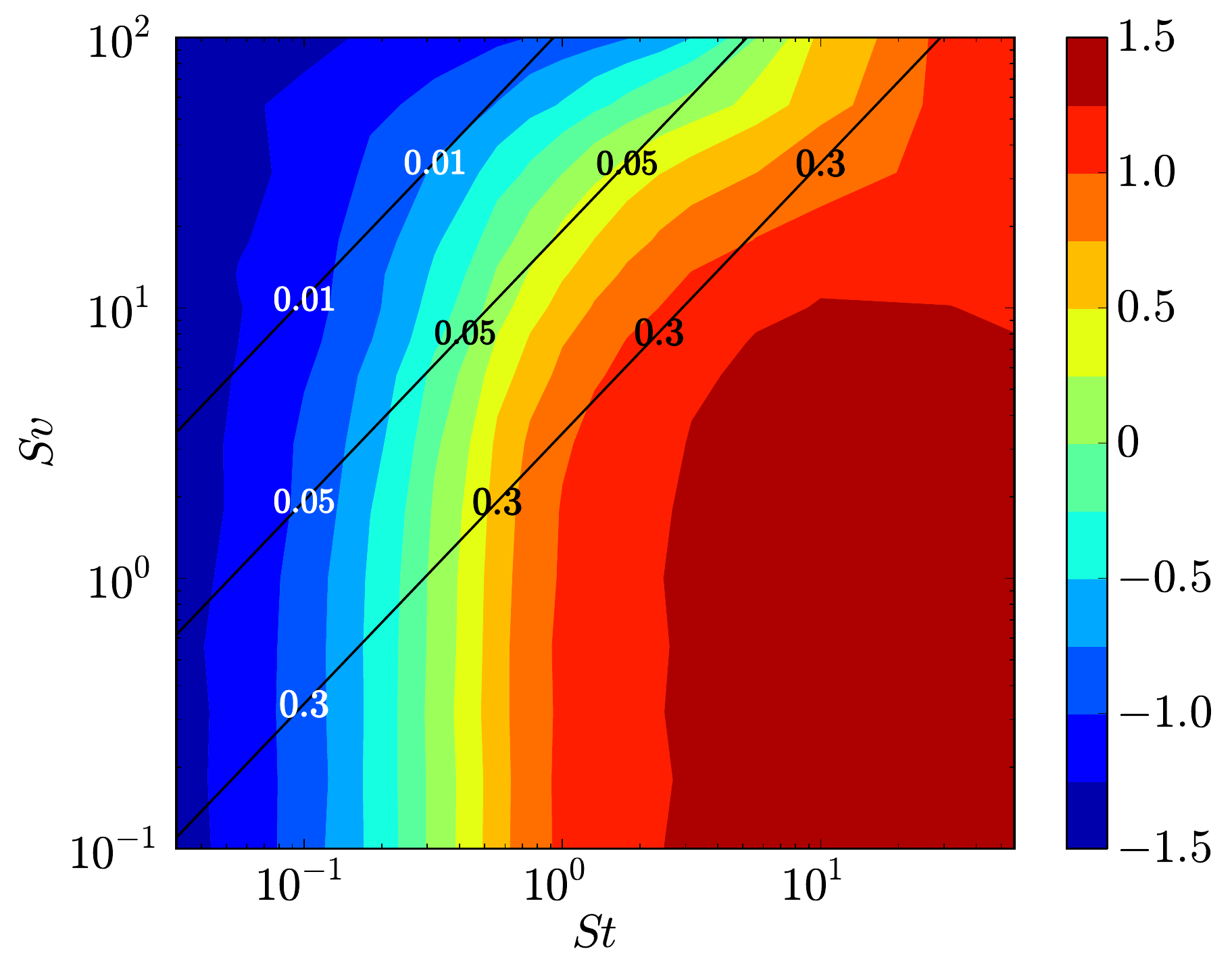}
 \caption{Filled contours of $\hat{K}(d)$  for different values of $St$ and $Sv$ at $R_\lambda = 227$
 for $\rho_p/\rho_f = 1000$.
 The contours are logarithmically scaled, and the colorbar labels indicate the exponents of the decade.
 The diagonal lines denote three different
 values of $Fr$, corresponding to conditions representative of stratiform clouds
 ($Fr = 0.01$), cumulus clouds ($Fr = 0.05$), and cumulonimbus clouds ($Fr = 0.3$).}
 \label{fig:collision_kernel_StSv}
\end{figure}

Next, we expand our parameter space to consider the collision kernel for $0 < St \leq 56.2$ and 
$0 < Sv \leq 100$ at $R_\lambda = 227$ in figure~\ref{fig:collision_kernel_StSv}. 
Here, we limit our analysis to cases where $\rho_p/\rho_f = 1000$. 
Once again, we caution that the statistics at the highest values of $St$ and $Sv$ 
may be affected by the periodic boundary conditions in the vertical direction 
(see Appendix~\ref{sec:periodicity}), the use of a linear drag model 
(cf. \textsection \ref{sec:particle_phase}), and the extrapolations based on power-law fits 
that become questionable for $St > 3$. Nevertheless, the contour plot shown in 
figure~\ref{fig:collision_kernel_StSv} captures the qualitative trends of the collision kernel.
In agreement with the discussion above, we 
see that the collision kernel generally increases with increasing $St$ 
and decreases with increasing gravity. 
A practical implication is that at a given value of $St$, cumulonimbus clouds ($Fr \approx 0.3$) 
will have more frequent droplet collisions and growth than stratiform clouds ($Fr \approx 0.01$).

\begin{figure}
 \centering
 \includegraphics[height=1.8in]{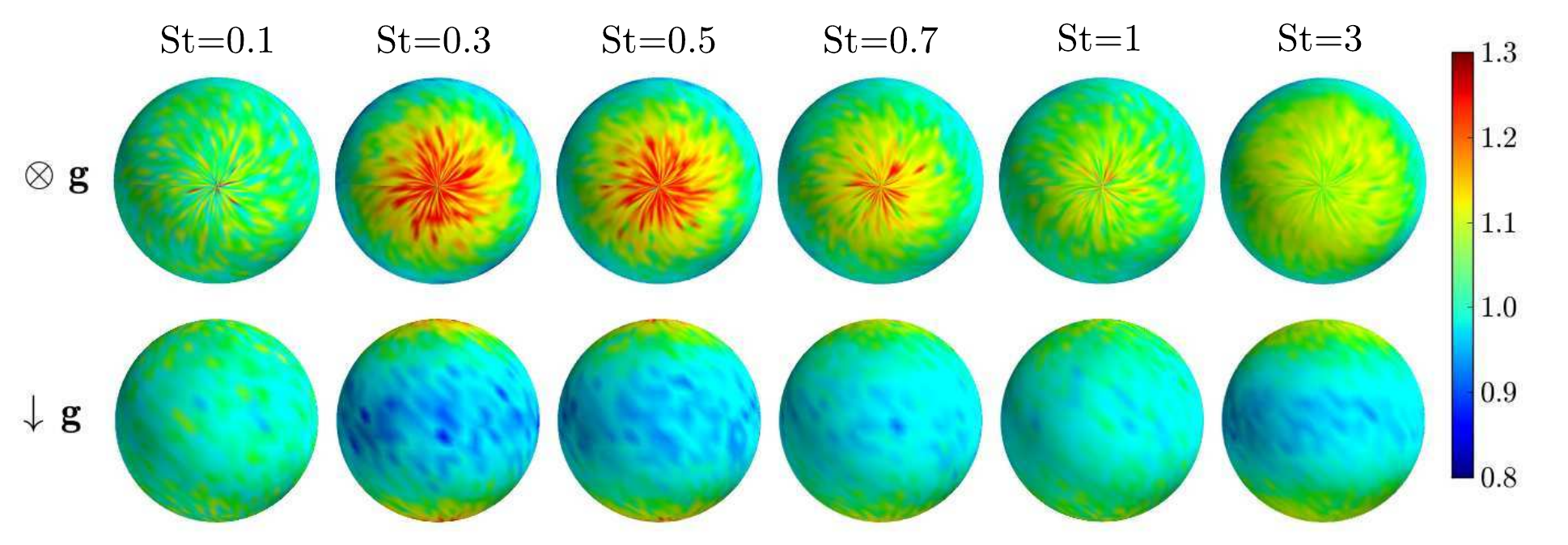}
 \caption{The anisotropic collision kernel $K(r,\theta,\phi)$ (normalized by the spherical average $K(r)$)
 shown on a unit sphere for $R_\lambda = 398$ and separations $r < \eta$ with gravity ($Fr = 0.052$).
 The different columns correspond to different values of $St$.  
 The top row shows the projection where gravity is directed
 into the page, and the bottom row shows the projection where gravity is directed downward.}
 \label{fig:collision_kernel_3d}
\end{figure}

We now consider the anisotropic collision kernel, $K(r,\theta,\phi)$, defined as
\begin{equation}
 K(d,\theta,\phi) \equiv 4 \pi r^2 g(d,\theta,\phi) S^p_{-\parallel}(d,\theta,\phi)  \mathrm{.}
\end{equation}
$K(d,\theta,\phi)$ provides a measure of the rate at which particles with diameter $d$ 
collide along an orientation defined by the radial angle $\theta$ and azimuthal angle $\phi$.
We plot the ratio between $K(d,\theta,\phi)$ and its spherical average in figure~\ref{fig:collision_kernel_3d}
at $R_\lambda = 398$. (We are unable to show the anisotropic collision kernel at $r = d$
due to inadequate statistics at these small separations. We instead show data
at $0 \leq r < \eta$, the smallest separation range over which adequate statistics are available.)

At low $St$, the collision kernel is approximately isotropic, since gravitational effects are weak.
Interestingly, the collision kernel also tends toward isotropy at large $St$, due to the opposing trends
in anisotropy of the relative velocities (see figure~\ref{fig:wr_3d})
and ADFs (see figure~\ref{fig:adfs_3d}).
For $St$ between $0.1$ and $1$, particles are more likely to collide along the vertical direction, 
since both the ADFs and the relative velocities 
are strongest in the vertical direction for these values of $St$.

Before closing this section, we emphasize two practical implications of these collision results 
for the cloud physics community. The first is that since the DNS indicates that the collision rates 
of particles with low and moderate $St$ are independent of $R_\lambda$, 
it is likely that the collision kernels computed here will be useful for predicting droplet collisions in 
high-Reynolds-number atmospheric clouds. The second is that gravity significantly reduces 
the collision kernels for $St \gtrsim 0.1$, which implies that simulations without gravity 
can significantly over-predict the collision rates of droplets in atmospheric clouds, 
and highlights the need to include gravity in the analysis of collisional droplet growth in turbulent clouds.

\section{Conclusions}
\label{sec:conclusions}

In this study, we explored the influence of gravity on inertial particle statistics in isotropic turbulence,
with the goal of understanding the turbulence mechanisms contributing to droplet collisions.
Our simulations were performed over
the largest Reynolds-number range ($90 \leq R_\lambda \leq 597$),
domain lengths ($\mathcal{L}/\ell \lesssim 40$), and range of particle
classes ($0 \leq St \leq 56.2$, $0 \leq Sv \leq 100$) to date. 
We showed that such large domain sizes are essential to obtain accurate statistics of settling inertial particles,
suggesting that earlier published DNS studies may have errors due to the periodic boundary conditions
in the vertical direction (see Appendix~\ref{sec:periodicity}).

Our results indicate that preferential sampling affects the dynamics of particles with 
$St \ll 1$, both with and without gravity. 
Gravity decreases the degree of preferential 
sampling by reducing the time the particles have to interact with the underlying turbulent fields. 
In particular, gravity reduces the Lagrangian timescales for strain and rotation along particle trajectories.
As gravitational forces increase, the particles fall more rapidly through the flow, 
leading to smaller velocities and larger accelerations than in the case without gravity. 
We developed models for the Lagrangian strain and rotation timescales along particle trajectories 
and the particle acceleration variances that are valid in the limit $Sv \gg u'/u_\eta$. 
The model predictions are in reasonable agreement with the DNS in this limit. 
We also find that the mean settling velocity is independent of $R_\lambda$ at low $St$, 
in agreement with \cite{bec14}.

We then applied this understanding of the gravity effect on single-particle statistics to 
two-particle statistics relevant for predicting the collision kernel. 
At high $St$, we observed that gravity reduces the particle relative velocities 
from their values without gravity by reducing the path-history effect. 
At low $St$, gravity acts primarily to reduce the degree of preferential sampling, 
causing the relative velocities at small separations to be closer to those of fluid particles.

Next, we relate the trends in the relative velocities to those in the particle clustering 
by considering the effect of gravity on a derivative model of the theoretical work of \cite{zaichik09}.
With gravitational effects included, the model is able to predict the spherically averaged RDFs 
very accurately when DNS data is used to prescribe the relative velocities and the 
Lagrangian strain timescales.  By analyzing the model at low $St$, 
we see that the primary effect of gravity is to decrease the inward drift, 
leading to a decrease in the RDFs. 
At higher $St$, in contrast, gravity modulates both the inward drift 
\emph{and} outward diffusion by diminishing the path-history effect in such a way 
as to increase the drift-to-diffusion ratio, 
causing an increase in clustering with gravity. 
We also find that the degree of clustering is only weakly dependent on 
$R_\lambda$ at low $St$, but becomes increasingly sensitive to 
$R_\lambda$ at higher $St$. The model 
captures both of these trends through its internal description of the path-history effects.

We also quantify the degree of anisotropy in these two-particle statistics using a
spherical harmonic decomposition. Our results indicate that the particle angular distribution
functions and radial relative velocities can have anisotropies on the order of 25\%,
with the degree of anisotropy generally peaking at small $r/\eta$. At larger separations,
the relative velocities induced by turbulence become comparatively larger,
limiting the effect of gravity on particle dynamics.

We used these data for the RDFs and relative velocities to compute the particle collision
kernel. As in Part I, we find that the collision kernel is generally independent of $R_\lambda$
at low $St$, while it increases with increasing $R_\lambda$ at high $St$.
We analyze the collision kernel using spherical harmonic decompositions, and find
that the collision kernel is generally more isotropic than the ADFs and the mean
inward relative velocities, since the anisotropies
in the ADFs and the relative velocities have opposing trends at large $St$.

We conclude by highlighting some practical implications of this work for the cloud physics
and turbulence communities and suggesting promising research directions.
As in Part I, the fact that the collision rates are generally independent of $R_\lambda$
for droplet sizes representative of those in warm, cumulus clouds suggests that the
collision rates predicted here (for $R_\lambda \lesssim 600$) may be representative
of those found in atmospheric clouds ($R_\lambda \sim 10,000$). Of course, more
simulations at even higher Reynolds numbers would help to verify this conjecture.
Also, as noted in \textsection \ref{sec:collision_kernel}, we observed that
collision rates with gravity are considerably lower than without gravity, suggesting that
earlier work that neglected gravity may have over-predicted the collision kernels.

While our study has focused on like-sized particles,
our results suggest that gravity will have a significant effect on the collision kernels
of unlike (different-sized) particles. For example, we observe that gravity tends
to enhance the settling speeds of particles with low and intermediate values of $St$.
The coupling between turbulence and gravity 
may thereby increase the relative velocities of unlike particles,
leading to more frequent collisions. In addition, we find that gravity causes
inertial particles in turbulence to experience large accelerations.
\cite{chun05} found that large accelerations can contribute
to higher relative velocities between unlike particles,
leading to higher collision frequencies.

Finally, one especially promising finding of this study is that by extending
the model of \cite{zaichik09} and using DNS data to specify the 
Lagrangian strain timescales and relative velocities,
we are able to accurately predict the RDFs 
both without gravity \citep[as was also found in][]{bragg14} and with gravity.
As noted in \cite{bragg14b}, however, this model provides poor predictions
of the relative velocities of particles with moderate-to-large inertia.
Future work should therefore be directed at improving these relative velocity 
predictions. In addition, while we developed a model for the strain and rotation timescales
in the limit of large $St$ with gravity, we generally require a model for all values of $St$. 
If accurate models can be developed for the relative velocities and the strain and rotation timescales
over the entire range of $St$ in atmospheric clouds, we would be able to use the theory of \cite{zaichik09}
to model the collision kernel accurately, both with and without gravity, without requiring any inputs from DNS.
\section*{Acknowledgments}
\label{sec:acknowledgements}
The authors gratefully acknowledge Garrett Good, Stephen Pope, and Parvez Sukheswalla for many helpful discussions.
The work was supported by the National Science Foundation through CBET grants 0756510 and 0967349,
and through a graduate research fellowship to PJI. Additional funding was provided
by Cornell University. Computational simulations were performed on Yellowstone
(ark:/85065/d7wd3xhc) at the U.S. National Center for Atmospheric Research \citep{yellowstone}
under grants ACOR0001 and P35091057, and on resources at the Max Planck Institute for Dynamics and Self-Organization.
We are grateful to Denny Flieger for assistance with the computational resources at the Max Planck Institute.
\appendix
\setcounter{equation}{0}
\renewcommand{\theequation}{\Alph{section}\arabic{equation}}
\section{Periodicity effects}
\label{sec:periodicity}

As noted in \cite{woittiez09}, particle statistics in DNS may be artificially
influenced by the periodicity of the domain when gravitational forces are strong.
In particular, \cite{woittiez09} estimated that periodicity effects become significant
when the time it takes a particle to fall through the domain ($\sim \mathcal{L} \tau_p^{-1} g^{-1}$) 
is less than the large-eddy turnover time $T_L \equiv \ell / u'$, or equivalently,
when 
\begin{equation}
 Sv = \frac{St}{Fr} \gtrsim \frac{\mathcal{L}}{\ell} \frac{u'}{u_\eta} \mathrm{.}
 \label{eq:critical_Sv}
\end{equation}
In this case, a particle can artificially encounter the same large eddy multiple times
due to the finite domain length. We define $St_\mathrm{crit}$ as the value of $St$
at which $St / Fr = \mathcal{L} u' / (\ell u_\eta)$. This roughly
corresponds to the largest Stokes number at which we can expect the results
to be unaffected by periodicity.

To study the effects of periodicity, we systematically increased the domain size
for the cases at the three lowest Reynolds numbers, while keeping the large scales,
small scales, and the forcing parameters the same.  The simulation parameters are summarized in 
table~\ref{tab:periodicity}. 
For convenience,
we will hereafter refer to the simulations from groups I, II, and III
by their nominal Reynolds numbers of $R_\lambda \approx 90$, $147$, and $230$, respectively.
In all cases, $0 \leq St \leq 3$ and $Fr = 0.052$.

\begin{table}
 \centering
 \caption{Simulation parameters for the periodicity study.}
 \label{tab:periodicity}
 \begin{tabular}{l  l  l  l  l  l  l  l  l  l  l  l  l  l  l  l}
 & & ~ & \multicolumn{4}{c}{I} & ~ & ~ & \multicolumn{3}{c}{II} & ~ & ~ & \multicolumn{2}{c}{III}\\
 $R_\lambda$ & ~ & ~ & 88 & 90 & 90 & 90 & ~ & ~ & 140 & 145 & 147 & ~ & ~ & 226 & 230  \\
 $N$ & ~ & ~ & 128 & 256 & 512 & 1024 & ~ & ~ & 256 & 512 & 1024 & ~ & ~ & 512 & 1024  \\
 $\mathcal{L}$ & ~ & ~ & $2 \pi$ & $4 \pi$ & $8 \pi$ & $16 \pi$ & ~ & ~ & $2 \pi$ & $4 \pi$ & $8 \pi$ & ~ & ~ & $2 \pi$ & $4 \pi$ \\
 $St_\mathrm{crit}$ & ~ & ~ & 1.06 & 2.06 & 4.26 & 8.55 & ~ & ~ & 1.39 & 2.69 & 5.55 & ~ & ~ & 1.78 & 3.39 \\
 $\ell$ & ~ & ~ & 1.46 & 1.52 & 1.47 & 1.47 & ~ & ~ & 1.40 & 1.49 & 1.44 & ~ & ~ & 1.40 & 1.49 \\
 $\ell/\eta$ & ~ & ~ & 55.8 & 57.7 & 55.7 & 55.6 & ~ & ~ & 106 & 111 & 107 & ~ & ~ & 202 & 213 \\
 $u'$ & ~ & ~ & 0.914 & 0.912 & 0.912 & 0.912 & ~ & ~ & 0.914 & 0.916 & 0.914 & ~ & ~ & 0.915 & 0.914 \\
 $u'/u_\eta$ & ~ & ~ & 4.77 & 4.81 & 4.82 & 4.82 & ~ & ~ & 6.01 & 6.12 & 6.15 & ~ & ~ & 7.60 & 7.70 \\
 \end{tabular}
\end{table}

\begin{figure}
 \centering
 \includegraphics[width=2.6in]{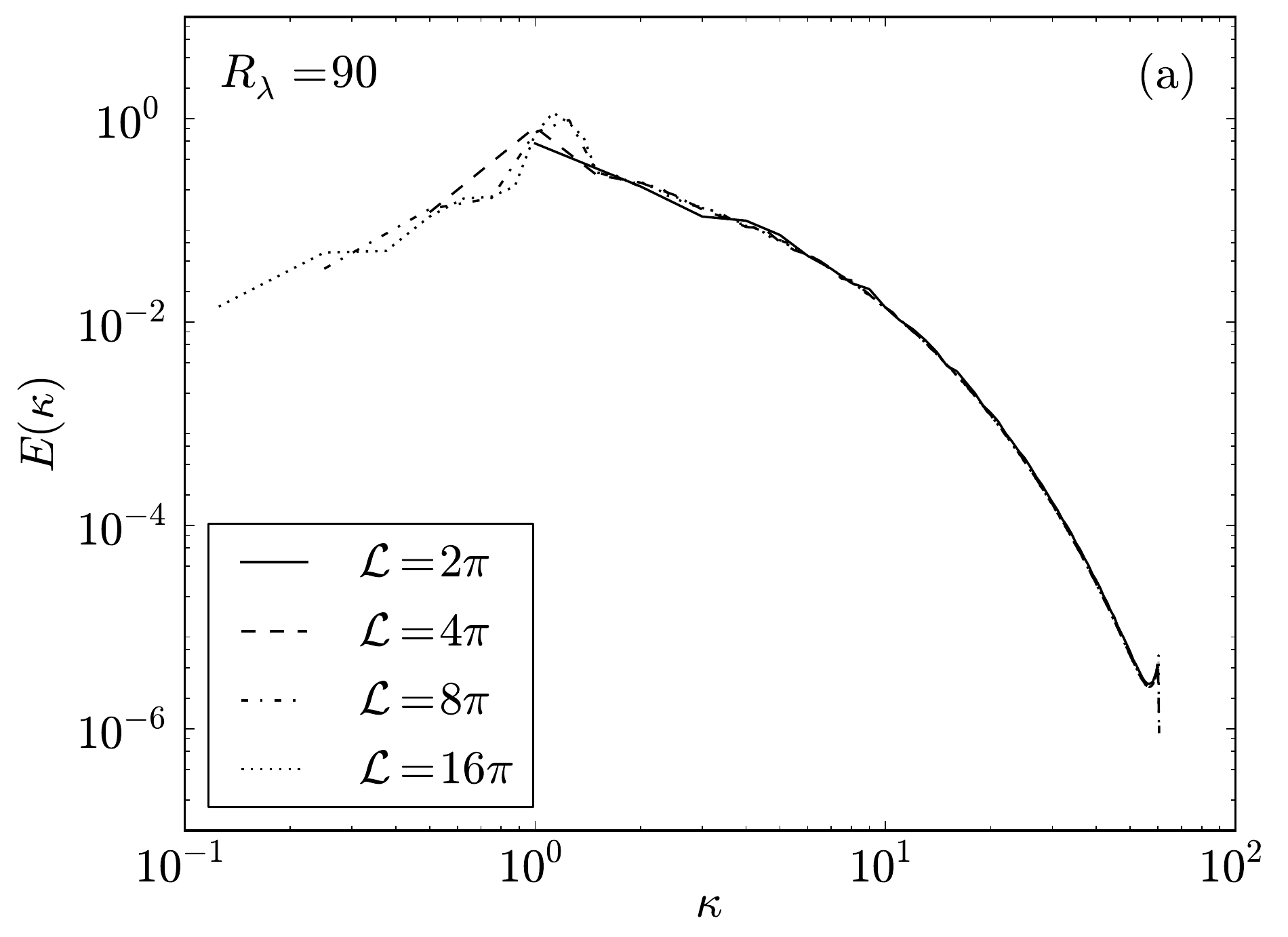}
 \includegraphics[width=2.6in]{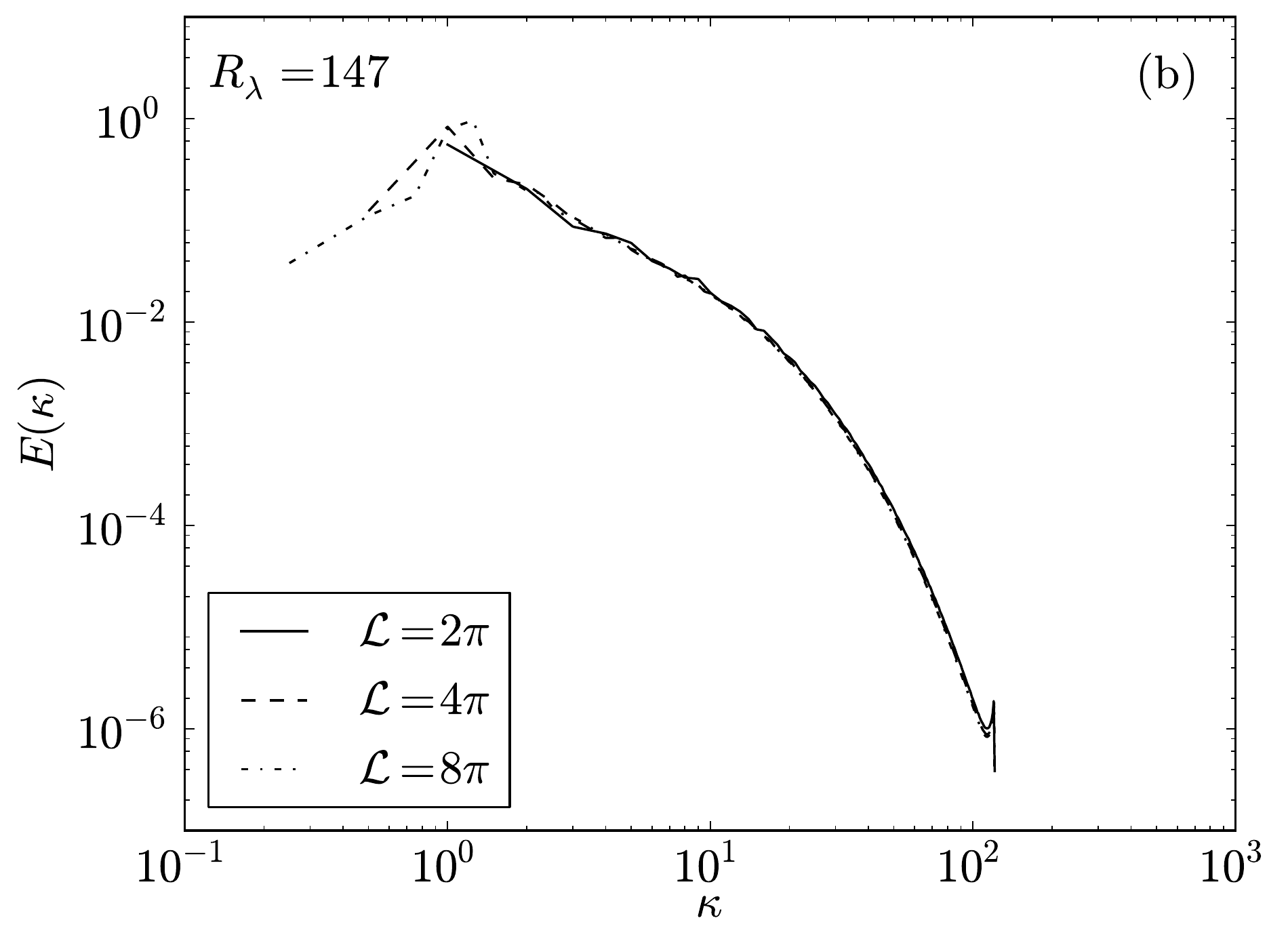}
 \includegraphics[width=2.6in]{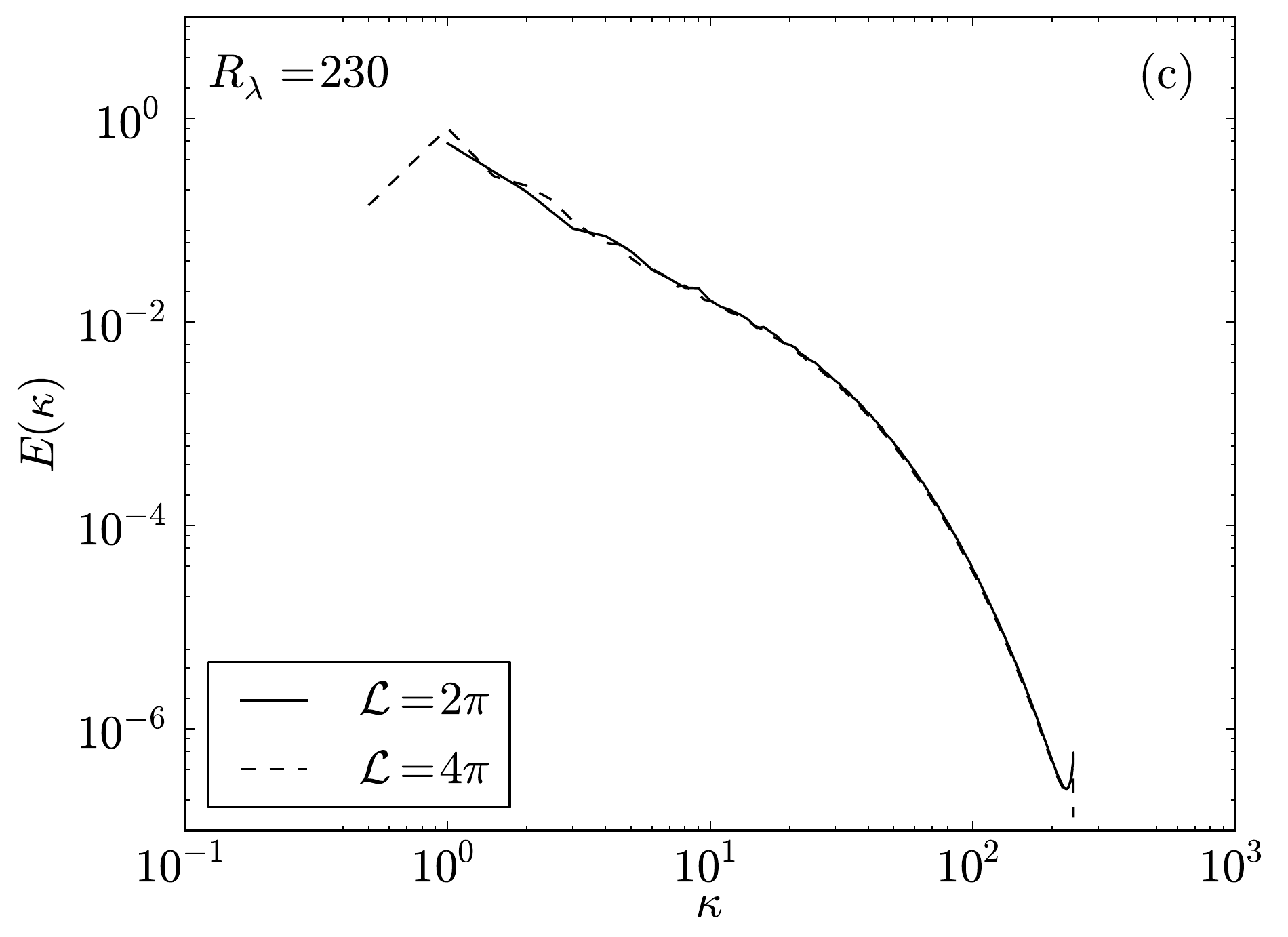}
 \caption{Energy spectra for different domain lengths $\mathcal{L}$ and
 nominal Reynolds numbers $R_\lambda = 90$ (a),
 $R_\lambda=147$ (b), and $R_\lambda=230$ (c). All values are in arbitrary units.}
 \label{fig:spectra_per}
\end{figure}

The resulting energy spectra are shown in figure~\ref{fig:spectra_per}.
In all cases, forcing is applied deterministically to wavenumbers with magnitude $\kappa=\sqrt{2}$,
and thus the location of the peak in the spectra is approximately 
the same for all domain sizes. The spectra are nearly identical to the right of the peak,
indicating that the inertial- and dissipation-range flow scales remain unchanged.
We notice that as the domain size increases, larger-scale (i.e., lower-wavenumber)
flow features become present. These features, however, account for
very little of the overall energy, and thus they do not significantly affect
the fluid statistics. 

\begin{figure}
 \centering
 \includegraphics[width=2.6in]{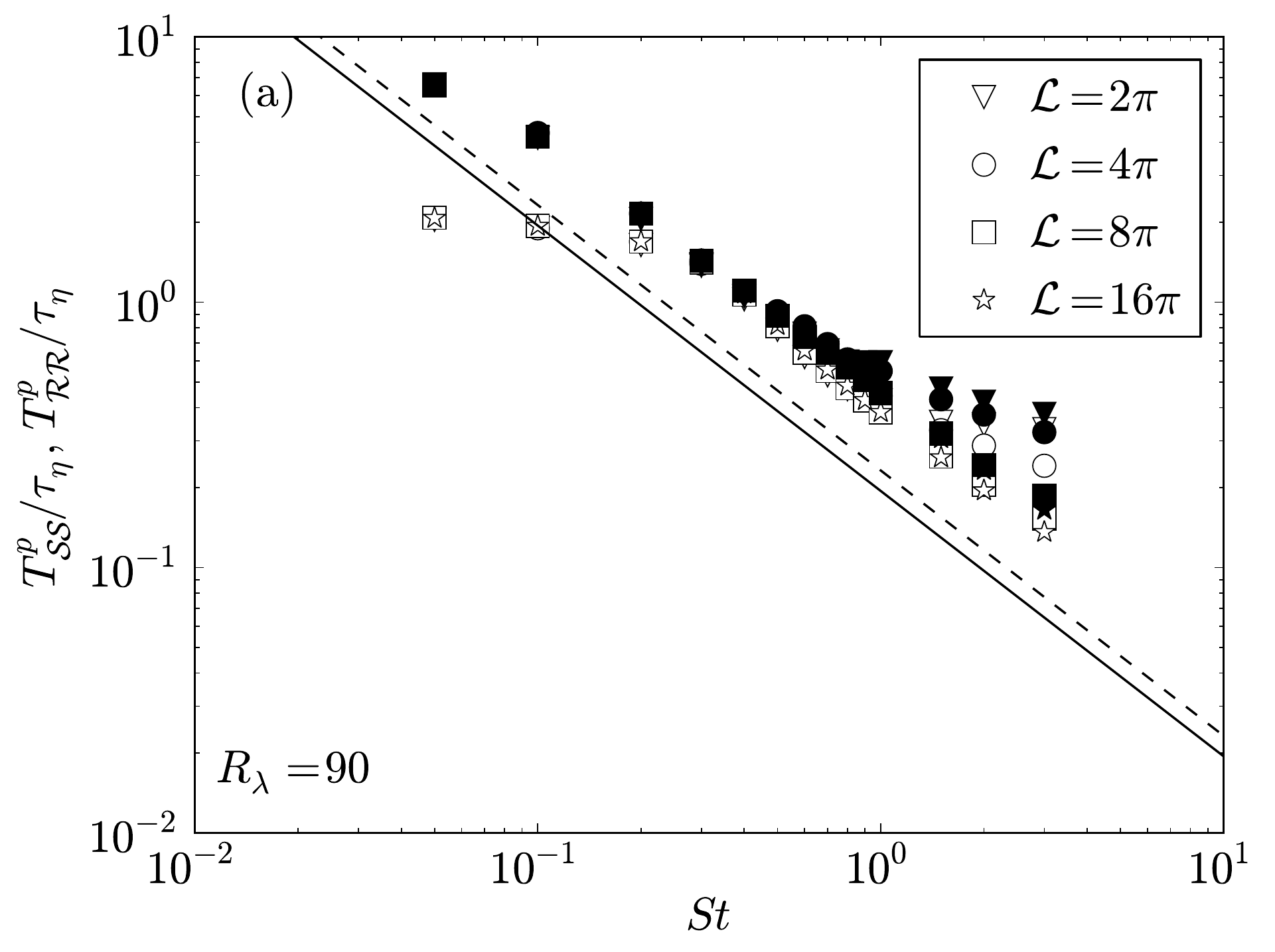}
 \includegraphics[width=2.6in]{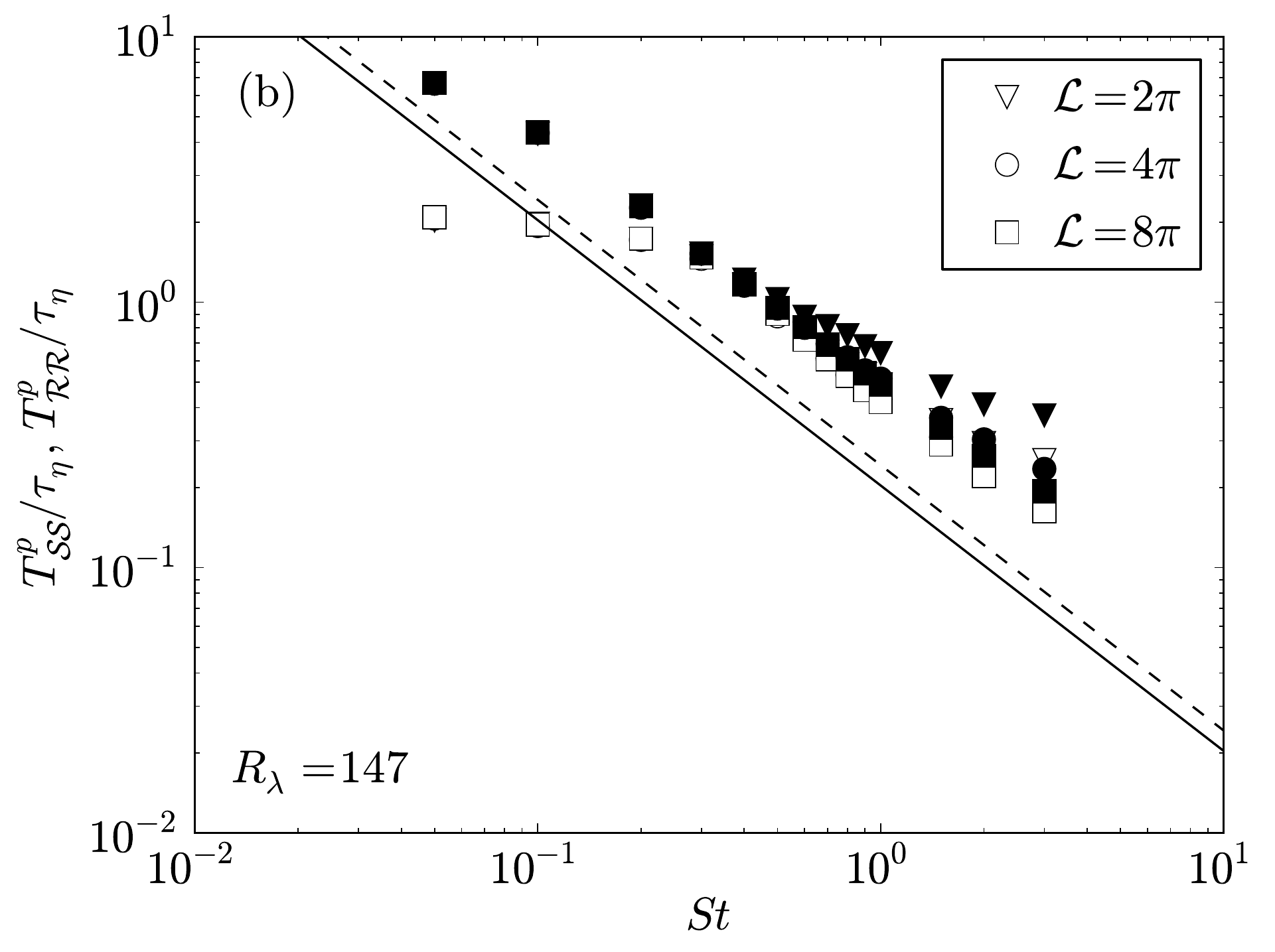}
 \includegraphics[width=2.6in]{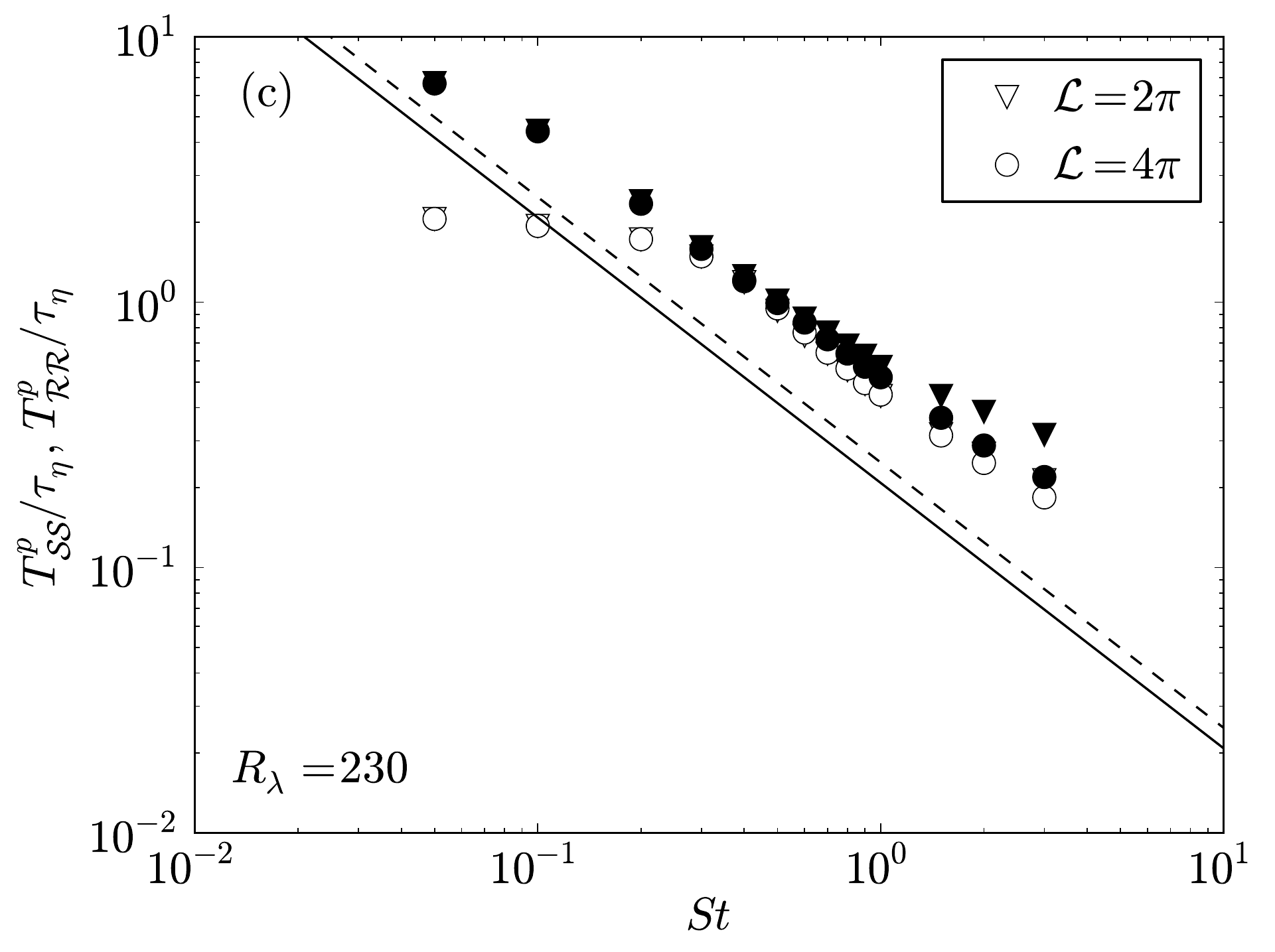}
 \caption{The Lagrangian strain (open symbols) and rotation (filled symbols)
 timescales for different domain lengths $\mathcal{L}$ and
 nominal Reynolds numbers $R_\lambda = 90$ (a),
 $R_\lambda=147$ (b), and $R_\lambda=230$ (c).
 The theoretical predictions for the strain and rotation timescales for $Sv \gg u'/u_\eta$
 are shown with solid and dashed lines, respectively.}
 \label{fig:tauSR_per}
\end{figure}

We first consider the effect of the domain size on the strain and rotation rates
along particle trajectories.
$\langle \mathcal{S}^2 \rangle^p$ and $\langle \mathcal{R}^2 \rangle^p$
(not shown) are generally invariant with changes in the domain size,
since they are small-scale, single-time quantities.
The Lagrangian strain and rotation timescales 
in figure~\ref{fig:tauSR_per}, however, appear to be
over-predicted on the smaller domain sizes at large $St$. The explanation is that these timescales
involve integrals of the autocorrelations of the strain and rotation components
(see Equation~\ref{eq:timescales_strain}), and artificially
high correlations can result when a particle is wrapped around a domain boundary, increasing the timescales.
At the highest values of $St$, our results suggest that the strain and rotation timescales
have not yet converged to a grid-independent result,
even though our scaling argument in (\ref{eq:critical_Sv}) indicates
that the Stokes numbers considered are generally well-below $St_\mathrm{crit}$.
Since these timescales are very small ($\sim 0.1 \tau_\eta$),
their values can be significantly affected by very weak correlations
induced by periodicity, and thus extremely large domain sizes may be necessary for
these artificial correlations to vanish.
We also see that our DNS timescales are considerably larger than the theoretical
predictions from \textsection \ref{sec:high_g_timescales}. While part of this discrepancy
at large $St$ is due to periodicity effects, it is also possible that the theory
in \textsection \ref{sec:high_g_acceleration} requires larger values of $Sv$ than
we are able to simulate here.

We next discuss the effect of periodicity on velocity and acceleration statistics.
The turbulence-induced changes in the 
mean particle settling speeds (not shown)
are independent of the domain size to within statistical noise.
This is presumably because the turbulence-induced settling speed modifications 
of large-$St$ particles 
(which experience the strongest periodicity effects) are negligible, and thus
the domain periodicity will not lead to significant changes in this statistic.
It is possible, however, that the settling speeds with a nonlinear drag model
\citep[which are reduced at large $St$, see][]{good14} will be affected by the domain size.

\begin{figure}
 \centering
 \includegraphics[width=2.6in]{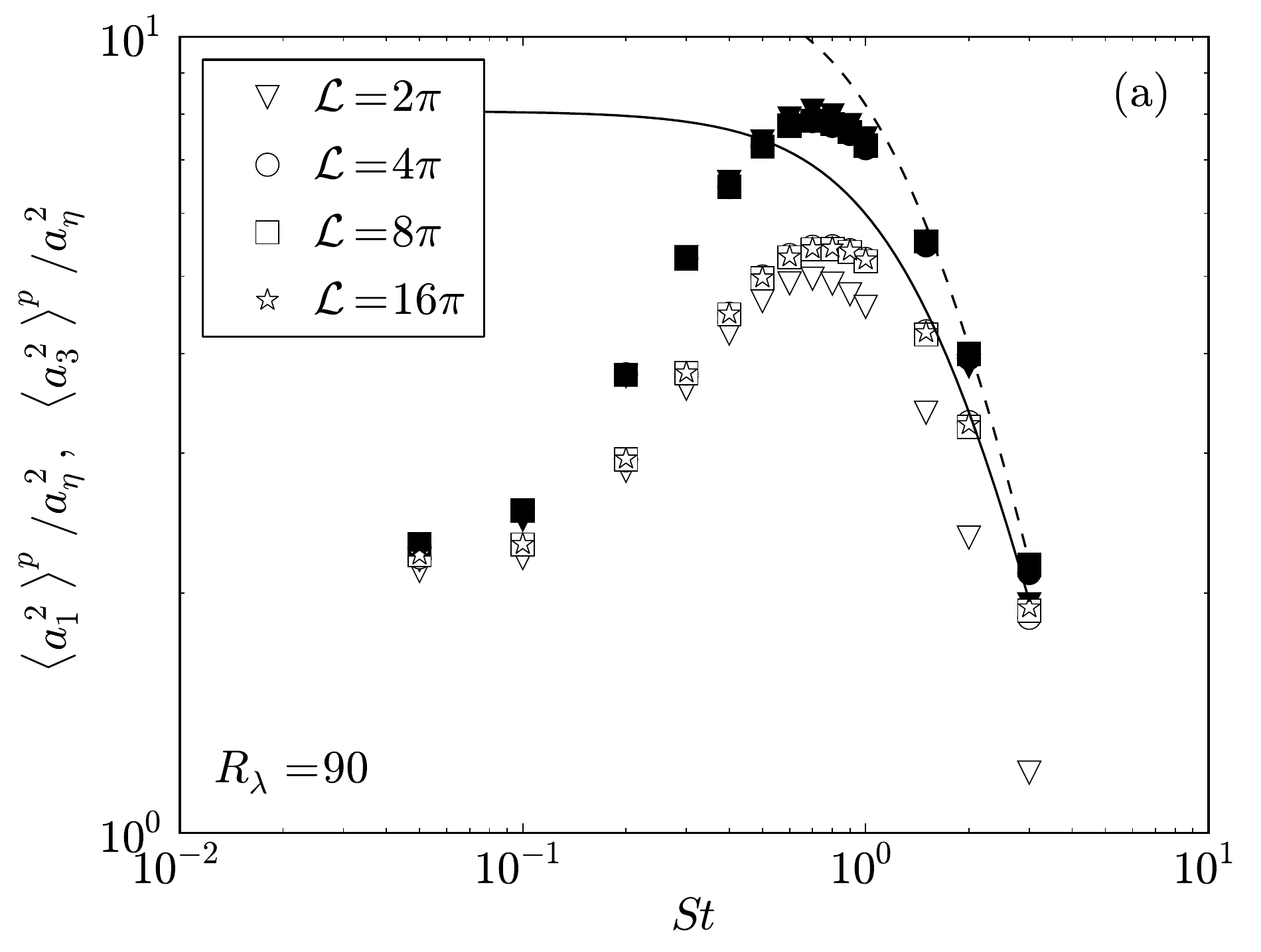}
 \includegraphics[width=2.6in]{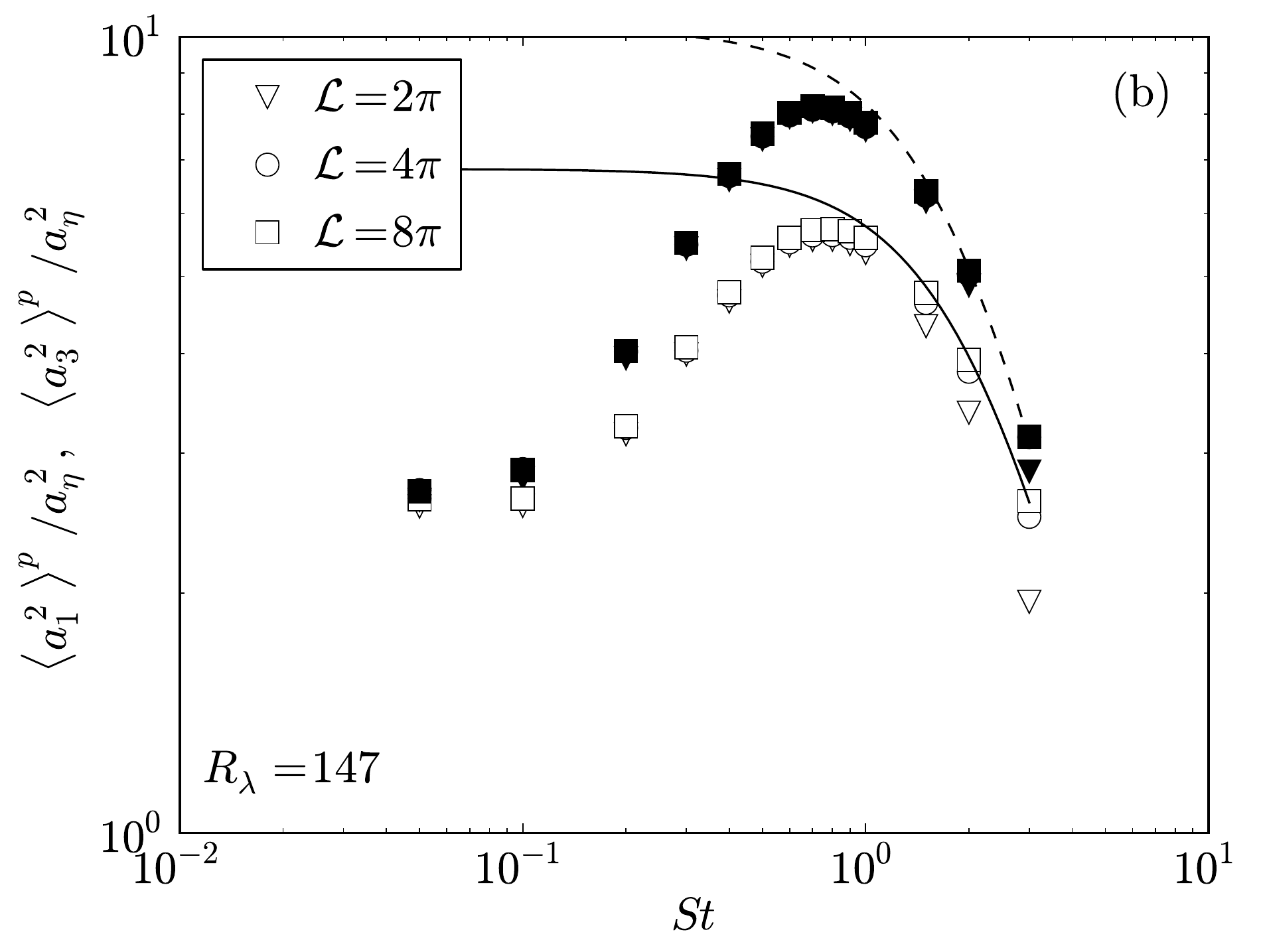}
 \includegraphics[width=2.6in]{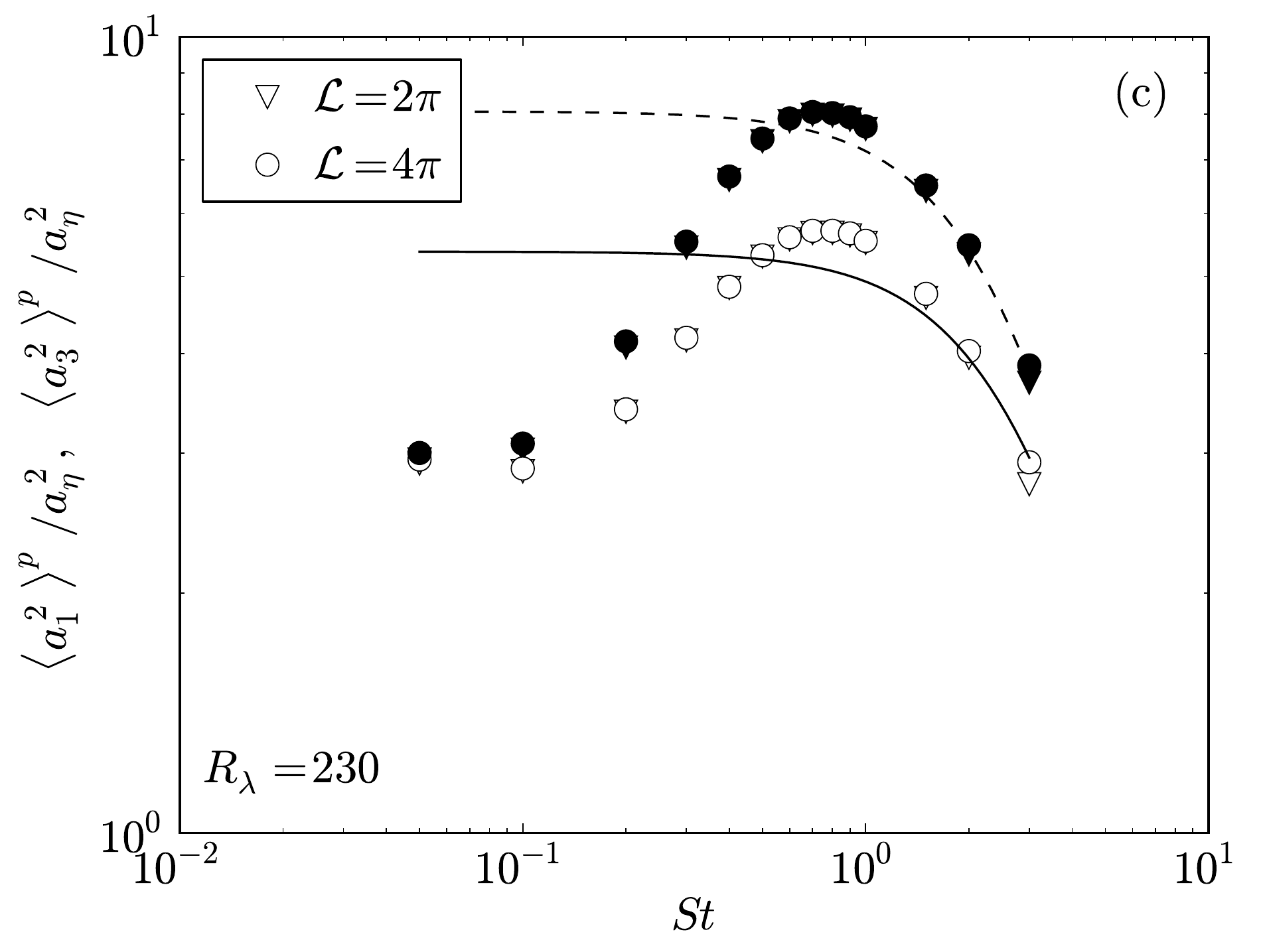}
 \caption{DNS data for $\langle a_1^2 \rangle^p/a_\eta^2$ (filled symbols)
 and $\langle a_3^2 \rangle^p/a_\eta^2$ (open symbols)
 for different domain lengths $\mathcal{L}$ and
 nominal Reynolds numbers $R_\lambda = 90$ (a), $R_\lambda=147$ (b), and $R_\lambda=230$ (c).
 The theoretical predictions for the horizontal (Equation~\ref{eq:accel_variance_perp})
 and vertical (Equation~\ref{eq:accel_variance_parallel}) acceleration variances for $Sv \gg u'/u_\eta$ 
 are shown with dashed and solid lines, respectively.}
 \label{fig:acceleration_variances_per}
\end{figure}

We show acceleration variance statistics in figure~\ref{fig:acceleration_variances_per}.
The acceleration variances
are under-predicted at the smallest domain sizes,
and converge to the theoretical predictions at high $St$ (see \textsection \ref{sec:high_g_acceleration})
as the domain size is increased. 
The vertical acceleration variances are more affected by the periodic
boundary conditions than the horizontal acceleration variances, since the vertical velocities
are correlated over longer lengthscales (as explained in \textsection \ref{sec:accelerations}).

\begin{figure}
 \centering
 \includegraphics[width=2.6in]{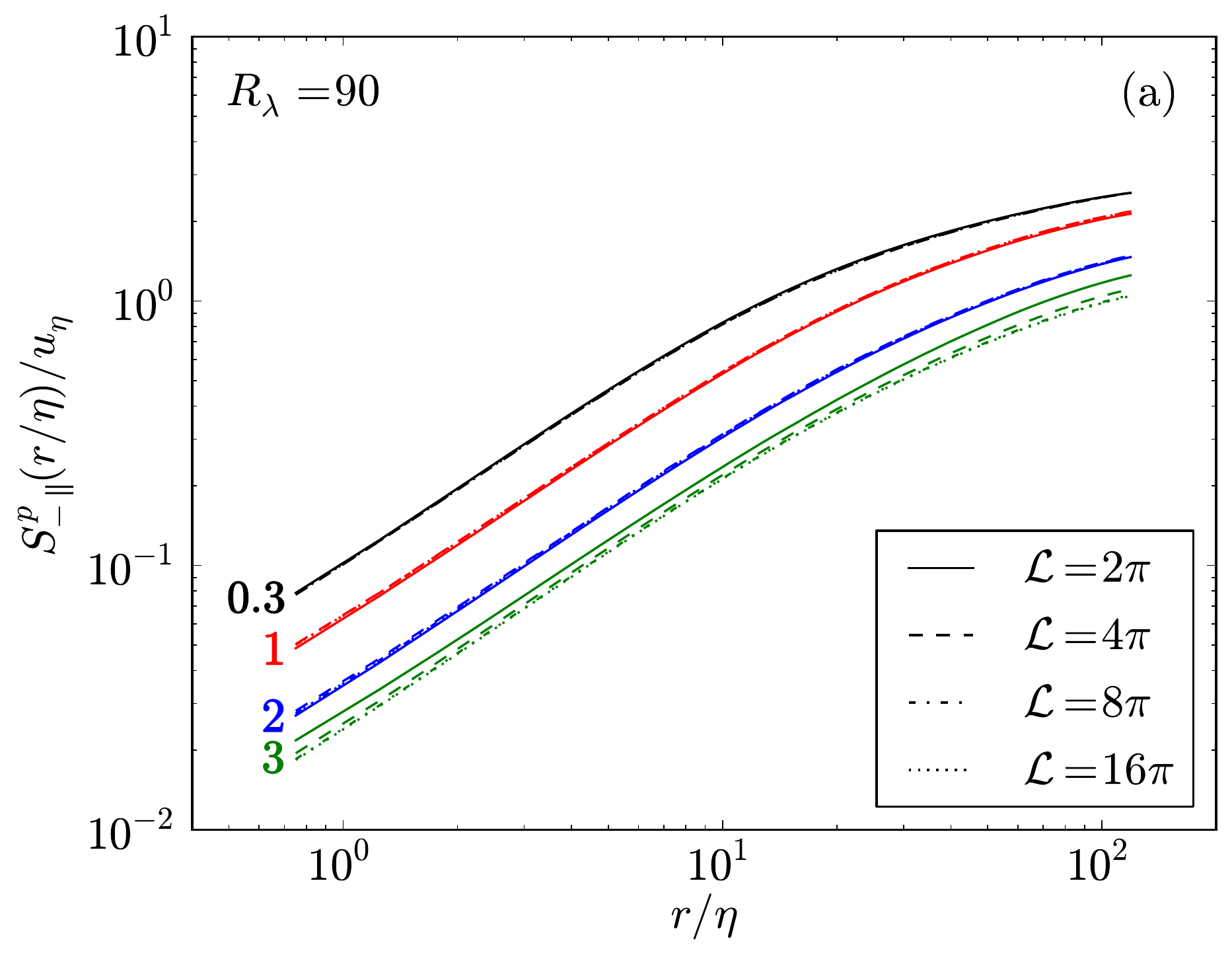}
 \includegraphics[width=2.6in]{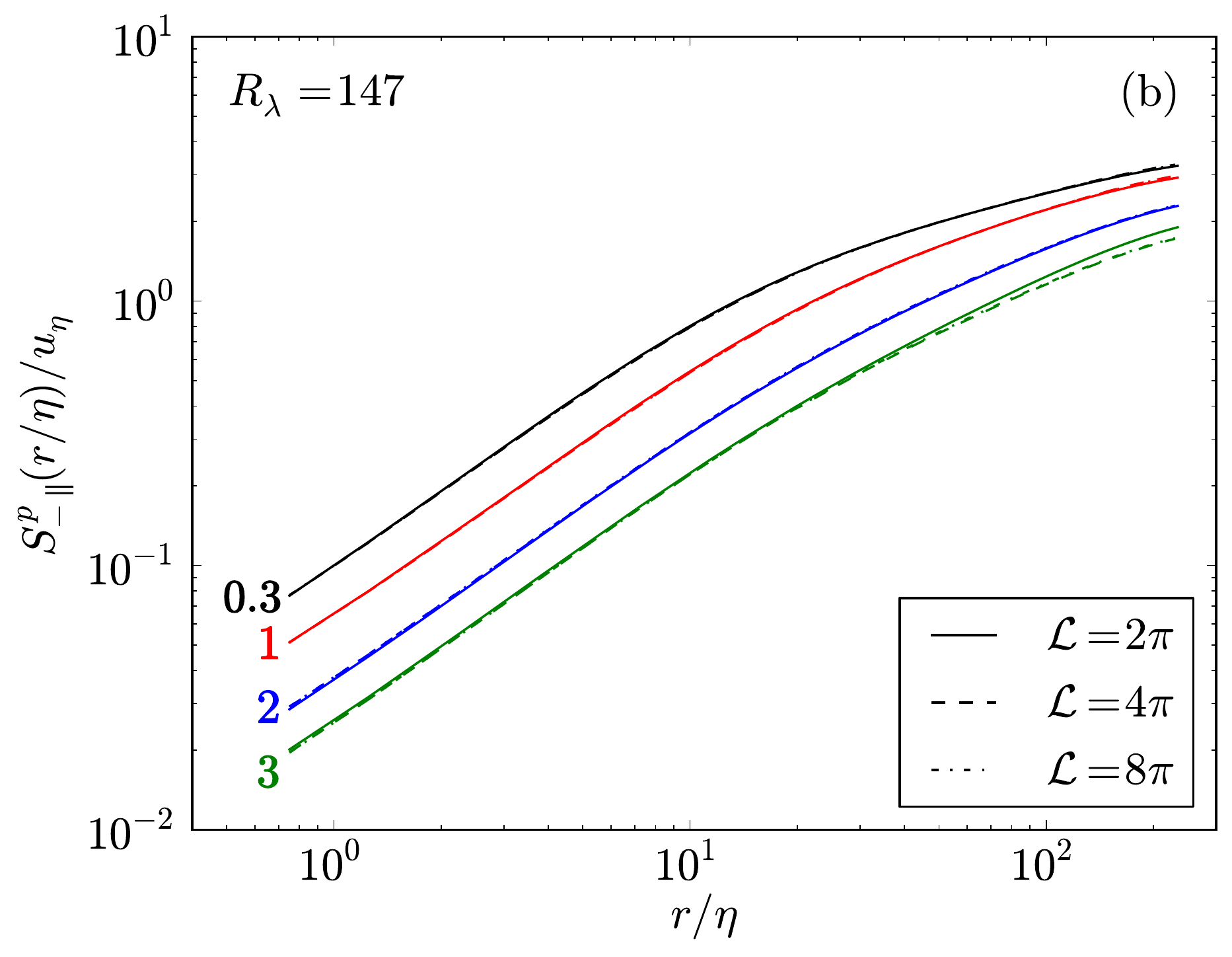}
 \includegraphics[width=2.6in]{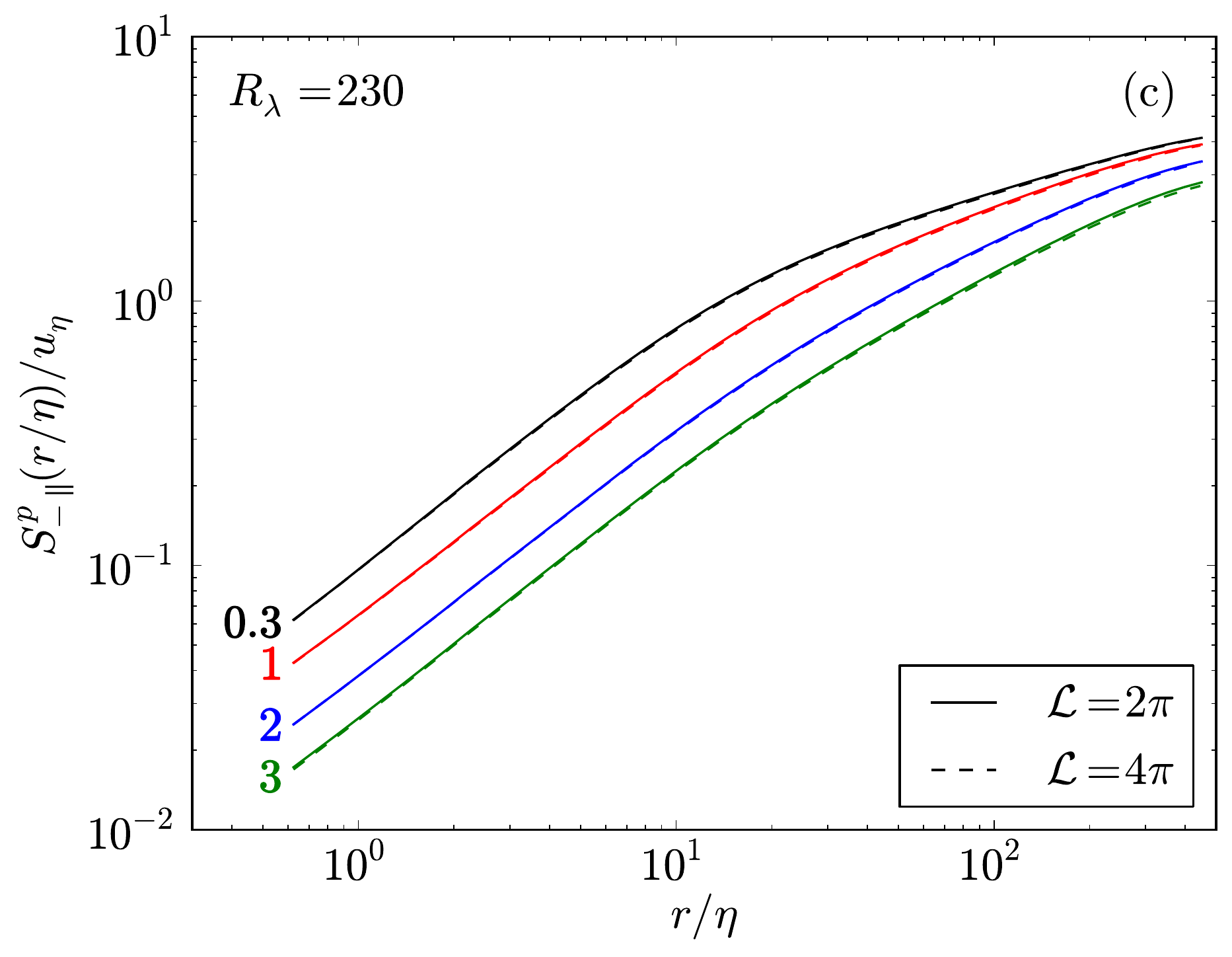}
 \caption{The longitudinal mean inward relative velocities (normalized by $u_\eta$)
 for different domain lengths $\mathcal{L}$ for
 nominal Reynolds numbers $R_\lambda = 90$ (a), $R_\lambda=147$ (b), and $R_\lambda=230$ (c).
 The different Stokes numbers considered ($St=0.3,\ 1,\ 2,\ 3$) are shown in black, red, blue, and green,
 respectively, and the Stokes numbers are indicated by the line labels.}
 \label{fig:wrel_mean_per}
\end{figure}

\begin{figure}
 \centering
 \includegraphics[width=2.6in]{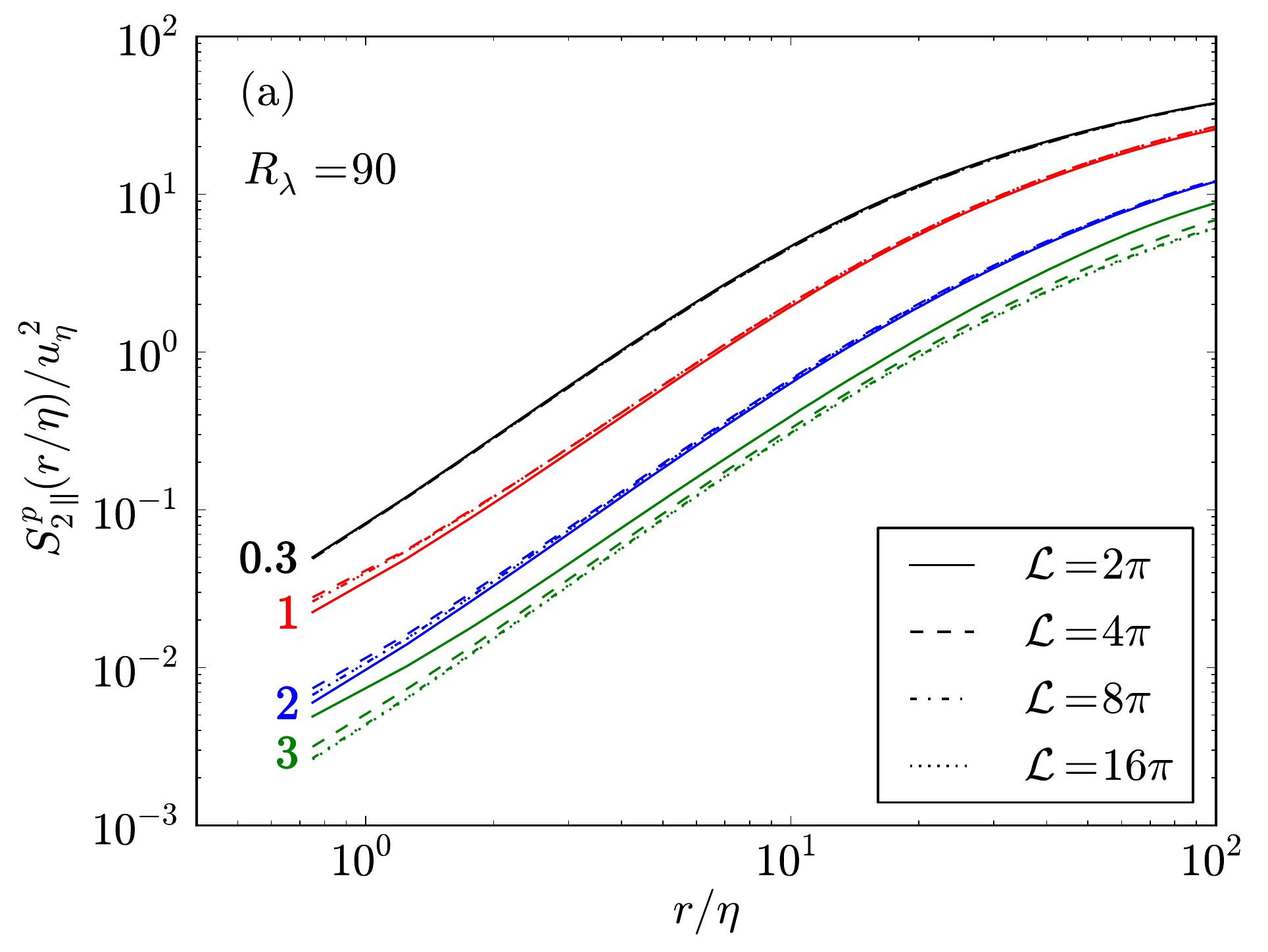}
 \includegraphics[width=2.6in]{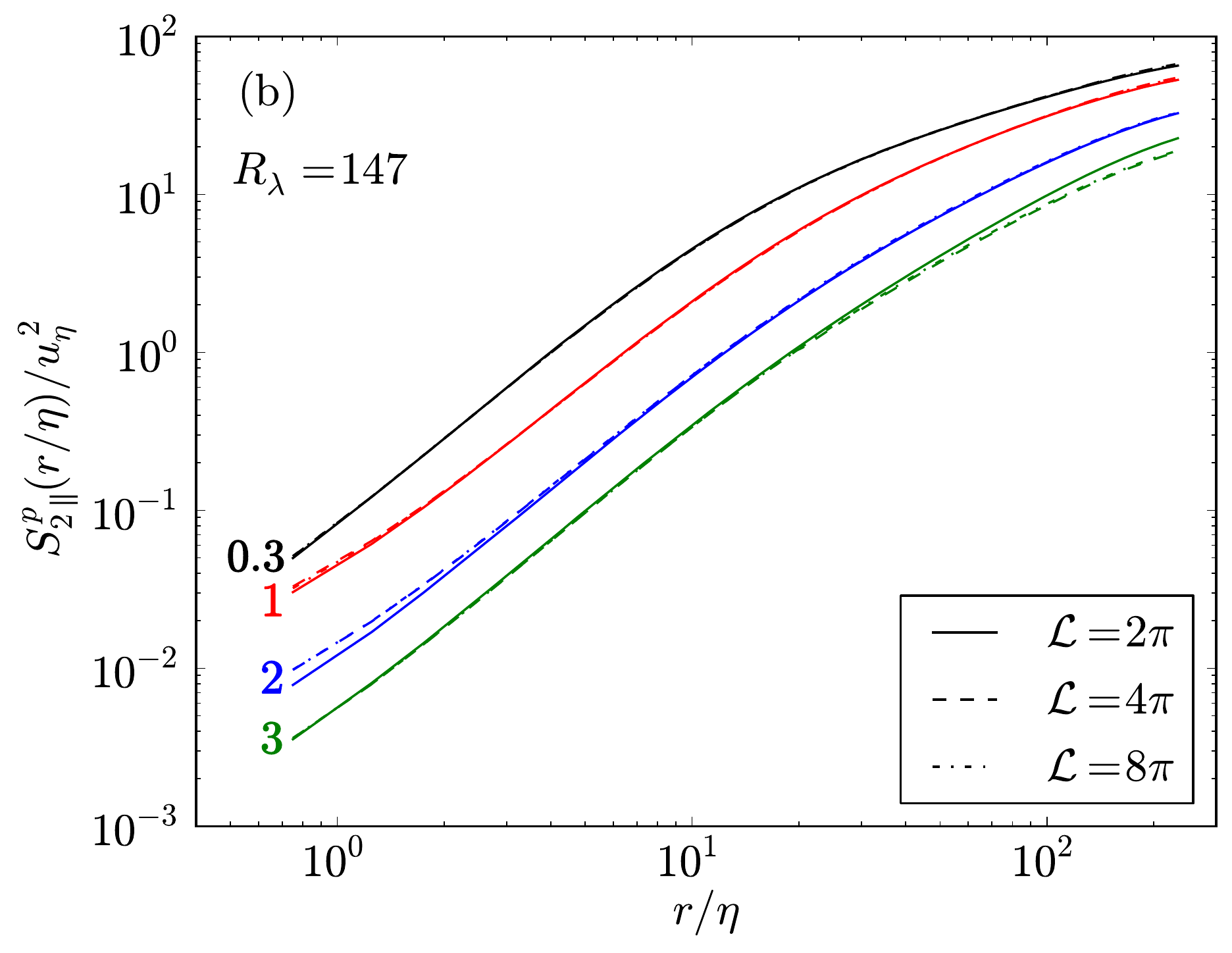}
 \includegraphics[width=2.6in]{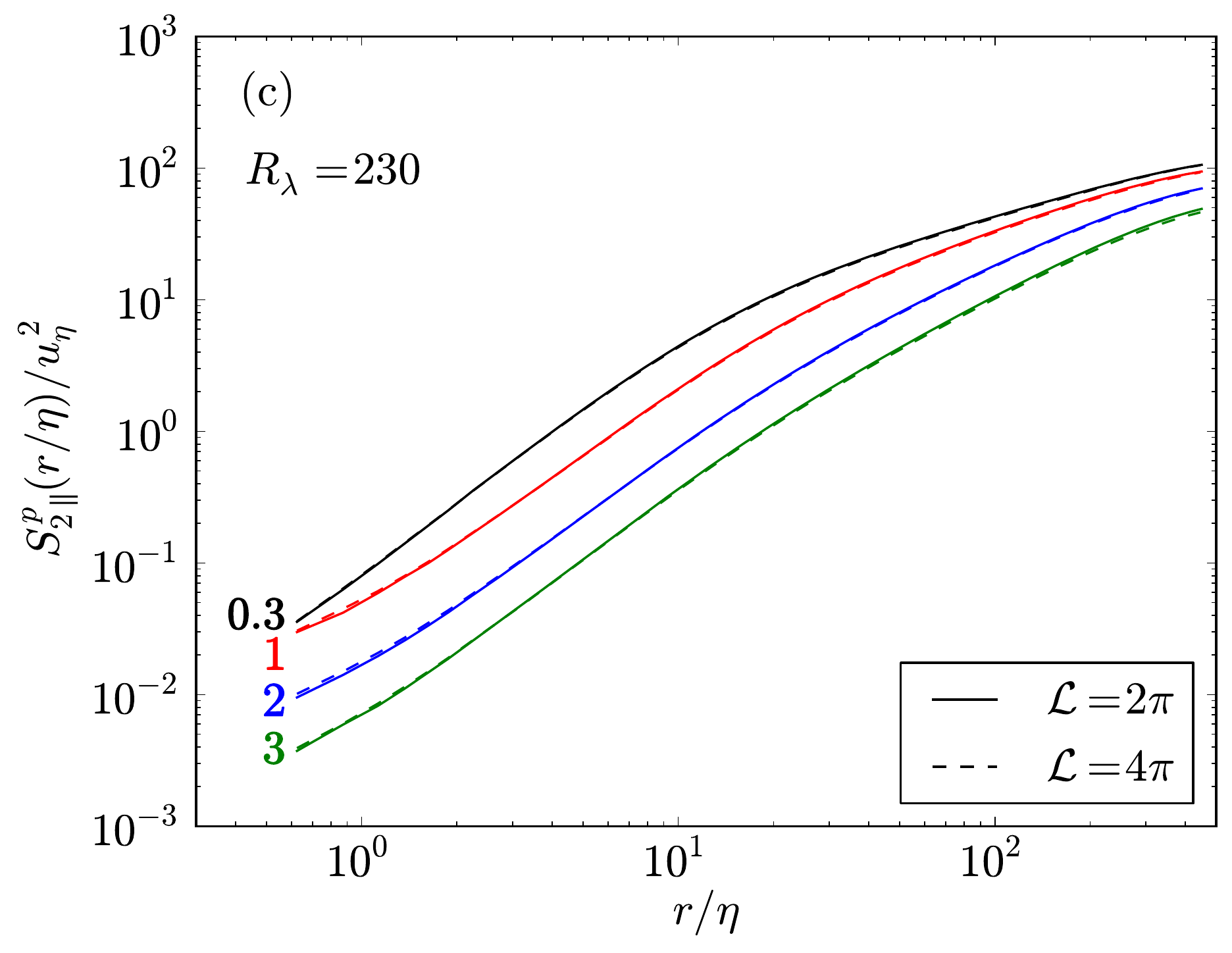}
 \caption{The longitudinal relative velocity variances (normalized by $u_\eta^2$)
 for different domain lengths $\mathcal{L}$ for
 nominal Reynolds numbers $R_\lambda = 90$ (a), $R_\lambda=147$ (b), and $R_\lambda=230$ (c).
 The different Stokes numbers considered ($St=0.3,\ 1,\ 2,\ 3$) are shown in black, red, blue, and green,
 respectively, and the Stokes numbers are indicated by the line labels.}
 \label{fig:wr2_per}
\end{figure}

We conclude this section by examining the relative velocity statistics and the RDFs.
Figure~\ref{fig:wrel_mean_per} and figure~\ref{fig:wr2_per} show the 
longitudinal mean inward relative
velocities and relative velocity variances, respectively. These statistics
are only weakly sensitive to the domain size for all $St$ and converge
to a grid-independent result. The relative velocity variances appear to be slightly
more sensitive to the domain size than the mean inward relative velocities,
presumably because the former are higher-order statistics and are thus sensitive
to larger and more intermittent flow features.

\begin{figure}
 \centering
 \includegraphics[width=2.6in]{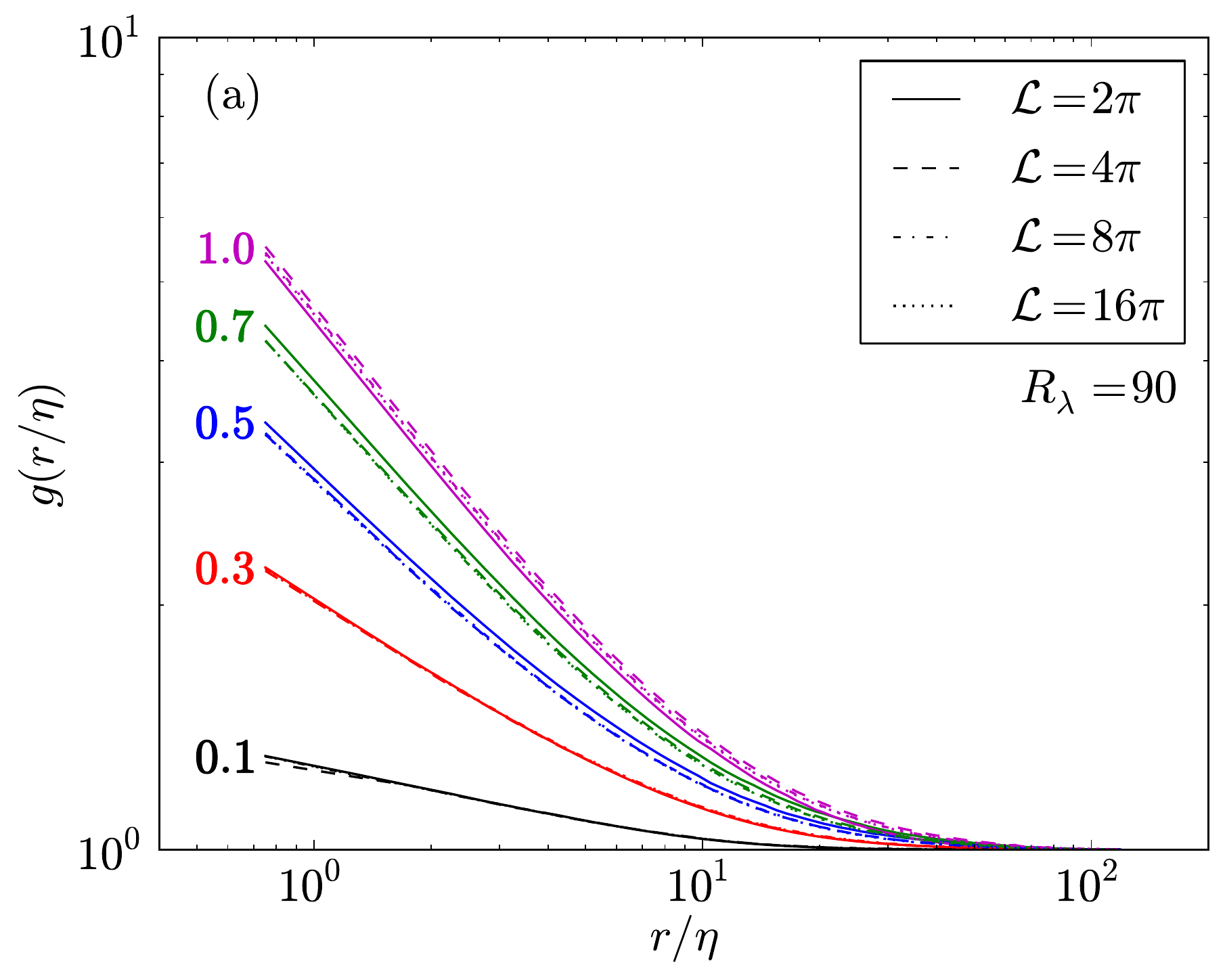}
 \includegraphics[width=2.6in]{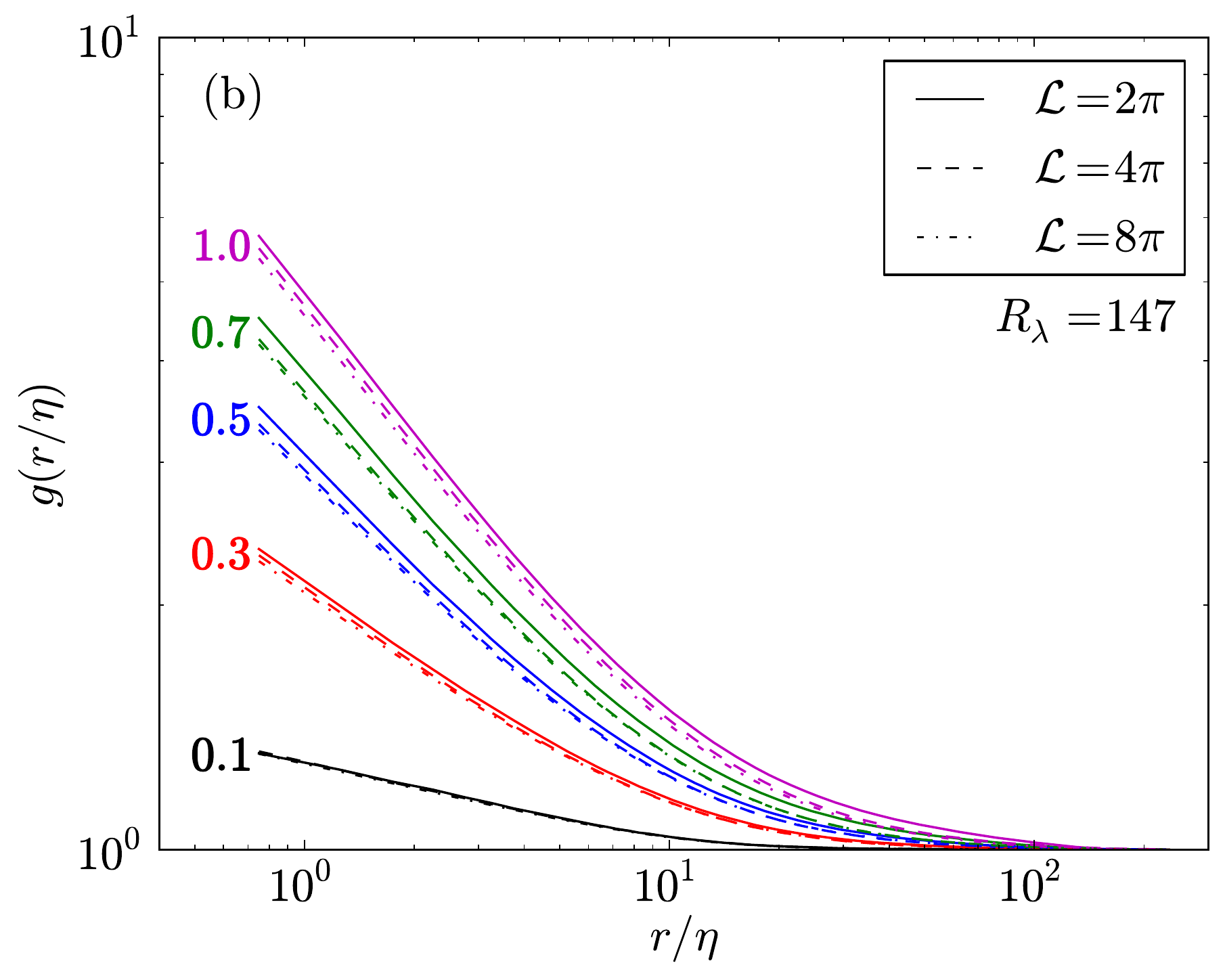}
 \includegraphics[width=2.6in]{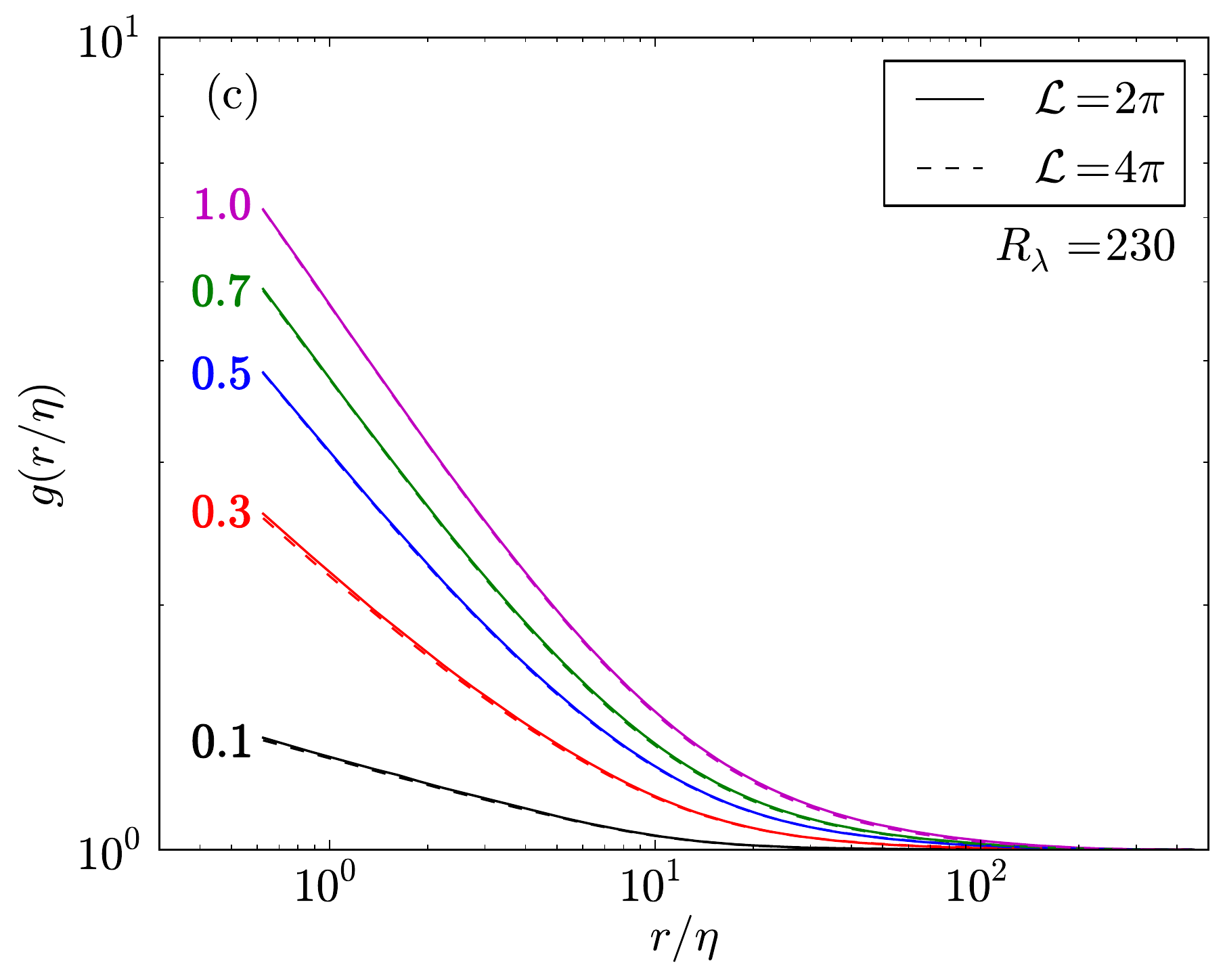}
 \caption{RDFs for different domain lengths $\mathcal{L}$ for
 nominal Reynolds numbers $R_\lambda = 90$ (a), $R_\lambda=147$ (b), and $R_\lambda=230$ (c).
 Data are shown for $St \leq 1$, with the Stokes numbers indicated by the line labels.
 The different Stokes numbers considered ($St=0.1,\ 0.3,\ 0.5,\ 0.7,\ 1$) are shown in black, red, blue, green,
 and magenta, respectively, and the Stokes numbers are indicated by the line labels.}
 \label{fig:gofr_lowSt_per}
\end{figure}

\begin{figure}
 \centering
 \includegraphics[width=2.6in]{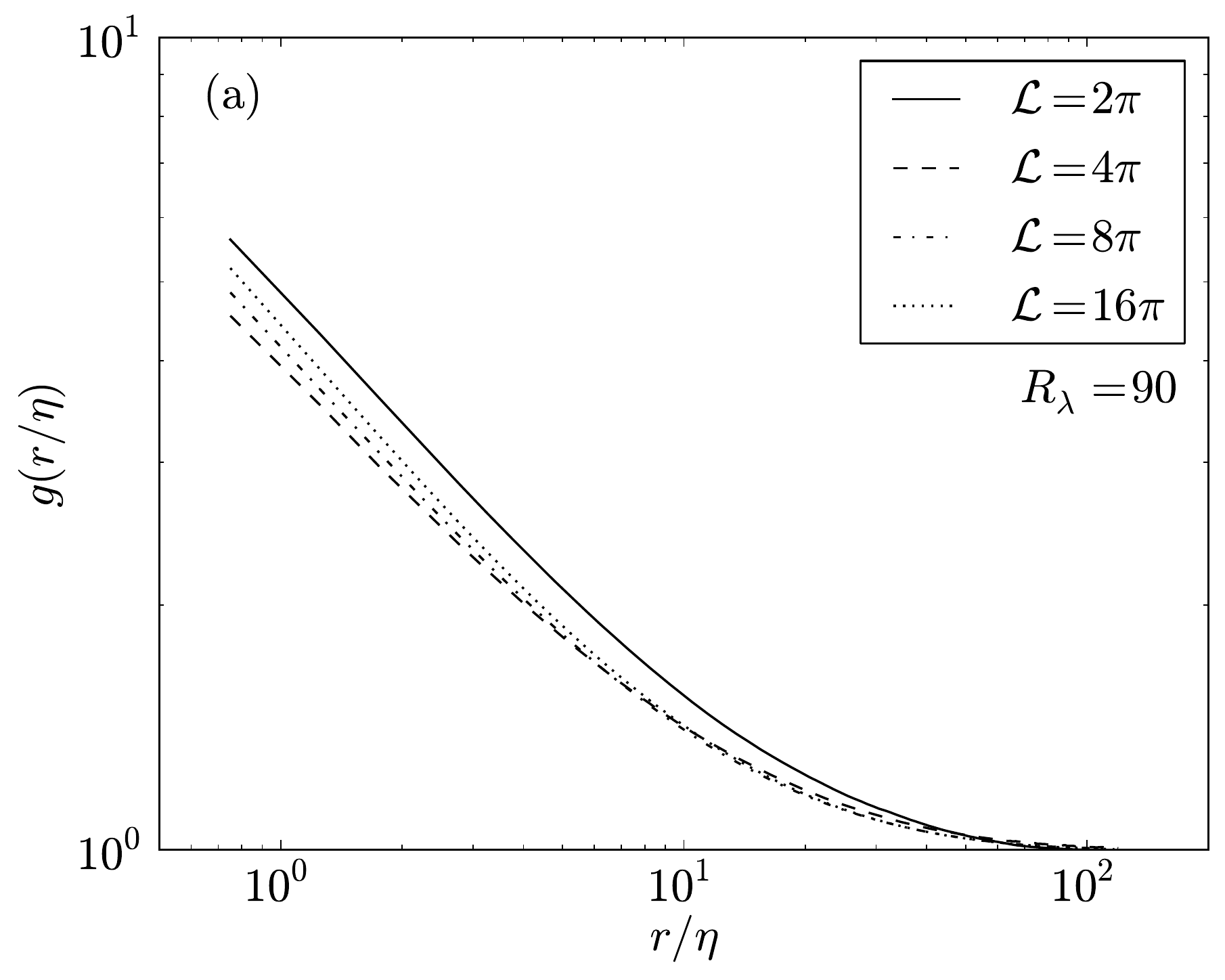}
 \includegraphics[width=2.6in]{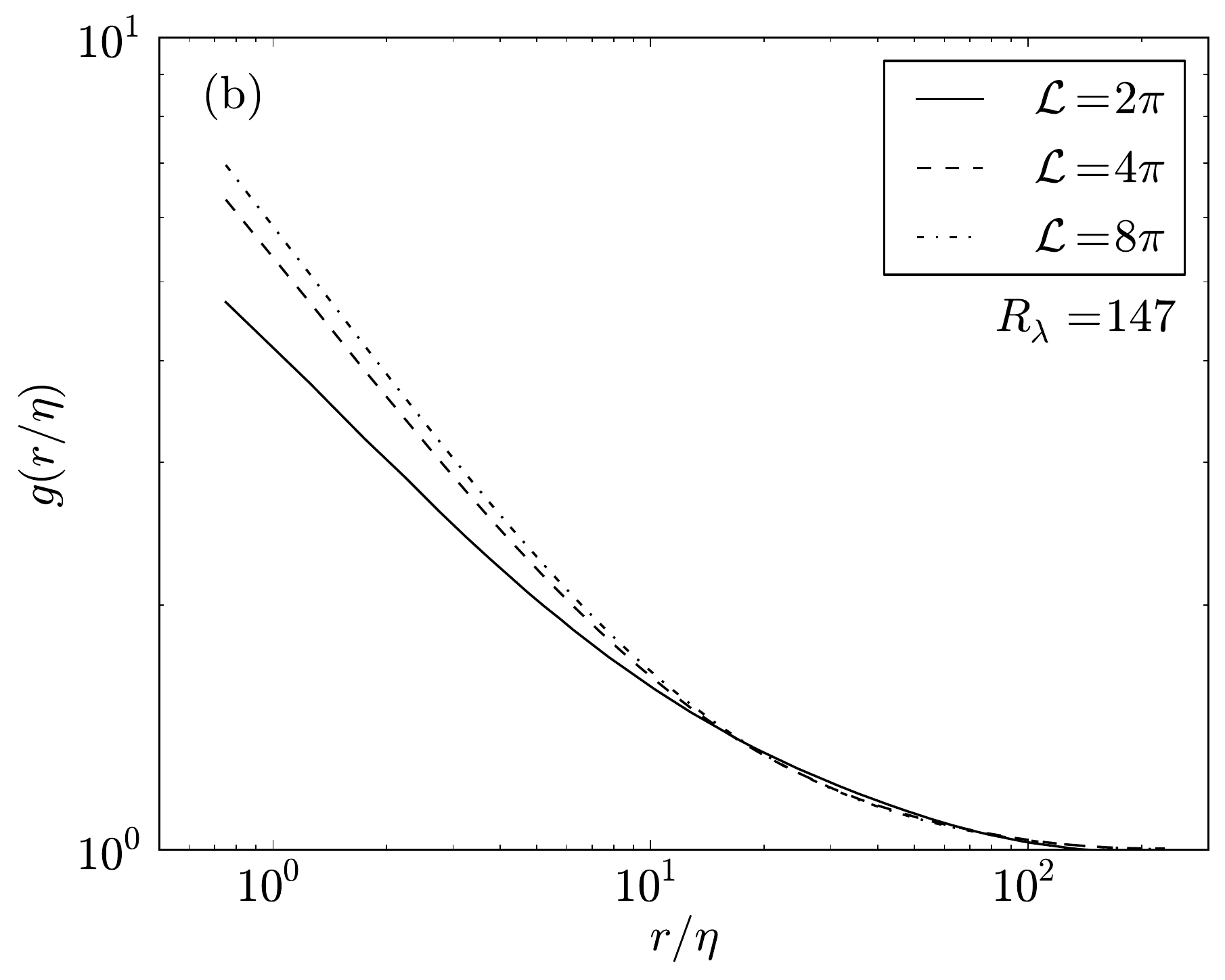}
 \includegraphics[width=2.6in]{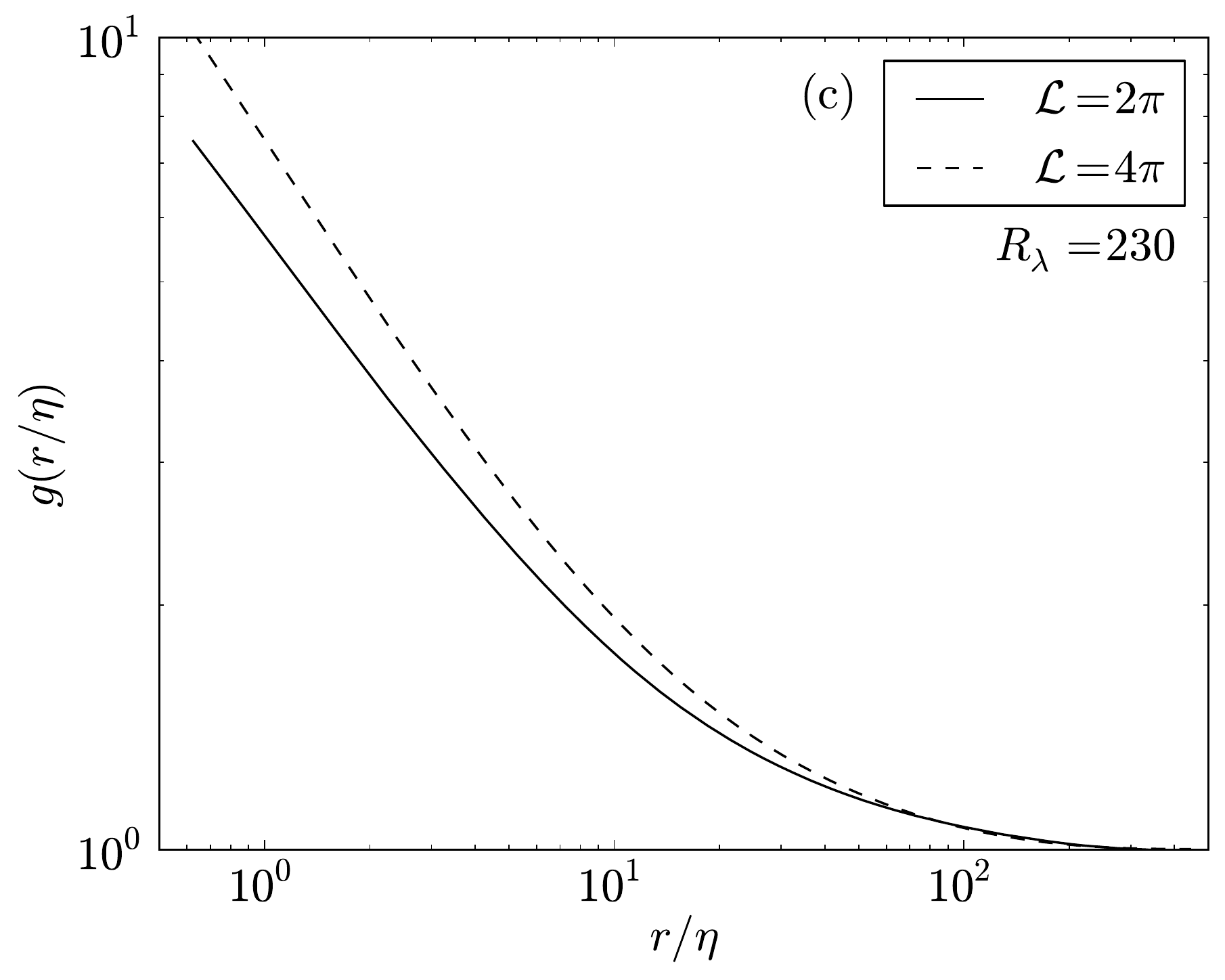}
 \caption{RDFs for different domain lengths $\mathcal{L}$ for
 nominal Reynolds numbers $R_\lambda = 90$ (a), $R_\lambda=147$ (b), and $R_\lambda=230$ (c).
 Data are shown for $St = 3$.}
 \label{fig:gofr_highSt_per}
\end{figure}

Figure~\ref{fig:gofr_lowSt_per} indicates that the RDFs
are almost entirely unaffected by the finite domain sizes at low $St$.
At $St=3$ (figure~\ref{fig:gofr_highSt_per}), however, 
the RDFs are evidently quite sensitive to the domain size,
and do not converge to a grid-independent result. It is possible
that these RDFs are influenced by the small fraction of larger-scale features which appear
as grid size is increased (see figure~\ref{fig:spectra_per}),
or that the scaling argument in (\ref{eq:critical_Sv}) is not sufficiently 
stringent for the RDF statistics. In any case, the results at large $St$ should
be interpreted with caution, since the linear drag model and the point-particle approximation
are less accurate here (see \textsection \ref{sec:particle_phase}).

\setcounter{equation}{0}
\section{Theoretical models in the strong-gravity limit}
\label{sec:high_g_models}

In this section, we derive theoretical models which are expected to be 
valid in the limit of strong gravitational forces.
We require two conditions for these models to hold.
First, particles must fall so quickly that the spatial decorrelation of the turbulence along their trajectories
occurs much faster than the 
Eulerian timescale of the flow.
Second, a particle's displacement over a time $s$ is to leading order equal to $\tau_p \mathfrak{g} s$.

The conditions can be shown to be satisfied when $Sv \gg u'/u_\eta$.
It is important to note that $u'/u_\eta \sim R_\lambda^{1/2}$ 
in isotropic turbulence, which implies that our models will
only hold for very large droplets (i.e., very large values of $Sv$) 
in high-Reynolds-number atmospheric clouds. 
In table~\ref{tab:cloud_parameters}, we provide an estimate of the droplet
diameters $d$ at which our model is expected to be valid
for three different cloud types:  stratiform clouds ($\epsilon \sim 10^{-3}$ m$^2$/s$^3$),
cumulus clouds ($\epsilon \sim 10^{-2}$ m$^2$/s$^3$), and cumulonimbus clouds
($\epsilon \sim 10^{-1}$ m$^2$/s$^3$).

\begin{table}
\centering
 \caption{Estimated atmospheric-cloud conditions at which our strong-gravity timescale model is expected to hold.}
 \label{tab:cloud_parameters}
 \begin{tabular}{llll}
  Cloud type & Stratiform & Cumulus & Cumulonimbus \\
  $\epsilon$ (m$^2$/s$^3$) & $10^{-3}$ & $10^{-2}$ & $10^{-1}$ \\
  $R_\lambda$ & 10,800 & 15,900 & 23,300 \\
  $u'/u_\eta$ & 52.8 & 64.0 & 77.6 \\
  $d$ ($\mu m$) & $\gg 127$ & $\gg 186$ & $\gg 273$ \\
 \end{tabular}
\end{table}

From table~\ref{tab:cloud_parameters}, we see that 
the droplet diameters at which these models are expected to be valid 
(on the order of hundreds to thousands of microns) are well above
those of droplets in the `size-gap' in atmospheric clouds (between about 30 and 80 microns--see Part I),
where turbulence is expected to play a key role in droplet growth.
However, while the models are not expected to yield accurate predictions of droplet
dynamics in the size-gap, they provide an important asymptotic limit in which to understand heavy-particle motion. 
Furthermore, since the Reynolds numbers in our simulations
are well below those in atmospheric clouds, we expect the models to agree well with our simulation
data at the largest values of $Sv$.

In \textsection \ref{sec:high_g_timescales}, we develop a model for
the Lagrangian strain and rotation timescales of the flow seen by inertial particles,
and in \textsection \ref{sec:high_g_acceleration}, we model the particle accelerations.

\subsection{Lagrangian strain and rotation timescales along particle trajectories}
\label{sec:high_g_timescales}

In the limit $Sv \gg u'/u_\eta$, we can model the Lagrangian strain timescale $T^p_{\mathcal{S}_{ij} \mathcal{S}_{km}}$ as
\begin{equation}
\label{eq:timescale_model}
 T^p_{\mathcal{S}_{ij} \mathcal{S}_{km}} = 
 \frac{1}{\Big \langle \mathcal{S}_{ij} (\bm{0},0) \mathcal{S}_{km} (\bm{0},0) \Big \rangle}
 \int\limits_0^{\infty} \int\limits_{\bm{X}}
 \Big \langle \mathcal{S}_{ij} (\bm{X},s) \mathcal{S}_{km} (\bm{0},0) \Big \rangle
 \varrho(\bm{X},s) \, d \bm{X} ds \mathrm{,}
\end{equation}
where $\varrho(\bm{X},s) \equiv \delta(\bm{X}-\tau_p \boldsymbol{\mathfrak{g}} s)$.
Here and through the rest of this section, no summation is implied by repeated indices in
the timescale expressions.

Note that (\ref{eq:timescale_model}) is constructed by assuming that the 
turbulence evaluated along the particle trajectories is uncorrelated with the trajectory itself
\citep[i.e., `Corrsin's hypothesis,' see][]{corrsin63}.
In the limit $Sv \gg u'/u_\eta$, Corrsin's hypothesis is expected to become exact, since
the particle motions are almost independent of the underlying flow field here.

Taking the integral of (\ref{eq:timescale_model}), we obtain
\begin{equation}
\label{eq:timescale_model_simplified}
 T^p_{\mathcal{S}_{ij} \mathcal{S}_{km}} = 
 \frac{1}{\Big \langle \mathcal{S}_{ij} (\bm{0},0) \mathcal{S}_{km} (\bm{0},0) \Big \rangle}
 \int\limits_0^\infty  \Big 
 \langle \mathcal{S}_{ij} (\tau_p \boldsymbol{\mathfrak{g}}s,0) \mathcal{S}_{km} (\bm{0},0) \Big \rangle
 \, ds \mathrm{.}
\end{equation}

To derive an analytical expression for the timescales, 
we assume that 
$\langle \mathcal{S}_{ij} (\tau_p \boldsymbol{\mathfrak{g}}s,0) \mathcal{S}_{km} (\bm{0},0) \rangle$
decorrelates exponentially, which gives us
\begin{equation}
\label{eq:timescale_model_simplified_exp}
 T^p_{\mathcal{S}_{ij} \mathcal{S}_{km}} = {\ell_{\mathcal{S}_{ij} \mathcal{S}_{km},3}}
                                                     /\left( \tau_p \mathfrak{g} \right) \mathrm{.}
\end{equation}
$\ell_{\mathcal{S}_{ij} \mathcal{S}_{km},3}$ is the integral lengthscale
of $\mathcal{S}_{ij} \mathcal{S}_{km}$ evaluated along the $x_3$-direction.
We express (\ref{eq:timescale_model_simplified_exp}) in non-dimensional form,
\begin{equation}
\label{eq:timescale_model_simplified_nondimensional}
 \hat{T}^p_{\mathcal{S}_{ij} \mathcal{S}_{km}}
 = \frac{\hat{\ell}_{\mathcal{S}_{ij} \mathcal{S}_{km},3}}{Sv} \mathrm{,}
\end{equation}
where we have used the top-hat symbol to denote a variable normalized by Kolmogorov scales.
The rotation timescales $T^p_{\mathcal{R}_{ij} \mathcal{R}_{km}}$ are defined analogously,
\begin{equation}
\label{eq:timescale_model_simplified_nondimensional_rotation}
\hat{T}^p_{\mathcal{R}_{ij} \mathcal{R}_{km}}
 = \frac{\hat{\ell}_{\mathcal{R}_{ij} \mathcal{R}_{km},3}}{Sv} \mathrm{.}
\end{equation}

\subsection{Inertial particle acceleration variances}
\label{sec:high_g_acceleration}
We now develop a model for the particle acceleration variances in the limit $Sv \gg u'/u_\eta$.
Following \cite{bec06a}, we write the formal solution for the particle acceleration as
\begin{equation}
 \bm{a}^p(t) = \frac{1}{\tau_p^2} \int_{-\infty}^t \exp\left[ \frac{s-t}{\tau_p} \right]
 \left[ \bm{u}(\bm{x}^p(t),t)-\bm{u}(\bm{x}^p(s),s) \right] ds \mathrm{.}
\end{equation}
Without loss of generality, we take $t=0$ and write the particle acceleration covariance
tensor $\langle \bm{a}^p(0) \bm{a}^p(0) \rangle$ as
\begin{equation}
\label{eq:acceleration_covariance}
\begin{split}
\langle \bm{a}^p(0) \bm{a}^p(0) \rangle = & 
\frac{1}{\tau_p^4} \left \langle \bm{u}(\bm{x}^p(0),0) \bm{u}(\bm{x}^p(0),0) 
\int_{-\infty}^0 \int_{-\infty}^0 
\exp\left[ \frac{s}{\tau_p} \right] \exp\left[ \frac{S}{\tau_p} \right] ds dS \right \rangle \\
& - \frac{2}{\tau_p^4} \left \langle \bm{u}(\bm{x}^p(0),0) 
\int_{-\infty}^0 \int_{-\infty}^0  
\exp\left[ \frac{s}{\tau_p} \right] \exp\left[ \frac{S}{\tau_p} \right] 
\bm{u}(\bm{x}^p(s),s)
ds dS \right \rangle \\
& + \frac{1}{\tau_p^4} \left \langle
\int_{-\infty}^0 \int_{-\infty}^0  
\exp\left[ \frac{s}{\tau_p} \right] \exp\left[ \frac{S}{\tau_p} \right] 
\bm{u}(\bm{x}^p(s),s) \bm{u}(\bm{x}^p(S),S) ds dS \right \rangle \mathrm{.}\\
 \end{split}
\end{equation}
By assuming that preferential concentration effects are weak, we can approximate the
first term on the right-hand-side of (\ref{eq:acceleration_covariance}) as
\begin{equation}
\label{eq:accel_first_term}
 \tau_p^{-2} \Big \langle \bm{u}(\bm{x}^p(0),0) 
 \bm{u}(\bm{x}^p(0),0) \Big \rangle \approx \frac{{u'}^2}{\tau_p^2} \bm{I} \mathrm{.}
\end{equation}
Based on PDF theory \citep{reeks93}, the second term 
on the right-hand-side of (\ref{eq:acceleration_covariance}) is
equivalent to the particle velocity covariance multiplied by $-2 \tau_p^{-2}$,
while the third term equivalent to the particle velocity covariance multiplied by $\tau_p^{-2}$.

For $Sv \gg u'/u_\eta$, we can simplify the relations for the velocity covariances
presented in \cite{wang93b} to obtain
\begin{equation}
 \frac{\langle v_1^p(t) v_1^p(t) \rangle}{u'^2} = \frac{2 + St Sv (\eta/\ell)}{2 \left[1 + St Sv (\eta/\ell)\right]^2} \mathrm{,}
\end{equation}
and
\begin{equation}
 \frac{\langle v_3^p(t) v_3^p(t) \rangle}{u'^2} = \frac{1}{1 + St Sv (\eta/\ell)} \mathrm{.}
\end{equation}

From these relations, we write the horizontal and vertical particle acceleration variances as
\begin{equation}
 \label{eq:accel_variance_perp}
 \frac{\langle a^2_1 \rangle^p}{a_\eta^2} = \left[ \frac{u'}{u_\eta} \right]^2
 \left[ \frac{Sv}{St} \right]
  \left[ \frac{2 St Sv + 3 \left( \frac{\ell}{\eta} \right)}
 {2 \left( St Sv + \frac{\ell}{\eta} \right)^2} \right] \mathrm{,}
\end{equation}
and
\begin{equation}
\label{eq:accel_variance_parallel}
 \frac{\langle a^2_3 \rangle^p}{a_\eta^2} =
 \left[ \frac{u'}{u_\eta} \right]^2 \left[ \frac{Sv}{St} \right]
 \left[ \frac{1}{St Sv + \frac{\ell}{\eta}} \right] \mathrm{,}
\end{equation}
respectively.
\setcounter{equation}{0}

\section{Kinematic collision kernel for an anisotropic particle phase}
\label{sec:collision_kernel_theory}

In this section, we extend the derivation of \cite{sundaram4} to determine
the kinematic collision kernel formulation for an anisotropic particle phase.
From \cite{sundaram4}, we write the average number of collisions per unit volume 
over a time period $\tau$ as
\begin{equation}
 Z_c(\tau) = \frac{n^2}{2} \int_{\bm{w}} \int_{\bm{r}}
 \boldsymbol{\Phi} (\bm{r},\bm{w},\tau) p (\bm{r}, \bm{w}) d\bm{r} d\bm{w} \mathrm{,}
\end{equation}
where $n$ is the particle-number density,
$\boldsymbol{\Phi} (\bm{r},\bm{w},\tau)$ is the collision operator,
and $p (\bm{r}, \bm{w})$ is the joint PDF of the particle-pair separation and relative velocity.
At this stage, it is simpler to leave the PDF as is rather than split it into a conditional 
average and the RDF as was done in \cite{sundaram4}. The collision kernel is given as
\begin{equation}
 K(d) = \frac{2}{n^2} \lim_{\tau \rightarrow 0} \frac{Z_c(\tau)}{\tau} = 
 \frac{2}{n^2} \frac{d}{d\tau} Z_c(0) \mathrm{,}
\end{equation}
or equivalently,
\begin{equation}
\begin{split}
 K(d) & = \int_{\bm{w}} \int_{\bm{r}} -\frac{\bm{r} \cdot \bm{w}}{r(0)} H(t^*) 
 \delta[ d-r(0) ] p (\bm{r}, \bm{w}) d \bm{r} d\bm{w} \\
 & = \int_{\bm{r}} -H(t^*) \delta[ d-r(0)] \frac{\bm{r}}{r(0)} \cdot
 \int_{\bm{w}} \bm{w} p(\bm{r},\bm{w}) d \bm{w} d \bm{r} \\
 & = \int_{\bm{r}} -H(t^*) \delta[ d-r(0)] \frac{\bm{r}}{r(0)} \cdot
 \varrho \langle \bm{w}^p(t) \rangle_{\bm{r}} d\bm{r} \bm{.}
\end{split}
\end{equation}
In the above equations, $t^*$ denotes the time required to reach the minimum particle separation,
$H$ is the Heaviside function, $\delta$ is the Dirac delta function, and $\varrho$ is the particle-pair PDF.

From the definition of $t^*$ in \cite{sundaram4}, we require $\bm{r} \cdot \bm{w} \leq 0$
for $H(t^*) \geq 0$. Since
\begin{equation}
 \bm{r} \cdot \langle \bm{w}^p(t) \rangle_{\bm{r}} 
 \equiv \langle \bm{r}^p (t) \cdot \bm{w}^p(t) \rangle_{\bm{r}} \mathrm{,}
\end{equation}
then in order to satisfy $H(t^*) \geq 0$, we replace
\begin{equation}
 \frac{\bm{r}}{r(0)} \cdot \varrho \langle \bm{w}^p(t) \rangle_{\bm{r}} \mathrm{,}
\end{equation}
with
\begin{equation}
 -\frac{\bm{r}}{r(0)} \cdot \varrho \langle \bm{w}^{p-}(t) \rangle_{\bm{r}} \mathrm{,}
\end{equation}
where
\begin{equation}
 \varrho \langle \bm{w}^{p-} (t) \rangle_{\bm{r}} \equiv -\int\limits_{\boldsymbol{-\infty}}^{\bm{0}}
 \bm{w} p(\bm{r},\bm{w}) d \bm{w} \mathrm{.}
\end{equation}
Furthermore, since
\begin{equation}
  \frac{\bm{r}}{r(0)} \cdot \varrho \langle \bm{w}^{p-}(t) \rangle_{\bm{r}} =
  \frac{1}{r(0)} \varrho \langle \bm{r}^p(t) \cdot \bm{w}^{p-}(t) \rangle_{\bm{r}} =
  \frac{r}{r(0)} \varrho S^p_{-\parallel}(\bm{r}) \mathrm{,}
\end{equation}
and we are considering the limit $\tau \rightarrow 0$ (in which $r(0) \rightarrow r$), we have
\begin{equation}
 \label{eq:Kd_cartesian}
 K(d) = \int_{\bm{r}} \delta[d-r] \varrho S^p_{-\parallel}(\bm{r}) d \bm{r} \mathrm{.}
\end{equation}
Since we are interested in collisions on a sphere, we write (\ref{eq:Kd_cartesian}) in spherical coordinates
by replacing $\bm{r}$ with $(r,\theta,\phi)$ and
$d\bm{r}$ with $r^2 \sin \theta d r d \theta d \phi$. 
This gives us
\begin{equation}
 \label{eq:Kd_spherical}
 \begin{split}
  K(d) & = \int_r \int_0^{2\pi} \int_0^\pi \delta[d-r] \varrho S^p_{-\parallel}(r,\theta,\phi)
         r^2 \sin \theta d \theta d \phi \mathrm{,} \\
         & = d^2 \int_0^{2\pi} \int_0^\pi g(d,\theta,\phi) S^p_{-\parallel}(d,\theta,\phi)
         \sin \theta d \theta d \phi \mathrm{.}
 \end{split}
\end{equation}
For an isotropic particle phase, we recover the expected result,
\begin{equation}
\label{eq:Kd_isotropic}
 K(d) = 4 \pi d^2 g(d) S^p_{-\parallel}(d) \mathrm{,}
\end{equation}
where
\begin{equation}
 g(d) = \frac{1}{4\pi} \int_0^{2\pi} \int_0^\pi g(d,\theta,\phi) \sin \theta d \theta d \phi \mathrm{,}
\end{equation}
and
\begin{equation}
 \label{eq:Kd_wr_isotropic}
 S^p_{-\parallel}(d) = \frac{1}{4\pi} \int_0^{2\pi} \int_0^\pi 
 S^p_{-\parallel}(d,\theta,\phi) \sin \theta d \theta d \phi \mathrm{.}
\end{equation}
For an anisotropic particle phase, however, (\ref{eq:Kd_wr_isotropic}) no longer holds.
Note that $S^p_{-\parallel}(d)$ is an average over particle pairs
on the entire surface of the sphere, while $S^p_{-\parallel}(d,\theta,\phi)$
is an average over particle pairs on a differential element of the sphere. 
When the particle phase is anisotropic, we must compute $S^p_{-\parallel}(d)$ by weighting each of the averages
on the differential elements $S^p_{-\parallel}(d,\theta,\phi)$
by the number of particle pairs contributing to that average.
In this case, we have
\begin{equation}
 \label{eq:Kd_wr_anisotropic}
 S^p_{-\parallel}(d) = \frac{1}{4\pi} \int_0^{2\pi} \int_0^\pi 
 S^p_{-\parallel}(d,\theta,\phi)
 \frac{g(d,\theta,\phi)}{g(d)}
 \sin \theta d \theta d \phi \mathrm{,}
\end{equation}
which reduces to (\ref{eq:Kd_wr_isotropic}) for an isotropic particle phase.
From (\ref{eq:Kd_wr_anisotropic}), we can therefore show that
\begin{equation}
 4 \pi d^2 g(d) S^p_{-\parallel}(d)
 = d^2 \int_0^{2\pi} \int_0^\pi 
  g(d,\theta,\phi) S^p_{-\parallel}(d,\theta,\phi) \sin \theta d \theta d \phi \mathrm{,}
\end{equation}
which is precisely the result we derived in (\ref{eq:Kd_spherical}) for the anisotropic collision
kernel. We have therefore demonstrated mathematically that (\ref{eq:Kd_isotropic}) holds for
both isotropic and anisotropic particle phases.
This explains the empirical observations in \cite{ayala08a}, from which the authors argue
that (\ref{eq:Kd_isotropic}) holds even when the particle distribution is anisotropic.

\bibliographystyle{jfm2}
\bibliography{refs}

\end{document}